\documentclass[12pt]{iopart}

\usepackage{graphicx}
\usepackage{iopams} 
\eqnobysec
\begin{document}
\date{\today}
\title[Nonlinear Aspects of the RG Flows of Dyson's Hierarchical Model]{Nonlinear Aspects of the Renormalization Group Flows of Dyson's Hierarchical Model}

\author{Y. Meurice\dag \S
\footnote[3]{To
whom correspondence should be addressed (yannick-meurice@uiowa.edu)}
}
\address{\dag\ Department of Physics and Astronomy\\ The University of Iowa\\
Iowa City, IA 52242 USA }
\address{\S\ Also at the Obermann Center for Advanced Study, University of Iowa}
\date{\today}




\begin{abstract}

We review recent results concerning the renormalization group (RG) transformation of Dyson's hierarchical model (HM). This model can be seen as an approximation of a scalar field theory on a lattice. 
We introduce the HM and show that its large group of symmetry simplifies drastically the blockspinning procedure.
Several equivalent forms of the recursion formula are presented with unified notations. Rigorous and numerical 
results concerning the recursion formula are summarized. It is pointed out that 
the recursion formula of the HM is inequivalent to both Wilson's approximate recursion formula and Polchinski's  equation in the local potential approximation (despite the very small difference with the exponents of the latter).
We draw a comparison between the RG of 
the HM and other RG equations in the local potential approximation.  
The construction of the linear and nonlinear scaling variables is discussed in an operational way. We describe 
the calculation of non-universal critical amplitudes in terms of the scaling variables of two fixed points. 
This question appears as a problem of interpolation 
between these fixed points. Universal amplitude ratios are calculated.
We discuss the large-$N$ limit and the complex singularities of the critical potential calculable in this limit. 
The interpolation between the HM and more conventional lattice models is presented as a symmetry breaking problem. We briefly 
introduce models with an approximate supersymmetry. One important goal of this review article is to present 
a configuration space counterpart, suitable for lattice formulations, of other RG 
equations formulated in momentum space (sometimes abbreviated as ERGE).

\end{abstract}


\newpage
\contentsline {section}{\numberline {1}Introduction}{6}
\contentsline {section}{\numberline {2}General and practical aspects of the RG method}{9}
\contentsline {subsection}{\numberline {2.1}Statement of the problem}{9}
\contentsline {subsection}{\numberline {2.2}Basic aspects of the RG method}{11}
\contentsline {subsection}{\numberline {2.3}Practical aspects of blockspinning}{14}
\contentsline {section}{\numberline {3}Dyson's hierarchical model}{15}
\contentsline {subsection}{\numberline {3.1}The Model}{15}
\contentsline {subsection}{\numberline {3.2}The RG transformation}{17}
\contentsline {subsection}{\numberline {3.3}The Gaussian (UV) fixed point}{18}
\contentsline {subsection}{\numberline {3.4}The HT fixed point}{20}
\contentsline {section}{\numberline {4}Equivalent forms of the recursion formulas}{20}
\contentsline {subsection}{\numberline {4.1}Baker's form}{20}
\contentsline {subsection}{\numberline {4.2}Gallavotti's form}{21}
\contentsline {subsection}{\numberline {4.3}Summary}{22}
\contentsline {section}{\numberline {5}Inequivalent extensions of the recursion formula}{23}
\contentsline {subsection}{\numberline {5.1}Relation with Wilson's approximate recursion formula}{23}
\contentsline {subsection}{\numberline {5.2}Gallavotti's recursion formula}{23}
\contentsline {section}{\numberline {6}Motivations, rigorous and numerical results}{24}
\contentsline {subsection}{\numberline {6.1}Motivations}{24}
\contentsline {subsection}{\numberline {6.2}Rigorous results}{25}
\contentsline {subsection}{\numberline {6.3}Numerical results}{26}
\contentsline {section}{\numberline {7}Numerical implementation}{26}
\contentsline {subsection}{\numberline {7.1}Polynomial truncations}{26}
\contentsline {subsection}{\numberline {7.2}Volume effects in the symmetric phase}{27}
\contentsline {subsection}{\numberline {7.3}The Eigenvalues of the Linearized RG Transformation}{28}
\contentsline {subsection}{\numberline {7.4}The critical potential}{29}
\contentsline {subsection}{\numberline {7.5}The low temperature phase}{30}
\contentsline {subsection}{\numberline {7.6}Practical aspects of the hierarchy problem}{32}
\contentsline {section}{\numberline {8}Perturbation theory with a large field cutoff}{33}
\contentsline {subsection}{\numberline {8.1}Feynman rules and numerical perturbation theory}{33}
\contentsline {subsection}{\numberline {8.2}Perturbation theory with a large field cutoff}{33}
\contentsline {subsection}{\numberline {8.3}Improved perturbative methods}{35}
\contentsline {subsection}{\numberline {8.4}Large field cutoff in ERGE}{35}
\contentsline {section}{\numberline {9}Relation with the ERGE in the LPA approximation}{36}
\contentsline {subsection}{\numberline {9.1}Polchinski equation in the LPA}{36}
\contentsline {subsection}{\numberline {9.2}Infinitesimal form of Gallavotti's recursion formula}{36}
\contentsline {subsection}{\numberline {9.3}The critical exponents of Polchinski's equation}{37}
\contentsline {subsection}{\numberline {9.4}Infinitesimal form of Wilson approximate recursion formula}{38}
\contentsline {subsection}{\numberline {9.5}Finite time singularities}{38}
\contentsline {subsection}{\numberline {9.6}Improvement of the LPA}{39}
\contentsline {section}{\numberline {10}The nonlinear scaling fields}{39}
\contentsline {subsection}{\numberline {10.1}General ideas and definitions}{39}
\contentsline {subsection}{\numberline {10.2}The small denominator problem}{39}
\contentsline {subsection}{\numberline {10.3}The linear scaling variables of the HT fixed point}{40}
\contentsline {subsection}{\numberline {10.4}The nonlinear scaling variables of the HT fixed point}{41}
\contentsline {subsection}{\numberline {10.5}Argument for the cancellation to all orders}{43}
\contentsline {subsection}{\numberline {10.6}The nonlinear scaling variables of the nontrivial fixed point}{46}
\contentsline {subsection}{\numberline {10.7}Convergence issues}{47}
\contentsline {subsection}{\numberline {10.8}The scaling variables of the Gaussian fixed point}{47}
\contentsline {section}{\numberline {11}Interpolation between fixed points and critical amplitudes}{48}
\contentsline {subsection}{\numberline {11.1}Global RG flows}{48}
\contentsline {subsection}{\numberline {11.2}Critical amplitudes and RG invariants}{48}
\contentsline {subsection}{\numberline {11.3}Overlapping regions of convergence}{50}
\contentsline {subsection}{\numberline {11.4}Approximately universal ratios of amplitudes}{51}
\contentsline {subsection}{\numberline {11.5}More about log-periodic corrections}{51}
\contentsline {section}{\numberline {12}Nontrivial Continuum limits}{53}
\contentsline {subsection}{\numberline {12.1}The infinite cutoff limit}{53}
\contentsline {subsection}{\numberline {12.2}Numerical estimates of the universal ratios}{54}
\contentsline {subsection}{\numberline {12.3}Other universal ratios}{55}
\contentsline {subsection}{\numberline {12.4}The critical potential of the symmetric phase}{56}
\contentsline {section}{\numberline {13}The large-N limit}{57}
\contentsline {subsection}{\numberline {13.1}Calculations at finite $N$}{57}
\contentsline {subsection}{\numberline {13.2}Ma's equation}{58}
\contentsline {subsection}{\numberline {13.3}Singularities of the critical potential $U_0^{\star }$}{61}
\contentsline {subsection}{\numberline {13.4}Open problems}{62}
\contentsline {section}{\numberline {14}The improvement of the hierarchical approximation}{62}
\contentsline {subsection}{\numberline {14.1}Scalar models on ultrametric spaces}{62}
\contentsline {subsection}{\numberline {14.2}Improvement of the hierarchical approximation}{65}
\contentsline {subsection}{\numberline {14.3}The Hierarchical Approximation and its Systematic Improvement}{67}
\contentsline {subsection}{\numberline {14.4}The Improvement of the Hierarchical Approximation As a Symmetry Breaking Problem}{68}
\contentsline {subsection}{\numberline {14.5}Other applications}{69}
\contentsline {section}{\numberline {15}Models with approximate supersymmetry}{69}
\contentsline {section}{\numberline {16}Conclusions}{73}

\newpage
\section*{Frequently used abbreviations}

$\ $

RG: Renormalization Group

ERGE: Exact Renormalization Group Equations

LPA: Local Potential Approximation

HM: (Dyson's) Hierarchical Model

HT: High-Temperature

UV: Ultra-Violet

IR: Infra-Red

\def\scale{\ell}
$\scale$: the change in linear scale after one RG transformation

\def\mainf{\Phi}
$\mainf$: the sum of all the fields

\def\nsites{\Omega}
$\nsites$: the total number of sites

\def\magden{\phi _c}
$\magden$: the classical field or magnetization density

\def\phin{\phi_n} 
$\phin$: the sum of the fields in a block of size $2^n$

\def\myw{{\cal W}}
\def\kw{{\cal H}}
\def\gal{{\cal G}}

$W_n$ the ``raw'' local measure after blockspinning $2^n$ sites

$\myw _n$: the rescaled local measure after $n$ RG transformations

$\kw _n$: Baker's form of $\myw _n$

$\gal_n$: the Gallavotti's form of $\myw _n$

$c\equiv 2^{1-2/D}$

\newpage

\section{Introduction}

Quantum field theorists face the arduous task of figuring out the large scale implications of 
models defined by interactions at a small scale. In general, the large distance behavior of the theory 
can be encoded in an effective action $S_{eff}$ which describes the interactions of the zero-momentum 
modes of the fields. 
Knowing $S_{eff}$, we can answer important questions regarding, for instance, the stability and triviality of the theory.
The renormalization group (RG) method \cite{wilson71a,wilson71bone,wilson71b,wilson74} was designed to 
calculate $S_{eff}$ by a sequence of small steps where the high energy modes are integrated progressively.
This procedure generates a sequence of $S_{eff,\Lambda}$ where $\Lambda$ is the scale above which we have integrated 
all the modes. In the process, many new interactions are introduced and with them functions also evolving with $\Lambda$. This can be seen as a flow in the space of theories. 
We call this type of flows the RG flows. 
Important simplifications may occur near fixed 
points that have only a few unstable directions and universal properties may emerge.
Despite the conceptual beauty of the construction, the practical calculation of the RG flows remains very difficult. 
The behavior near fixed points can usually be handled by some expansion (weak coupling, strong coupling, etc...), but interpolating between two fixed points is in general a major difficulty. Unfortunately, this is essential 
to calculate $S_{eff}$. 

In the following, we review our present understanding of the interpolation between fixed points for Dyson's hierarchical model \cite{dyson69,baker72}, a model for which 
the calculation of $S_{eff,\Lambda}$ can be reduced to the calculation of the effective potential $V_{eff,\Lambda}$ and can performed numerically with great accuracy. 
Other type of hierarchical models have been discussed in the literature (see, for instance, reference \cite{kaufman81}), however, in the following, 
we only consider Dyson's model. The Hierarchical Model (HM) will only refer to Dyson's model in this review. 
The RG transformation for the HM can be expressed as 
a simple one-variable integral equation very similar to the so-called approximate recursion formula proposed 
by Wilson \cite{wilson71b}. In many respects, the techniques involved in the 
solution of this model can be compared to those used in elementary quantum mechanics. 
The reason for this remarkable simplicity is that the kinetic term of the model is not renormalized. 
In other words, it is a model for which the Local Potential Approximation (LPA) is exact.

In recent years, the LPA has been widely used in the context of  
Exact Renormalization Group Equations (ERGE)  and has generated a lot of interest.
We recommend references \cite{bagnuls00,berges00,pawlowski05} for reviews of the recent progress. ERGE allow in principle 
the study of global and nonlinear aspects of the RG flows of field theoretical or statistical models.
However, truncation methods are necessary 
in order to make practical calculations. One particularly popular choice is to combine 
Polchinski's ERGE \cite{polchinski84} with the  
LPA \cite{hasenfratz86}. This results in a simple partial differential equation (called ``Polchinski's equation'' below) for the 
effective potential.  Polchinski's equation can also be obtained \cite{felder87} as an infinitesimal 
version of the RG equation for the HM. 
This suggests that the linearized theories should be close to each other. Accurate calculations of the critical exponents \cite{gottker99,litim00,oktayphd,bervillier04,on06,bervillier07} show that the exponents differ only in the fifth significant digit. In reference \cite{on06}, it was believed that the 
two exponents should coincide, however this is not the case (this will be explained 
in section \ref{sec:erge}).

Originally, the HM was introduced by Dyson \cite{dyson69} as a long-range ferromagnetic Ising model with 
couplings weaker than the one-dimensional Ising model with long-range couplings 
falling off like a power. The existence of a phase transition and of the infinite 
volume limit for some range of the parameter controlling the decay of the interactions,  
can be proved rigorously.
Historical motivations and rigorous results are discussed in more 
detail in section \ref{sec:motrig}. The model was rediscovered a few years later by 
Baker \cite{baker72} in an attempt to construct models for which Wilson's approximate 
recursion formula or an integral equation of the same form become the exact RG transformation. In this context, the HM appears rather like an approximation of 
a scalar field theory on a $D$-dimensional lattice. Later, we call the process of  
interpolation between the HM and lattice models with nearest neighbor interaction 
the ``improvement of the hierarchical approximation''.
 
One important goal of the review article is to stress the similarities between Polchinski equation 
(a generic computational tool) and  the HM (often 
perceived as a toy model). The motivations for using the HM model are:
\begin{itemize}
\item
It is a lattice model right from the beginning. 
\item
High-accuracy methods exist to calculate numerically the critical exponents and the RG flows for arbitrary potentials. 
\item
Conventional expansions (weak and strong coupling, $\epsilon$-expansion) 
can be implemented easily and large order series can be obtained.
\item
The hierarchical approximation can in principle be improved \cite{meurice93,marseille93}.
\end{itemize}
The study of critical phenomena has reached a stage where many methods have been 
refined to a point where they provide numbers very close to each other \cite{kbook,pelissetto00b}. In the case of the HM, all the approximations can be compared to very accurate numerical answers. The ability to construct the RG flows very 
accurately means that we can study general features of these flows far away from fixed points. This type of study is also possible using the LPA \cite{bagnuls90,bagnuls97,bagnuls00,bagnuls00b}.

An interesting feature of the HM is its discrete scale invariance which, 
depending on the context, can be seen either as  
an annoyance \cite{osc1,osc2} or an interesting intrinsic property \cite{jogi98}. 
The review is focused on doing calculations directly in 3 dimensions. The $\epsilon$-expansion 
near 4 dimensions,which is reviewed in reference \cite{collet78}, is not discussed here. 

The paper is organized as follows. In section \ref{sec:block}, we briefly review the basic concepts of the 
RG method, the scaling hypothesis and the practical difficulties of blockspinning. This section motivates approximations that 
simplify the blockspinning procedure.
Dyson's HM is introduced in section \ref{sec:model}, 
with the notation used later in the review. 
Several equivalent forms of the recursion formula are presented in section \ref{sec:recursion}. 
More general recursion formulas which coincide with the HM's one for a particular choice of what we call the scale parameter,  
are introduced in section 
\ref{sec:inequivalent}. This includes Wilson's approximate recursion formula \cite{wilson71b} which is inequivalent 
to the recursion formula of the HM. In the following, we denote the scale parameter $\scale$. In the literature, the same parameter is 
often denoted $L$, but we preferred to keep that symbol for the linear size of the whole system. 

Having introduced the basic concepts, motivations for the HM, numerical and rigorous results are reviewed in section \ref{sec:motrig}.  
The numerical treatment of the recursion formula is discussed in section \ref{sec:num} and perturbation theory with a large field cutoff in section \ref{sec:cutpt}. 
The relation to Polchinski's ERGE in the LPA is 
discussed in section \ref{sec:erge}. 

We introduce the non-linear scaling variables associated with a fixed point in section \ref{sec:scaling}. 
These quantities, originally introduced by Wegner \cite{wegner72}, transform multiplicatively under a RG transformation. They have features similar to the action-angle variables used in classical mechanics.  Small-denominator 
problems are in general present and need to be discussed in each case. In section \ref{sec:global}, we describe the non-universal critical amplitudes 
as RG invariants made out of two non-linear scaling fields. A practical calculation based on this method requires 
the ability to use both sets of scaling fields in a common intermediate region, in other words, to 
interpolate between the fixed points. 

The notion of a nontrivial continuum limit, originally introduced by Wilson \cite{wilson72} is reviewed and applied to the HM in section \ref{sec:nontrivial}. Calculations of critical 
amplitudes and their universal ratios are then discussed. In section \ref{sec:largen}, we introduce the extension of the HM for $N$ components and discuss the large-$N$ limit. We compare the results with those obtained with 
Polchinski equation emphasizing the difference in the 5th digit already mentioned above, which reflects the non-equivalence of the models.

The possibility of improving 
the hierarchical approximation by breaking its symmetries in a systematic way is discussed in section \ref{sec:imp}. 
Compared to the improvement of the LPA by the derivative expansion, the improvement of the hierarchical approximation is an underdeveloped subject.
On the other hand, it is clear that much progress remains to be done in the ERGE approach in order to match the accuracy of other methods \cite{brez,zinnjustinbook,pelissetto00b} for the calculation of the critical exponents \cite{cberge}. We hope that this review 
will facilitate the communication between the two approaches. 
Reference \cite{litim07} is a  very recent example of progress made in this direction.
Finally, the possibility of including fermions in approximately supersymmetric models is briefly 
discussed in section \ref{sec:susy}. 

One motivation to study global and nonlinear aspects of RG flows not covered in this review is to improve our understanding 
of quantum chromodynamics. In this theory of strongly interacting quarks and gluons, weakly interacting particles  are seen at short distance (asymptotic freedom), while nonperturbative effects cause the confinement 
of quarks and gluons at large distance. Understanding how these two behaviors can 
be smoothly connected is a single theory amounts to interpolate between two fixed points of the Renormalization Group 
(RG) transformation. Despite recent progress \cite{litim98,tomboulis03,pawlowski03,tomboulis05,pawlowski05,tomboulis06,dlerge} it remains a challenge to understand confinement in terms of weak coupling variables. This is work for the future.

\section{General and practical aspects of the RG method}
\label{sec:block}
In this section, we present the basic ideas behind the RG method and point out the practical difficulties associated 
with the so-called blockspin method. 
\subsection{Statement of the problem}

We are interested in the large distance (low momentum, long wavelength) behavior of scalar models. 
We consider a generic scalar model with a lattice regularization and a action $S$. The scalar field $\phi_x$ is 
coupled linearly to a constant source $J$. We call the total field 
\begin{equation}
\mainf=\sum_x\phi_x\ ,
\end{equation}
and the total number of sites $\nsites$. 
With these notations, the partition function reads 
\begin{equation}
	Z[J]=\int_{-\infty}^{+\infty}\dots\int_{-\infty}^{+\infty} \prod _x d\phi_x \exp(-S+\mainf J)\ ,
\end{equation}
We define the average at $J=0$ of an arbitrary function $A$ of the fields as
\begin{equation}
	<A>=\int_{-\infty}^{+\infty} \dots \int_{-\infty}^{+\infty} \prod _x d\phi_x A \exp(-S)/Z[0]
\end{equation}
It is clear that 
\begin{equation}
	Z[J]/Z[0]=1+\sum_{q=1}^\infty \frac{1}{q!}J^q <\mainf ^q>
\end{equation}
The connected parts can be obtained by taking the logarithm of this expression. 
At the lowest orders,  
\begin{eqnarray}
\label{eq:connected}
	<\mainf^2>^{c}&=&<\mainf^2>-(<\mainf>)^2\cr
	<\mainf^3>^{c}&=&<\mainf^3>-3<\mainf^2><\mainf> +2(<\mainf>)^3\cr
	<\mainf^4>^{c}&=&<\mainf^4>-4<\mainf^3><\mainf>-3(<\mainf^2>)^2\cr &\ &+12 <\mainf^2>(<\mainf>)^2-6(<\mainf>)^4
\end{eqnarray}
In general, we expect that 
\begin{equation}
	<\mainf^q>^{c}\propto \nsites \ ,
\end{equation}
and we define 
\begin{equation}
\label{eq:susceptibility}
	\chi^{(q)}=	<\mainf^q>^{c}/\nsites \
\end{equation}
even though the individual terms scale faster than $\Omega$. 
Unless we take the infinite volume limit, we should in principle write $\chi^{(q)}_{\Omega}$ in order to remind the 
dependence on the volume.
In the rest of this subsection, this dependence will be kept implicit. 
We then write the generating function of the connected densities
\begin{equation}
(1/\nsites) {\rm ln}(Z[J]/Z[0])=	\sum_{q=1}^\infty \frac{1}{q!} J^q \chi^{(q)}
\label{eq:conn}
\end{equation}
It is common to call
\begin{equation}
	G[J]\equiv-(1/\nsites) {\rm ln}(Z[J]/Z[0])\ ,
\end{equation}
the Gibbs potential and it is clear that 
\begin{equation}
	\chi^{(q)}=-\partial^q G[J]/\partial J^q	. 
\end{equation}

We now introduce 1 in the functional integral in the following way 
\begin{equation}
	1=\int_{-\infty}^{+\infty} d\magden \delta(\magden - \mainf/\nsites)\ .
\end{equation}
The partition function becomes \cite{fukuda74} 
\begin{equation}
	Z[J]=\int_{-\infty}^{+\infty} d\magden \exp\Bigg(-\Omega\Big( V_{eff}(\magden) -\magden J\Big)\Bigg)\ ,
\end{equation}
with 
\begin{equation}
\exp(-\Omega V_{eff}(\magden))\equiv \int_{-\infty}^{+\infty}\dots \int_{-\infty}^{+\infty}\prod _x d\phi_x \exp(-S)\delta(\magden - \mainf/\nsites)\ .
\end{equation}
If for any $J$, $V_{eff}(\magden) -\magden J$ has a unique minimum at $\magden =\bar{\magden}$, we have in the limit of arbitrarily large $\nsites$, 
that 
\begin{equation}
	Z[J]\propto\exp\Bigg(-\Omega\Big(V_{eff}(\bar{\magden}) -\bar{\magden} J\Big)\Bigg)\ .
\end{equation}
$\bar{\magden}$ is a function of $J$ defined implicitly by 
\begin{equation}
\partial V_{eff}(\magden)/\partial \phi_c|_{\magden =\bar{\magden}}=J\ .
\label{eq:jofphi}
\end{equation}
$G[J]$ and $V_{eff}(\magden)$ are related by a Legendre transform 
\begin{equation}
G[J]=	V_{eff}(\bar{\magden}) -\bar{\magden} J
\label{eq:gibbs}
\end{equation}
In the ferromagnetic language, $\phi_c$ is called the magnetization and $J$ the magnetic 
field. 
In analogy with the gas-liquid transition, $V_{eff}$ plays the role of the Helmholtz potential, $\bar{\phi_c}$  
the role of minus the volume and $J$ the role of the pressure. 
When we are in the gas phase below the critical 
temperature (but above the triple point temperature), if we increase the pressure keeping the temperature fixed, the volume decreases until a critical pressure is reached 
where the gas and the liquid can coexist at an equilibrium pressure but with different specific volumes. 
The discontinuity in the magnetization is analog to the specific volume discontinuity. 
The sign is justified by the fact that as we increase $J$, the magnetization increases, but if we increase the 
pressure, the volume decreases. 

In the simple case where $V_{eff}$ is an even function with a unique minimum at zero, we can expand 
\begin{equation}
	V_{eff}(\magden)=\sum_{q=1}^\infty \frac{1}{2q!} \magden^{2q} \Gamma^{(2q)}
\end{equation}
Using the derivative of this expansion to express $J$ in terms of $\bar{\magden}$, plugging into 
equation (\ref{eq:gibbs}), using equation (\ref{eq:conn}) and solving order by order in $\bar{\magden}$, we 
obtain the well-known relations
\begin{eqnarray}
\Gamma^{(2)}&=&1/\chi^{(2)}\ ,\cr
\Gamma^{(4)}&=&-\chi^{(4)}/(\chi^{(2)})^4\ , \cr	
\Gamma^{(6)}&=&10(\chi^{(4)})^2/(\chi^{(2)})^7
-\chi^{(6)}/(\chi^{(2)})^6\ , 
\end{eqnarray}
and so on. The calculation of the $V_{eff}$ is an important objective the RG method. 
\subsection{Basic aspects of the RG method}
\label{subsec:basic}
We now 
consider a scalar model on a 
$D$-dimensional cubic lattice with a lattice spacing $a$. The RG transformation proceeds in two steps. First, we 
integrate the $\scale^D$ fields in blocks of linear size $\scale a$ while keeping
the sum of the fields in the block constant.
We then divide the sums of the fields by a factor 
$\scale^{(2+D-\eta)/2}$ and treat them as our new field variables. 
The exponent $\eta$ is introduced in order to keep the canonical form of the kinetic term and its calculation is non-trivial.  
We then obtain a new theory in terms of a new field variable which is equivalent to 
the previous one as long as we only consider processes involving energies smaller than the new ultraviolet cutoff 
$\sim 1/(\scale a)$. 
Given an original action, we assume that the RG transformation provides a new effective action expressed in 
terms of the new field variable. We postpone the discussion of the practical aspects of the partial integration to subsection \ref{subsec:practical}. 

The information that is kept during the RG transformation is encoded in the average values of all
the integer powers of the sum of the fields in the blocks. We call these
average values  
the ``zero momentum Green's functions at finite volume''. 
This set of values can be thought of as 
an element of an infinite vector space. In the following we call the trajectories in this space the RG flows. 
If we start with an infinite volume lattice, the RG transformation can in principle be 
repeated an infinite number of times. If this can be done, the resulting effective theory is limited to the zero-momentum 
Green's functions. We are particularly interested in finding the fixed points of the RG transformation, since the above limit can be simplified in the neighborhood of a fixed point.

The RG transformation has several obvious fixed points. If the interactions are limited to quadratic ones and there is no restriction on the range of the fields, the partition 
function can be calculated exactly using diagonalization and Gaussian integration. The model describes non-interacting particles of a 
given mass. We call this mass in cutoff units $m$. It can be determined in cutoff units using the zero-momentum two point function.

After one RG transformation, the mass for the new effective theory is $m\scale$. The Gaussian fixed point (also called the trivial fixed point) 
corresponds to the theory 
with $m=0$. There are usually other fixed points which corresponds to stable phases. For instance, the high-temperature fixed point which can be thought as an infinitely massive theory where the fluctuations about the 
zero field value are entirely suppressed. 

In addition to these obvious fixed points, we expect non-trivial fixed points when $D<4$. These are characterized by 
one or more unstable direction.
The critical hypersurface is given as the stable manifold (e.g. the basin of 
attraction) of this
non-trivial fixed point. 
Its codimension is the number of unstable directions. 
In the rest of this section, we specialize the discussion to the case where there is 
only one unstable direction. 
The stable manifold can then be reached by considering 
a family of models indexed by a parameter which can be tuned in order
to cross the stable manifold. In field theory context, we usually pick 
the bare mass to accomplish this purpose. In the statistical mechanics 
formulation, the inverse temperature $\beta$ can be tuned to its 
critical value $\beta_c$ which is a function of the other interactions. 

Near the non-trivial fixed point, we can use the 
eigenvectors of the linearized RG transformation as a basis. 
As far as we are close to the fixed point, the 
average values of the powers of the {\it rescaled} total field stay 
approximately unchanged after one transformation. 
However at each iteration, the components in the eigendirections are 
multiplied by the corresponding eigenvalue. In the following, we 
denote the eigenvalue larger than 1 as $\lambda_1$. As we assumed that there is only one unstable direction, there is 
only one such eigenvalue and it is the largest one. 

After repeating the renormalization group transformation $n$ times,
we have replaced $\scale^{Dn}$ sites by one site
and associated a block variable with it. We define the finite volume susceptibility 
$\chi_{_n}^{(2)}$ as the average value of the square of the sum of all the 
(unrescaled) fields inside the block
divided by the number of sites $\scale^{Dn}$. We can estimate $\chi_{_n}^{(2)}$ near the non-trivial 
fixed point. The average of the square of the rescaled variables is approximately a constant that 
we call $K_1$. To get the susceptibility, we need to go back to the original variables (so we multiply $K_1$ 
by $\scale^{(2+D-\eta)n}$) and divide by the volume $\scale^{Dn}$. We can also take into account the motion along 
the unstable direction in the linear approximation. In summary, near the fixed point, 
\begin{equation}
\chi_n^{(2)}\simeq \scale^{n(2-\eta)}(K_1+K_2\lambda_1^n(\beta_c-\beta))\ .
\label{eq:chi}
\end{equation}
The constant $K_2$ depends on the way the critical 
hypersurface is approached when $\beta$ is varied close to $\beta_c$.
equation (\ref{eq:chi}) is valid only if the linearization procedure is 
applicable, in other  words if $\lambda_1^n(\beta_c-\beta)<<1$. 
On the other hand, when $n$ reaches some critical value $n^\star$ such that
\begin{equation}
\lambda_1^{n^\star}(\beta_c-\beta)\simeq 1\ ,
\label{eq:nstar}
\end{equation}
non-linear effects become important
and the sign of $(\beta_c-\beta)$ becomes important. 
In the following we consider the case of the 
symmetric phase ($\beta<\beta_c$) which is simpler.
The order of magnitude of $\chi$ starts stabilizing when $n$ gets 
of the order of $n^\star$ and 
\begin{equation}
\chi_{\infty}^{(2)}\approx \scale^{n^{\star}(2-\eta)}\approx (\beta_c-\beta)^{-\gamma}\ ,
\end{equation}
with
\begin{equation}
\gamma=(2-\eta)\nu , 
\end {equation}
and 
\begin{equation}
\nu ={\rm ln}\scale/{\rm ln}\lambda_1 \ .
\end{equation}
When $n >> n^\star$, the trajectories fall into the completely 
attractive HT fixed point. The approach  of the fixed point is characterized by 
corrections proportional to negative powers of the volume.

We can also take into account the corrections due to the so-called irrelevant directions which correspond to 
eigenvalues less than 1 and eigenvectors along the stable manifold. We call the largest of the irrelevant eigenvalues $\lambda _2<1$. As  
$n$ increases, the distance along the corresponding eigenvector shrinks as $\lambda_2^n$. The relative size of the correction should be proportional to 
\begin{equation}
	\lambda_2^{n^\star}\approx(\beta_c-\beta )^{\Delta_s} \ ,
\end{equation}
with 
\begin{equation}
\Delta_s=-{\rm ln}\lambda_2/{\rm ln}\lambda_1 \ .
\end{equation}
$\Delta_s$ has a lower index $s$ which is short for subleading and 
should not be confused with the gap exponent that will be denoted $\Delta_g$.
In summary, in the large volume limit and for $\beta \rightarrow \beta_c^-$, we have the parametrization
\begin{equation}
\label{eq:suspar}
\chi ^{(2)}\simeq ( \beta_c-\beta )^{-\gamma} [ A_0+A_1
(\beta_c-\beta )^{\Delta_s} +\dots ].  
\end{equation}
Linearized calculations near the non-trivial fixed point allow us to calculate the exponents $\gamma$ and $\Delta_s$, 
but not the amplitudes $A_0$ and $A_1$.

It should be emphasized that often, the lattice spacing does not appear explicitly but 
is considered as a function of the bare parameters. The relation between the two is usually established by 
relating lengths calculated in lattice spacing to physical lengths. The continuum limit consist in taking a 
trajectory in the space of bare parameters which corresponds to the limit of zero lattice spacing.

When $\beta > \beta_c$, we are in the low temperature phase and $\chi ^{(1)}\neq 0$. The sign of $\chi ^{(1)}\neq 0$ 
is positive (negative) if we take the limit $J\rightarrow 0$ by positive (negative) values. Additional subtractions are then required as shown in equation (\ref{eq:connected}). The case of the low temperature will be discussed in subsection 
\ref{subsec:lt} for the HM.

Relations among exponents can be obtained from the scaling hypothesis  \cite{wegner76,niemeijer76,cardy96}. 
Making explicit the dependence of the non-analytical part of the Gibbs potential, denoted $G_s$, on the reduced temperature $t\equiv(\beta_c/\beta)-1$, the scaling hypothesis amounts to have 
\def\ye{{\mathcal Y}}
\begin{equation}
\label{eq:scalinghyp}
G_s[\scale ^{\ye_t}t,\scale ^{\ye_h}J]=\scale ^D G_s[t,J]\ .
\end{equation}
The exponents $\ye$ are connected to the critical exponents discussed above by the 
relations 
\begin{eqnarray}
\label{eq:scalingexp}
	\ye_t&=&\frac{1}{\nu} \nonumber \\
	   &\ & \ \\ 
	\ye_h&=&\frac{D+2-\eta}{2} \nonumber \ .
\end{eqnarray}
It is also common to introduce exponents which describe the correlations as a function 
of the inverse distance for 
the conjugated variables at criticality, namely 
\begin{eqnarray}
x_h &\equiv& D-\ye_h \\
x_t &\equiv& D-\ye_t \ ,
\end{eqnarray} 
for the field and for the energy respectively. 
The scaling hypothesis will be further discussed in subsections \ref{subseq:uv}, \ref{subsec:critical} and 
\ref{subsec:lt}. 

\subsection{Practical aspects of blockspinning}
\label{subsec:practical}
In the previous subsection, we have assumed that it was possible to integrate over the variables in a block while keeping the sum of the field constant. This procedure is usually called ``blockspinning'', an idea which can be traced back to Kadanoff \cite{kadanoff66}.  In practice, blockspinning is usually quite complicated. However, for models with actions quadratic in the 
fields, it is possible to do it analytically. Of course, for such models, all the correlation functions 
can be calculated exactly, but the procedure can be considered as the first step in a perturbative expansion. 
To fix the ideas, we can consider a one-dimensional lattice model where the Fourier transform of the 
two point function (the propagator) is $G(k)$. Assuming that the lattice sites are labeled by integers, we have the periodicity $G(k+2\pi)=G(k)$. If we now partition the lattice into blocks of even-odd pairs  of 
neighbor sites and blockspin 
within these blocks, we obtain a new two-point function  on a new lattice with a lattice spacing twice larger. 
If we denote the Fourier transform of the two-point function after $n$ steps $G_n(k)$, the iteration formula is: 
\begin{equation}
G_{n+1}(k)=(1+{\rm cos}(k/2))G_n(k/2)+(1-{\rm cos}(k/2))G_n(k/2	+\pi)\ .
\end{equation}
It is clear that the $2\pi$ periodicity is preserved and no sharp edges are introduced at least in a finite number of iterations. The construction can be extended in arbitrary dimensions and was used as the starting point for the 
finite-lattice approximation \cite{bell75}. 

In order to get an idea of the difficulty to extend the procedure when higher order interactions are introduced, 
it is instructive to consider the simple case of a 4-sites ring with nearest-neighbor quadratic interactions 
and local quartic interactions. 
\begin{equation}
Z=\int_{-\infty}^{+\infty}\dots\int_{-\infty}^{+\infty} d\phi_1 d\phi_2 d\phi_3 d\phi_4 {\rm e}^{-S_2^{A}-S_2^{B}-S_4}\ ,
\end{equation}
with nearest neighbor quadratic terms
\begin{equation}
S_2^{A}=(\phi_1-\phi_2)^2+	(\phi_3-\phi_4)^2\ ,
\end{equation}
and 
\begin{equation}
S_2^{B}=(\phi_2-\phi_3)^2+(\phi_4-\phi_1)^2 \ ,
\end{equation}
and local quartic interactions
\begin{equation}
S_4=\phi^4_1 + 	\phi^4_2 + \phi^4_3 + \phi^4_4 \ .
\end{equation}
For further convenience, we also define the next to nearest neighbor quadratic interactions
\begin{equation}
S_2^{NNN}=(\phi_1-\phi_3)^2+(\phi_2-\phi_4)^2
\end{equation}

We now try to blockspin. We pick (1,2) and (3,4) as our basic blocks and then combine them into the block containing all the sites. More specifically, it consists 
in introducing 1 in the integral in the following form:
\begin{eqnarray}
1= \int_{-\infty}^{+\infty} \int_{-\infty}^{+\infty}\int_{-\infty}^{+\infty}&\ & d\Phi d\Phi_{(1,2)} d\Phi_{(3,4)}
\delta(\Phi-\Phi_{(1,2)}-\Phi_{(3,4)})\nonumber\\
&\times&\delta(\Phi_{(1,2)}-\phi_1-\phi_2)\delta(\Phi_{(3,4)}-\phi_3-\phi_4)
\end{eqnarray}
If we can perform the integration over $\phi_1-\phi_2$ and $\phi_3-\phi_4$, and then over $\Phi_1-\Phi_2$, 
we will be able to write
\begin{equation}
Z=\int_{-\infty}^{+\infty} d\Phi {\rm e}^{-S_{eff.}(\Phi)} \ .
\end{equation}
In this illustrative model, $S_{eff.}(\Phi)$ is the main quantity of interest. 
Because of the quadratic terms $S_2^{B}$, we cannot 
perform the integral over $\phi_1-\phi_2$ independently of the integral over $\phi_3-\phi_4$ and blockspinning is potentially more difficult than evaluating the integral without intermediate steps. 
It is interesting to notice that the terms of $S_2^{B}$ connect the fields across the blocks and are not invariant 
under independent interchanges $(1\leftrightarrow2)$ and $(3\leftrightarrow4)$. These two transformations are of order 2 and commute, they generate a group of order 4. If we average $S_2^{B}$ over the 4 elements $g$ of this group, 
we obtain 
\begin{equation}
(1/4)\sum_g S_2^{B}(g)=(1/2)(S_2^{B}+S_2^{NNN})
\end{equation}
If we replace $S_2^{B}$ by this average, we have twice more terms but with half of the strength. 
Also, the new terms have a longer range. At first sight, it is not clear that the situation is better 
than before. However, if we combine this average with one half of the harmless $S_2^{A}$, which is 
invariant under the above mentioned symmetry group, we obtain 
\begin{equation}
(1/2)(S_2^{A}+S_2^{B}+S_2^{NNN})=4(\phi_1^2 +\phi_2^2 +\phi_3^2 +\phi_4^2 )-(\phi_1 +\phi_2 +\phi_3 +\phi_4 )^2\ .
\end{equation}
The first term affects only the local measure and the second can be incorporated directly into $S_{eff.}(\Phi)$.
This simple example illustrates how a  symmetry can be used as a guide to build a (modified) model where blockspinning is feasible.

Dyson's hierarchical model is a model where a hierarchical exchange symmetry among the sites is built-in and allows 
to blockspin the partition function by performing a sequence of one-dimensional integrals. 

\section{Dyson's hierarchical model}
\label{sec:model}
\subsection{The Model}
\label{subsec:dyson}

In this subsection, we describe Dyson's hierarchical model with the notations used in most of the 
rest of this review. The relationship  between this formulation and other ones found in the 
literature is discussed in sections \ref{sec:recursion} and \ref{sec:inequivalent}. 
The model requires $2^{n_{max}}$ sites. We label the sites with $n_{max}$
indices $x_{n_{max}}, ... , x_1$, each index being 0 or 1. We divide the $2^{n_{max}}$ sites into
two blocks, each containing $2^{n_{max}-1}$ sites. If $x_{n_{max}}=0$,
the site is in the first block, if $x_{n_{max}} = 1$, the site is in the
second block. Repeating this procedure $n$ times (for the two blocks,
their respective two sub-blocks , etc.), we obtain an unambiguous
labeling for each of the sites. The indices on the left provide the coarser division while the indices on the 
right provide a finer division. With an appropriate choice of origin, the indices can be interpreted as the 
binary representation of the site numbers. 
This is represented graphically in figure \ref{fig:boxes}:
%
%
%

\begin{figure}[b]
\begin{center}
\vskip10pt
\includegraphics[width=1\textwidth]{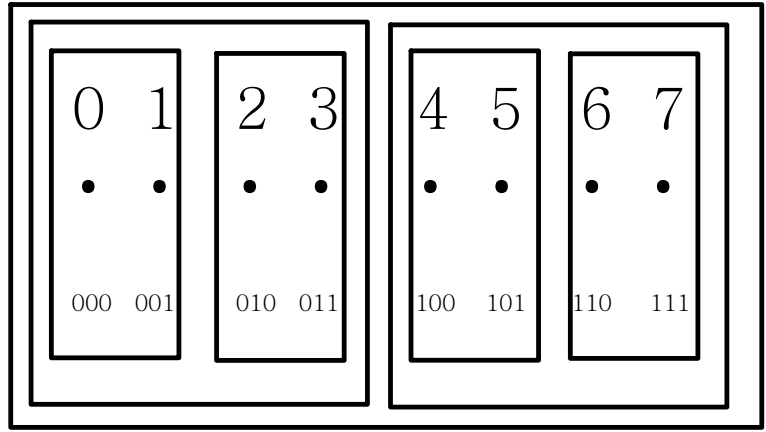}
\vskip-150pt
\caption{Divsion in blocks for 8 sites. The lower row is $x_3 x_2 x_1$. }
\label{fig:boxes}
\end{center}
\end{figure} 

The nonlocal part of the total energy reads
\begin{equation}
H =
-{1\over2}\sum_{n=1}^{n_{max}}({c\over4})^n\sum_{x_{n_{max}},...,x_{n+1}} 
(\sum_{x_n,...,x_1}\phi_{(x_{n_{max}},...x_1)})^2 \ .
\label{eq:ham}
\end{equation}
The index $n$, referred to as the ``level of interaction'' hereafter,
corresponds to the interaction of the total field in blocks of size $2^n$.
The constant $c$ is a free parameter assumed positive and which controls the decay of the 
iterations with the size of the blocks and 
can be adjusted in order to mimic a $D$-dimensional model. This
point is discussed in more detail below (see equation(\ref{eq:cdim})).

The field $\phi_{(x_{n_{max}},...,x_1)}$ is integrated over a local
measure which we need to specify. In the following, we will often work
with the Ising measure, $W_0(\phi) = \delta(\phi^2-1)$ or a 
Landau-Ginsburg measure of the form
$W_0(\phi) =
exp(-{1\over2}m^2\phi^2-\lambda\phi^{4})$. 

In the case of the Ising measure, the only free parameter is $c$. If all the $\phi=+1$, 
then the cost in energy for flipping one spin is $2\sum_{l=1}(c/4)^l(2^l-1)$ and is finite in the infinite volume limit only if $c<2$. In the following it will be assumed that
$0<c<2$.

The hierarchical structure of equation (\ref{eq:ham}), allows us to integrate iteratively 
the fields while keeping their sums in blocks with 2 sites constant. 
\begin{equation}
\hskip-60pt
W_{n+1}(\phi_{(1,2)}) =
\exp\Big({(\beta/2)(c/4)^{n+1}}\phi^2_{(1,2)}\Big) \times 
\int_{-\infty}^{+\infty}
d\xi W_n(\frac{\phi_{(1,2)}}{2}+\xi)W_n(\frac{\phi_{(1,2)}}{2}-\xi)\ ,
\end{equation}
where $\phi_{(1,2)}$ is understood as the sum of the two fields in the block. 
After $n_{max}$ integrations, we obtain 
\begin{equation}
\nsites	W_{n_{max}}(\mainf)={\rm e}^{-\Omega V_{eff}(\mainf/\nsites)}\ ,
\end{equation}
with 
\begin{equation}
	\nsites=2^{n_{max}}
\end{equation}
Remarkably, the symmetries of the model have allowed us to calculate the partition function 
and the effective potential by calculating 
only $n_{max}$ independent integrals instead of the $2^{n_{max}}$ coupled integrals that one would naively expect. 

The symmetry group of the HM is of order $2^{\Omega-1}$, which is one half of the number of configurations for a 
Ising measure. This can be seen as follows. In any of the blocks of size 2 we can interchange two sites. 
The exchange can be done independently in each of the blocks and this symmetry can be seen as {\it local}. There are $\Omega /2$ blocks of size 2 and so $2^{\Omega/2}$ distinct symmetry transformations. Similarly, we can interchange the blocks of size 2 inside any block of size 4 which generates $2^{\Omega/4}$ independent transformations. We can similarly generate independent symmetries until we reach the group of order 2 exchanging two blocks of size $\Omega/2$. Using 
\[1/2+1/4+\dots+(1/2)^{n_max}=1-1/\Omega\ ,\]
we obtain the announced result.
\subsection{The RG transformation}

As explained in subsection \ref{subsec:basic}, the RG transformation consists in a blockspinning and a rescaling. 
In the present example, the rescaling is fixed by the requirement of keeping $H$ unchanged in the 
infinite volume limit. 
More specifically, if  
we define 
\begin{equation}
\phi_{x_{n_{max}-1}',\dots,x_1'}'=\sqrt{\frac{c}{4}}(\phi_{x_{n_{max}},\dots,x_2,1}+\phi_{x_{n_{max}},\dots,x_2,0})
\end{equation}
with 
\begin{equation}
x_{n_{max}-1}'=x_{n_{max}}, \dots , x_1'=x_2 \ ,
\end{equation}
we can rewrite the energy as 
\begin{equation}
H = H'-\frac{1}{2}\sum_{x_{n_{max}-1}',...,x_{1}'} (\phi_{(x_{n_{max}-1}',...x_1')}')^2 \ ,
\label{eq:hamprime}
\end{equation}
with 
\begin{equation}
H'=-{\frac{1}{2}}\sum_{n=1}^{n_{max}-1}({c\over4})^n\sum_{x_{n_{max}-1}',...,x_{n+1}'} 
(\sum_{x_n',...,x_1'}\phi_{(x_{n_{max}-1}',...x_1')}')^2 \ .
\end{equation}
In other words, we can blockspin without thinking about $H$ and then include the second term of equation (\ref{eq:hamprime}) 
in the local measures. The problem is then identical to the original problem except for the fact that we have to reduce 
$n_{max}$ by 1 and to use a new local measure.

The change in the local measure can be expressed through the 
recursion relation  
\begin{equation}
\label{eq:wresc}
\myw_{n+1}(\phi') =
{N_{n+1}}\ \exp({(\beta/2)}\phi'^2) \ 
\int_{-\infty}^{+\infty}
d\xi \myw _n(\frac{\phi'}{\sqrt{c}}+\xi)\myw _n(\frac{\phi'}{\sqrt{c}}-\xi)\ ,
\end{equation}
where $N_{n+1}$ is a normalization factor which can be fixed at our
convenience. The symbol $\myw$ has been used instead of $W$ in order to specify that the field had been rescaled at every step. 
The explicit relationship is 
\begin{equation}
\label{eq:twow}
	\myw_n ((c/4)^{n/2} \phin)=W_n(\phin)\ ,
\end{equation}
where $\phin$ denotes the sum of all the fields in a block of size $2^n$.
Introducing the Fourier representation
\begin{equation}
\label{eq:ft}
\myw_n(\phi) = \int_{-\infty}^{+\infty}{dk\over2\pi}e^{ik\phi}R_n(k)\ ,
\end{equation}
the recursion formula becomes
\begin{equation}
R_{n+1}(k) = C_{n+1}\ 
{\rm e}^{-{1\over2}\beta{\partial^2\over\partial
  k^2}}\ \Big(R_n({\sqrt{c/4}\ k})\Big)^2 \ ,
\label{eq:rec}
\end{equation}
with $C_{n+1}$ another arbitrary normalization constant related to the previous one by the 
relation $C_{n+1}=(c/\sqrt{2})N_{n+1}$. 
We will fix the normalization constant $C_n$ is such a way that
$R_n(0)=1$. $R_n(k)$ has then a direct probabilistic interpretation. 
If we call $\phin$ the total unrescaled field $\sum\phi_x$ inside blocks
of side $2^n$ and $<...>_n$ the average calculated without taking into
account the interactions of level strictly larger than $n$, we can
write
\begin{equation}
R_n(k) =1+ \sum_{q=1}^{\infty}{(-ik)^{q}\over q!}(c/4)^{qn/2}<(\phin)^{q}>_n \ .
\label{eq:gen}
\end{equation}
We see that the Fourier transform of the local measure after $n$
iterations generates the zero-momentum Green's functions calculated
with $2^n$ sites. 

Everything that has been done in this section can be generalized in a straightforward manner for models with 
$N$ components. All we need to do is to replace $\phi$ by a $N$-dimensional vector ${\vec{\phi}}$, 
$k$ by $\vec{k}$ and $d\xi$ by $d^N\xi$. The $N$ component model is discussed in  section \ref{sec:largen}.
\subsection{The Gaussian (UV) fixed point}
\label{subseq:uv}
In the case where the initial measure is Gaussian
\begin{equation}
	W_0(\phi)=\myw_0(\phi)={\rm e}^{-A\phi^2}\ ,
\end{equation}
we have 
\begin{equation}
	\myw _n(\phi)\propto {\rm e}^{-A_n\phi^2}\
\end{equation}
with $A_0=A$ and the $A_n$ calculable from the recursion relation 
\begin{equation}
	A_{n+1}=-\beta/2+(2/c)A_n
\end{equation}
which follows from equation (\ref{eq:wresc}). The fixed point of this transformation is 
\begin{equation}
\label{eq:ast}
	A^{\star}=\frac{\beta c}{2(2-c)}\ .
\end{equation}
With our assumption $0<c<2$, $A^{\star}$ is positive and finite. 
After resumming the terms, we obtain
\begin{equation}
\label{eq:gscaling}
	A_n=A^{\star}+(A-A^{\star})(2/c)^n\ .
\end{equation}
Using equation (\ref{eq:twow}), we then find that the measure for the main field $\mainf$ is 
\begin{equation}
W_{n_{max}}(\mainf)\propto {\rm e}^{-A_{n_{max}}\mainf ^2(c/4)^{n_{max}}}\ .
\end{equation}
This implies that 
\begin{equation}
	V_{eff}(\magden)=((A-A^{\star})+A^{\star}(c/2)^{n_{max}})\magden ^2\ . 
\end{equation}
\def\ast{A^{\star}}
Given that $0<c<2$, the second term disappears in the infinite volume limit and if $A> \ast$, we can interpret 
$A - \ast$ as a quantity proportional to the square of the mass. On the other hand if $A = \ast$, we have a massless 
theory in the infinite volume limit. At finite volume, we expect that for $A=\ast$, 
\begin{equation}
	m^2\propto (c/2)^{n_{max}}\propto L^{-2}\ ,
\end{equation}
\def\nm{n_{max}}
with $L$ the linear size of the whole system defined by 
\begin{equation}
	L^D=\nsites=2^{\nm}\ .
\end{equation}
Putting these two equations together, we obtain 
\begin{equation}
\label{eq:cdim}
	c=2^{1-2/D}\ .
\end{equation}
In other words, the parameter $c$ can be tuned in such a way that a Gaussian massless field scales with the 
number of sites in the same way as a $D$ dimensional model. 

Following the same reasoning, we can determine the parameters introduced 
in subsection \ref{subsec:basic}. 
Since we integrate over two sites in one RG transformation we have
\begin{equation}
	\scale^D=2\ .
\end{equation}
 On the other hand, comparing the field rescaling, we obtain 
\begin{equation}
	(4/c)=\scale^{+D+2-\eta}
\end{equation}
which is consistent with equation (\ref{eq:cdim}) only if 
\begin{equation}
\eta =0 \ . 
\end{equation}

In summary, we have 
\begin{equation}
	c=\scale^{D-2} \ .
\end{equation}
We can reinterpret equation (\ref{eq:gscaling}) by noticing that it implies that at 
each RG transformation, $A_n-\ast$ is rescaled by a factor $2/c$. If we introduce some 
reduced variable $\rho_n\equiv (A_n-\ast)/\ast$, then 
\begin{equation}
	\rho_{n+1}=(2/c)\rho_n=\scale^2\rho_n \ .
\end{equation}
With the notations introduced in equation (\ref{eq:scalingexp}), we could say that the 
exponent $\ye$ at the Gaussian fixed point associated with the mass term squared is 2. 

It should also be said that the Gaussian fixed point is often called the UV fixed point 
because, it is common to use perturbation theory about the Gaussian fixed point to 
construct a RG flow from the Gaussian fixed point to the nontrivial fixed point (which 
is then called an IR fixed point). 

\subsection{The HT fixed point}

In the high temperature (HT) limit, 
$\beta =0$, equation (\ref{eq:rec}) becomes  
\begin{equation}
R_{n+1}(k) = 
\ (R_n({\sqrt{c/4}\ k}))^2 \ .
\label{eq:htrec}
\end{equation}
We set $C_{n+1}=1$ and require $R(0)=1$. 
It is easy to check that 
\begin{equation}
	R(k)={\rm e}^{Bk^{2 {\rm ln}2/{\rm ln}(4/c)}}={\rm e}^{Bk^{2D/(D+2)}}\ , 
\end{equation}
is a fixed point of equation (\ref{eq:htrec}) for arbitrary $B$. 
However if $B\neq 0$ and the exponent of $k$ is not a positive integer, we have a branch cut at $k=0$ and by taking sufficiently many derivatives 
with respect to $k$, we obtain expressions which blow up at $k=0$. For the range of values $0<c<2$, the only way to 
get a power of $k$ in the exponential which is an integer is to have $c=1$, which according to equation (\ref{eq:cdim}) corresponds to $D=2$. For $c=1$, the exponent of $k$ is 1, and the Fourier transform 
in equation (\ref{eq:ft}) is ill defined unless we replace $k$ by $|k|$ which in turn leads to singular derivatives at $k=0$. Consequently, the only choice that leads to a probability distribution with finite moments is 
$B=0$, or in other words, $R=1$. This fixed point remains a fixed point when $\beta \neq 0$ and is called the 
HT fixed point hereafter. 
The HT fixed point corresponds to an arbitrarily narrow probability distribution about 0 for the main field
and can be interpreted as the case of an arbitrarily massive free field.

\section{Equivalent forms of the recursion formulas}
\label{sec:recursion}
The recursion formula will be the main tool used for calculations hereafter. It is important to 
identify equivalent or inequivalent forms and to find accurate numerical implementations. 
In this section, we review equivalent forms of the recursion formulas used by Baker \cite{baker72}, Felder \cite{felder87} and Koch and Wittwer \cite{kw86,koch91}. This section follows closely reference \cite{koch91} but 
with different notations. 
In the following, 
the constants $N_n',\ N_n'',\dots$ need to be fixed by some additional requirement and play no essential role.  
\subsection{Baker's form}
It is sometimes convenient to factor out the Gaussian fixed point so that in a new ``system of coordinates'', 
the Gaussian fixed point is represented by a constant. If we define
\begin{equation}
\label{eq:baf}
	\myw _n (\phi) =\exp(-\ast \phi^2) \kw_n(\phi)\ , 
\end{equation}
the recursion formula becomes 
\begin{equation}
\label{eq:kwbaker}
	\kw_{n+1}(\phi) =
N'_{n+1}\ 
\int_{-\infty}^{+\infty}
d\xi \exp(-2 \ast \xi^2) \ \kw _n(\frac{\phi}{\sqrt{c}}+\xi)\kw _n(\frac{\phi}{\sqrt{c}}-\xi)\ ,
\end{equation}
It is clear from this equation that if $\kw_n$ is a constant then $\kw _{n+1}$ is also constant. 
What is remarkable about this redefinition is that we have replaced $\exp({(\beta/2)}\phi^2) d\xi$ in 
equation (\ref{eq:wresc}) by $d\xi \exp(-2 \ast \xi^2)$. Any other multiplicative redefinition, would in general 
lead to some hybrid form where the weight depends on $\phi$ and $\xi$. 

It should also be noted that the variance of the Gaussian weight can be replaced by another value by a simple change of variable. 
If we define 
\def\kwr{\kw ^{[B]}}
\begin{equation}
	\kwr(\phi)\equiv\kw (B\phi)\ ,
\end{equation}
the recursion formula becomes
\begin{equation}
	\kwr _{n+1}(\phi) =
N'_{n+1}B\ 
\int_{-\infty}^{+\infty}
d\xi \exp(-2 \ast B^2\xi^2) \ \kwr _n(\frac{\phi}{\sqrt{c}}+\xi)\kwr _n(\frac{\phi}{\sqrt{c}}-\xi)\ .
\end{equation}
The recursion used by Baker in reference \cite{baker72} is obtained by 
setting
\begin{equation}
	\kw _n  ^{[(K/2\ast)^{1/2}]}(\phi)=\exp(-\frac{1}{2}Q_n(\phi))
\end{equation}

Note that Baker allowed a non-zero $\eta$ in order to mimic the scaling near the non-trivial fixed point 
instead of the scaling near the Gaussian fixed point for a conventional lattice model. 
It is possible to take into account this modification by changing the value of $D$ used in equation (\ref{eq:cdim}).
Confusion can be avoided by making reference to the rescaling factor $c^{-1/2}$. More explicitly, if we call 
$D'$ the dimension used for a field rescaling that depends on $\eta$, then the correspondence is
\begin{equation}
	2^{(2-\eta -D')/2D'}=c^{-1/2}=2^{(2-D)/2D}\ .
\end{equation}
A short calculation shows that $D'=D(1-(\eta/2))$. As an example, if we want to have a scaling corresponding to 
$D'=3$ and $\eta=0.04$, we can simply work with $D=3.061...$, with $D$ defined in equation (\ref{eq:cdim}).
Using notation that we hope make clear the relation with reference \cite{baker72}, 
the recursion formula can then be written as 
\begin{equation}
\label{eq:baker}
\hskip-65pt
	\exp(-\frac{1}{2}Q_{n+1}(\phi))=N''_{n+1}\ 
\int_{-\infty}^{+\infty}
d\xi \exp\Big(- K\xi^2 -\frac{1}{2}Q_n(\frac{\phi}{\sqrt{c}}+\xi)-\frac{1}{2}Q_n(\frac{\phi}{\sqrt{c}}-\xi)\Big)
\end{equation}

\subsection{Gallavotti's form}
\label{subsec:gal}
An alternate way of formulating the recursion is to use a convolution with the Gaussian fixed point: 
\begin{equation}
\label{eq:galf}
	\gal _n (t)=\int_{-\infty}^{+\infty} d \phi \exp(-\ast (t-\phi)^2 )\kw _n (\phi)\ .
\end{equation}
By factoring out the two exponentials of the quadratic terms, it is possible to relate $\gal$ to the Fourier transform of $\myw$ introduced in 
equation (\ref{eq:rec}). More explicitly: 
\begin{equation}
\label{eq:galt}
	\exp(\ast t^2)\gal _n (t) \propto R_n(i2t\ast )\ ,
\end{equation}
and in particular the HT fixed point is now proportional to $\exp(-\ast t^2)$
Under the new transformation, the recursion formula takes the form:
\begin{equation}
\label{eq:galeq}
\gal _{n+1}(t)=	N'''_{n+1}\ 
\int_{-\infty}^{+\infty}
d\xi \exp(-2 \frac{\ast c}{2-c} \xi^2)\ \Big(\gal _n(\frac{t}{\sqrt{c}}+\xi)\Big)^2\ .
\end{equation}
This form of the recursion formula was used in references \cite{kw86,felder87,koch91} and its origin can be found in 
the work of Gallavotti \cite{gallavotti78}. More about this question can be found in subsection \ref{subsec:galext} 
below. In the following, we call this form of the recursion 
formula the Gallavotti's form. 

\def\myf{{\mathcal F}}
If we now define
\begin{equation}
\label{eq:myff}
\gal _{n} (t)=\exp(-\ast t^2)\myf _n (t)\ ,
\end{equation}
the recursion formula becomes after some algebra: 
\begin{equation}
\label{eq:myfrec}
\myf _{n+1}(t)=	N'''_{n+1}\ 
\int_{-\infty}^{+\infty}
d\xi \exp(-4 \frac{\ast}{2-c} (\xi-t\sqrt{c/4}\ )^2) \ \Big(\myf _n(\xi)\Big)^2\ .
\end{equation}
From the definitions of $\gal_n$ and $\myf _n$, we have 
\begin{equation}
	\myf _n (t) \propto R_n(i2t\ast )\ ,
\end{equation}
and we can prove the equivalence of (\ref{eq:myfrec}) and (\ref{eq:rec}) by using the 
identity
\begin{eqnarray}
\exp(-{1\over2}\beta{\partial^2\over\partial
  k^2})\ \Big(k\sqrt{\frac{c}{4}}\Big)^q=\sqrt{\frac{4\ast}{\pi(2-c)}} \nonumber \\ \times \int_{-\infty}^{+\infty}
d\xi \exp\Big(-4 \frac{\ast}{2-c} (\xi+i\frac{k}{2\ast}\sqrt{c/4}\ )^2\Big) (2i\xi\ast)^q 
\end{eqnarray}

\def\kwf{f}
Finally, if we write 
\begin{equation}
\label{eq:lff}
	\kwf_{n}(t) =\myf _n \Big(t\ \sqrt{\frac{2-c}{\ast (4-c)}} \ \ \Big)\ ,
\end{equation}
we obtain the form most often used in reference \cite{koch91}:
\begin{equation}
\label{eq:kwrec}
\kwf _{n+1}(t)=	N''''_{n+1}\ 
\int_{-\infty}^{+\infty}
ds \exp(-\frac{1}{1-c/4} s^2) \ \Big(\kwf _n(s+t\sqrt{c/4})\Big)^2\ .
\end{equation}

\subsection{Summary}
The form of the UV and HT fixed points in the various set of coordinates is summarized in table \ref{tab:FixedPoint}.
\begin{table*}[h]
	\caption{Form of the fixed point up to an overall constant in the various coordinates. 1 is short for a constant and $\delta$ short for a delta function; the last line is the 
	defining equation.}
	\label{tab:FixedPoint}
	\vskip20pt
	\centering
		\begin{tabular}{||c|c|c|c|c|c|c||}
		\hline
			\ & $\myw$ & $\kw$ & $\gal$& $\myf$&$R$&$\kwf$ \\
			\hline
			UV&  ${\rm e}^{-\ast \phi^2}$ & 1&1 & ${\rm e}^{\ast t^2}$&${\rm e}^{-\frac{k^2}{4\ast }}$& 
			${\rm e}^{\frac{2-c}{4-c} t^2}$\\
			\hline  
			HT&$\delta$ & $ \delta$& ${\rm e}^{-\ast t^2}$&1&1&1 \\
			\hline
	Eq.&\ref{eq:wresc}&\ref{eq:baf}&\ref{eq:galf}&\ref{eq:myff}&\ref{eq:ft}&\ref{eq:lff}\\
	\hline
		\end{tabular}
\end{table*}

All this looks quite reminiscent of quantum mechanics where we can look at a problem in a basis where the position 
operator is diagonal or in another basis where the momentum operator is diagonal. The two basis are related 
by a unitary transformation. More generally, in quantum mechanics, unitary transformations do not affect the spectrum of hermitian operators (which represent observables). Universality is a stronger notion than observability in the sense that some non-universal quantities may have an absolute physical meaning independent of our choice of integration variables in the functional integral.
The idea of transformations that leave universal properties unchanged have been discussed in reference \cite{wegner74} 
where this translate into a reparametrization invariance. 

\section{Inequivalent extensions of the recursion formula}
\label{sec:inequivalent}
In this section, we 
discuss more general recursion formulas which coincide with the HM's one for a particular choice of the parameter 
$\scale$ that controls the change in the linear scale after one RG transformation.
\subsection{Relation with Wilson's approximate recursion formula}
 Wilson's approximate recursion formula was the first simplified RG transformation that was proposed. 
 It appears in one of the basic RG papers \cite{wilson71b} as a result of a rather involved analysis of the 
 partition function of a scalar model with a UV cutoff. It played an important role, because it was realized that 
 the RG method could lead to a fully numerical treatment without any reference to expansions such as perturbation 
 in a weak coupling. It is intended to represent a situation where $\scale =2$, but instead of having $2^D-1$ integration variables, it has only one corresponding to an approximation where the $2^D$ fields take only two independent values \cite{wilson72}, one in each half of the block. 
Having decoupled the number of integration variables to $\scale$, we can now write for arbitrary $\scale$:
\begin{equation}
\label{eq:warf}
\hskip-40pt
H_{n+1}^{[ \scale ]}(\phi) =
N''''''_{n+1}\ 
\int_{-\infty}^{+\infty}
d\xi {\rm e}^{-\xi^2} \ \Big[ H ^{[ \scale ]}_n \left( \scale^{1-\frac{D}{2}}\phi+\xi\right) H^{[ \scale ]} _n \left( \scale^{1-\frac{D}{2}}\phi-\xi\right) \Big] ^{{ \scale^D/2}}\ ,
\end{equation}
For $\scale=2$, we obtain Wilson's approximate recursion formula, most often written in a form similar to equation (\ref{eq:baker}). For $\scale=2^{1/D}$, we obtain 
 the HM recursion formula in the form given in equation (\ref{eq:kwbaker}).

The general case $\scale =2^{\zeta}$ was discussed in reference \cite{fam} where it was shown that for $D=3$, as $\zeta$ increases 
from $\zeta=1/3$ to 1, the exponent $\gamma$ decreases monotonically from 1.30 to 1.22. Clearly, different values of 
$\zeta$ correspond to different classes of universality. 

\subsection{Gallavotti's recursion formula}
\label{subsec:galext}
The recursion formulas presented in subsection \ref{subsec:gal} can also be extended for arbitrary $\scale$. This is indeed easier because the number of sites integrated for the HM, namely 2, appears as the exponent for $\myf$, $\gal$ and $f$. The replacements are 
\begin{eqnarray}
	2&\rightarrow & \scale^D\\
	\frac{1}{\sqrt{c}}&\rightarrow & \scale ^{1-D/2}\\
	\frac{c}{4}&\rightarrow & \scale ^{-2 -D}
\end{eqnarray}
For instance equation (\ref{eq:kwrec}) becomes
\begin{equation}
\label{eq:extgal}
f ^{[ \scale ]} _{n+1}(t)=	N''''''_{n+1}\ 
\int_{-\infty}^{+\infty}
ds \ \exp\Big(-\frac{1}{1-\scale^{-D-2}} s^2\Big) \ \Big[ f^{[ \scale ]} _n(s+t\scale^{-D/2-1})\Big]^{\scale ^D}\ .
\end{equation}
Again for $\scale=2^{1/D}$ we recover the HM recursion formula in the form given in equation (\ref{eq:kwrec}). 
In reference \cite{gallavotti78}, Gallavotti has introduced equations that can be identified with the case $\scale =2$. The variable $s$ in equation (\ref{eq:extgal}) plays a role similar to the Gaussian variable  $z_{\Delta}$  in his notations, 
and the factor $\scale ^D=2^D$ is reabsorbed in the definition of the potential. 
The limit $\scale \rightarrow 1$ of equation (\ref{eq:galeq}) will be discussed in section \ref{sec:erge}.

For $\scale=2$ and $D=3$, the value $\gamma =1.30033\dots $ was obtained numerically in reference \cite{kw88}. 
This value is significantly different from the value $\gamma =1.2991407\dots$ obtained in references \cite{gam3rapid,gam3} 
for $\scale=2^{1/3}$ and $D=3$. For $D=3$, the limit $\scale \rightarrow 1$ was studied in 
references \cite{litim02,bervillier04} with the result $\gamma = 1.299124$. The difference in the fifth digit is significant 
and was confirmed by new calculations \cite{bervillier07} (see also our section \ref{sec:erge} below). 
This shows that again different values of $\scale$ correspond to different classes of universality and also that 
for a given $\scale\neq 2^{1/D}$, the extensions given in equations (\ref{eq:warf}) and (\ref{eq:extgal}) are 
inequivalent. In addition we see that the $\scale$ dependence is much weaker 
in equation (\ref{eq:extgal}) and the slope is apparently opposite 
to the slope found for equation (\ref{eq:warf}). We are lacking calculations at intermediate values but as far as we can see, when $\scale$ increases, $\gamma$ increases.

\section{Motivations, rigorous and numerical results}
\label{sec:motrig}
\subsection{Motivations}
\label{subsec:motivations} 

Dyson's hierarchical model was invented and reinvented several times with different motivations that 
we briefly review. Dyson's original motivation \cite{dyson69} was to construct models more weakly coupled 
than the one dimensional Ising models with long range hamiltonians of the form
\begin{equation}
	H=-J\sum _{m<n} |n-m|^{-\alpha}\phi_n\phi_m \ .
	\label{eq:kt}
\end{equation}
Dyson's was trying to figure out if the model has an ordered phase when $\alpha=2$. 
Dyson constructed a more general family of models where $c^l$ is replaced by $b_l$ in equation (\ref{eq:ham}).
He proved several theorems concerning the infinite volume limit and the existence of phase transitions 
at finite temperature that are discussed in the next subsection.

The fact that $\alpha = 2$ is borderline can be anticipated by refining \cite{thouless69} 
Landau's argument for the absence of ordered phase in one dimension.
If $\alpha<2$ and a system of size $L$ has an average magnetization $\mu>0$, then the cost in energy for flipping 
all the spins in a large subsystem,  is of the order of $L^{2-\alpha}$. 
If $\alpha > 2$, the cost should grow slower than ln($L$) when $L$ is increased.
On the other hand, there 
is of the order of $L$ ways to choose the subsystem and so the gain in entropy is of order 
$kT{\rm ln}L$. Consequently if $\alpha>2$, the entropy dominates and the  free energy is 
decreased after flipping, which is incompatible with the possibility of an equilibrium situation. 
On the other hand if $\alpha <2$, there is no incompatibility. 
In the case $\alpha =2 $, both contributions are logarithmic and a more careful estimate is required. 
Thouless \cite{thouless69} estimated that if the width of the magnetization distribution is proportional to $L^{1/2}$, 
the change in free energy is 
\begin{equation}
\Delta E= 2	\mu^2 J {\rm ln }L-(1/2)kT{\rm ln}L\ ,
\end{equation}
and an ordered state seems possible for $T$ small enough.
If this occurs at some strictly positive critical temperature $T_c$, then the magnetization changes 
abruptly to $(kT_c/4J)^{1/2}$ when the temperature is lowered to $T_c$. This is called the Thouless effect. 
Later, the existence of an ordered state at sufficiently low temperature for the model defined by equation (\ref{eq:kt})  with $\alpha =2$ was proved 
rigorously \cite{spencer82} as well as the Thouless effect \cite{simon81}.

Baker rediscovered the HM  \cite{baker72} in the context of the development of the RG ideas.
His goal was to construct models for which a simple recursion formula would be exact. He 
reinvented Dyson's HM and several variant of it. This has been partially reviewed in sections \ref{sec:recursion} 
and \ref{sec:inequivalent}.  

The hierarchical structure of the block variables can be naturally reconstructed using the $2$-adic numbers (see section \ref{sec:imp}). 
At the end of the eighties, physicists started reformulating models of classical, quantum and statistical 
mechanics over the fields of $p$-adic numbers \cite{freund}. In particular models of random walks over the
$p$-adic numbers were considered \cite{missarov88,cuerna,lucio90} and it was recognized that 
it was possible to reformulate the HM as a scalar model on the 2-adic fractions \cite{lerner89}.
This reformulation was used in high temperature (HT) expansions \cite{meuricejmp95}, helps understanding the absence of certain Feynman graphs \cite{wilson71b} 
and suggests ways to improve the hierarchical approximation \cite{marseille93}. This is discussed in more detail in 
section \ref{sec:imp}.

\subsection{Rigorous results}

Dyson proved several theorems for the HM with 
\begin{equation}
	c=2^{2-\alpha}\ ,
\end{equation}
where $\alpha$ is the same as in equation (\ref{eq:kt}).
With this notation it is clear that the ferromagnetic interactions are weaker than for the model of equation (\ref{eq:kt}). If we refer to figure \ref{fig:boxes}, the relative strength of the couplings between the 
 0th spin and its right neighbors are, from left to right, $1,2^{-\alpha}, 3^{-\alpha}, \dots,  7^{-\alpha}$ for 
equation (\ref{eq:kt}) and $2^{-\alpha},4^{-\alpha},4^{-\alpha},8^{-\alpha},8^{-\alpha},8^{-\alpha},8^{-\alpha}$ for 
the HM. Griffiths has proved that for Ising models with ferromagnetic interactions, the averages of two arbitrary spins 
variables are positive and increase with the strength of the ferromagnetic interactions. 
Consequently, if one can prove that $<\phi_n\sum_m\phi_m>$ blows up in the infinite volume limit below some temperature for the more weakly coupled model, it will also blow up for the other model. 
Dyson proved the existence of the infinite volume limit for $\alpha>1$ and that there was a phase transition 
at finite temperature if and only if $1<\alpha<2$. In other words, there is no phase transition at $\alpha=2$ 
for the HM, but this does not allow us to extend the result to the more strongly coupled model of equation (\ref{eq:kt}) 
that has indeed a phase transition at finite temperature \cite{spencer82}. 
With the notations used in section \ref{sec:model}, Dyson theorems mean that an infinite volume limit exists for $c<2$ 
($D>0$) and that a phase transition at finite temperature occurs if $1<c<2$ ($D>2$).

The HM has been studied near $D=4$ ($c=\sqrt{2}$) using the $\epsilon$ expansion. The existence of a non-trivial 
fixed point for $\epsilon$ small enough was proved \cite{bleher75}. 
The $\epsilon$-expansion was shown to be asymptotic \cite{collet77}. Many details regarding this approach can be found in reference \cite{collet78}. Extensions beyond the hierarchical 
model are discussed in reference \cite{abd06}.

The HM has also been studied directly  at $D=3$ ($c=2^{1/3}$). The existence of a non-trivial fixed point was proved for a large enough number of components \cite{kupiainen83}. Proofs of the existence of the non-trivial 
fixed point in $D=3$ for the one component model were given in references \cite{kw86,kw88,koch91,koch94,koch95}. 
They also put exponential bounds on the fixed point in various representations discussed in section \ref{sec:recursion}. In particular, for real $\phi$, some positive $C$ and $\kw ^\star$ the non trivial fixed point of equation (\ref{eq:kwbaker}), the following bound holds:

\begin{equation}
\label{eq:rigbound}
	|\kw^{\star}(\phi)|<\exp(-C\phi^6)\ .
	\end{equation}

\subsection{Numerical results}

The critical and tricritical behavior of the HM was investigated numerically in the presence of a magnetic field and a staggered magnetic field \cite{baker77}. The literature contains many numerical estimates of the 
critical exponent $\gamma$  for $D=3$. The first calculation was done 
by Wilson. The result reported in 
reference \cite{baker72} is $\gamma$=1.2991. 
In the following, unless specified differently, errors of order one should be assumed for the 
last printed digit. 
Values when $2/D$ is a multiple of 0.05 are given in reference \cite{kim77}. 
Interpolating linearly to $D=3$, we obtained 1.302. 
Using the $\epsilon$ expansion up to order 
34 and a Borel resummation method, the value 1.2986 was obtained in reference \cite{collet77b}. 
Analysis of the HT expansion \cite{osc1,osc2} yields 1.300(2) for $D=3$. The value 1.299141 
can be obtained from a footnote in reference \cite{kw88}. 
The value 1.29914 was obtained in reference \cite{pinn94}.
Using two independent methods discussed in 
section \ref{sec:num}, the value 1.299140730159 was found in references \cite{gam3rapid,gam3}. 
Less accurate calculations in the low-temperature phase were performed in reference \cite{hyper} 
and confirmed hyperscaling with three decimal points.

\section{Numerical implementation}
\label{sec:num}

\subsection{Polynomial truncations}
\label{subsec:pol}
The recursion formula can be implemented numerically using numerical integration methods 
in equations (\ref{eq:kwbaker}) or (\ref{eq:warf}) as was done for instance 
in Refs \cite{wilson71b,baker77,fam}. However, in the symmetric phase, it seems easier to get very accurate results by using 
polynomial approximations in forms based on the Fourier transform such as equations (\ref{eq:rec}) or (\ref{eq:kwrec}). 
This method was justified rigorously in \cite{kw88,koch95}, and used for numerical calculations for instance in references \cite{pinn94, guide,gam3,gottker99}.  
For definiteness, we start with the recursion formula for $R$ given in 
equation (\ref{eq:rec}). A 
finite dimensional approximations of degree
$l_{max}$ has the form:
\begin{equation}
R_n(k) = 1 + a_{n,1}k^2 + a_{n,2}k^4 + ... + a_{n,l_{max}}k^{2l_{max}}\ .
\end{equation}
This type of approximation can be justified in the context of the HT expansion and works extremely well 
in the symmetric phase.
We can reabsorb the inverse temperature $\beta$ in $k$ in such a way that it does not appear 
anymore in the exponential in equation (\ref{eq:rec}). This change would then be compensated by 
a transformation $a_{n,l}\rightarrow \beta^l a_{n,l}$ and the truncation at order $k^{2l_{max}}$ would be sufficient 
to calculate exactly the HT expansion of $R$ up to order $\beta ^{l_{max}}$. 
This technique was used in references \cite{osc1,osc2}. It was then realized (by accident) that large order coefficients 
in the HT expansion could be calculated in good approximation by using polynomial truncations at order 10 times smaller than the HT order. The same method can be applied to numerical calculations in the symmetric phase. 
The apparent convergence is studied empirically in reference \cite{guide}.

With the polynomial truncation, the recursion formula equation (\ref{eq:rec}) becomes a $l_{max}$-dimensional 
quadratic map. After squaring $R$, we obtain a polynomial of order $ 2l_{max}$ in $k^2$. We could in principle 
truncate at order $l_{max}$, however, the derivatives in the exponential in equation (\ref{eq:rec}) will lower the 
degree and the terms of order larger than $l_{max}$ will contribute to the orders smaller than $l_{max}$ after 
enough derivatives are applied.  Of course, a truncation at order $l_{max}$ is made after all the derivatives are 
performed, but it was realized empirically \cite{scalingjsp} that intermediate truncations reduce 
the accuracy of the calculation.
The explicit algebraic transformation reads.
\begin{equation}
\hskip-50pt
a_{n+1,m} = {
{\sum_{l=m}^{2l_{max}}(\sum_{p+q=l}a_{n,p}a_{n,q}){[(2l)!/(l-m)!(2m)!]}({c/4})^l[-(1/2)\beta]^{l-m}}\over{\sum_{l=0}^{2l_{max}}(\sum_{p+q=l}a_{n,p}a_{n,q}){[(2l)!/l!]}{(c/4)^l}[-(1/2)\beta]^l}} \ .
\label{eq:alg}
\end{equation}
The 
initial conditions for the Ising measure is $R_0(k)=cos(k)$. For
the LG measure, the coefficients in the $k$-expansion need to be
evaluated numerically. The susceptibility at finite volume and higher moments can then be obtained by rescaling the 
coefficients, for instance:
\begin{equation}
\chi_n^{(2)} = -2a_{n,1}(2/c)^n \ .
\label{eq:resc}
\end{equation}

\subsection{Volume effects in the symmetric phase}

When calculating the susceptibility at values of $\beta$ slightly below
$\beta_c$, we spend about $-{\rm ln}(\beta_c-\beta)/ln(\lambda_1)$ iterations near 
the fixed point.
During these iterations, the round-off errors
are amplified along the unstable direction (see next subsection). 
After that,  
the order of magnitude of the 
susceptibility stabilizes and the corrections get smaller by a factor 
${c\over 2}$ at each iterations. At some point, all the recorded digits
stabilize (irrespectively of the numerical errors which occurred in the 
first stage described above).
This gives the estimate \cite{guide} for the number of iterations
$n(\beta, P)$ to stabilize $P$ digits (in decimal notations)
\begin{eqnarray}
n(\beta,P)&=&\left(
{Dln(10)\over2ln(2)} \right) [P-\gamma
log_{10}(\beta_c-\beta)]\ .
\end{eqnarray}

\subsection{The Eigenvalues of the Linearized RG Transformation}

The 
critical exponents can be calculated by 
linearizing the RG
transformation near the fixed point $R^{\star}(k)$ specified by 
the coefficients ${a^\star}_ l$. 
We express the coefficients after $n$ iterations in terms of
small variations about the fixed point:
\begin{equation}
a_{n,l}=a^\star_l+\delta a_{n,l}\ .
\end{equation}
At the next iteration, we obtain the linear variations
\begin{equation}
\delta a_{n+1,l} = \sum_{m=1}^{l_{max}} M_{l,m} \delta a_{n,m} \ .
\end{equation}
The $l_{max}\times l_{max}$ matrix appearing in this equation is
\begin{equation}
M_{l,m} = {\partial a_{n+1,l}\over \partial a_{n,m} } \ ,
\label{eq:mat}
\end{equation}
evaluated at the  fixed point. 

Approximate fixed points can be found by approaching
$\beta_c$ from below and iterating until the ratio $a_{n+1,1}/a_{n,1}$ 
takes a value which is
as close as possible to 1. The determination of $\beta_c$ can be done by following the bifurcations 
in $a_{n+1,1}/a_{n,1}$ for sufficiently large $n$. When $\beta<\beta_c$, the susceptibility stabilizes at a 
finite value without subtraction and for $n$ large enough, $a_{n+1,1}/a_{n,1}\rightarrow c/2$. 
On the other hand if $\beta>\beta_c$, the unsubtracted susceptibility grows like the volume and for $n$ large enough, $a_{n+1,1}/a_{n,1}\rightarrow c$ (until the polynomial truncation breaks down). 

The approximated fixed points obtained with this procedure
depend on $\beta_c$. Using their explicit form 
which we denote $R^{\star}(k,\beta_c)$, one obtains
a universal 
\def\ru{{\mathfrak R}}
function $\ru(k)$
by absorbing $\beta$ into $k$
\begin{equation}
\ru(k)=R^{\star}(\sqrt{\beta_c}k,\beta_c)\ .
\label{eq:ufonc}
\end{equation}
It was shown that in very good approximation, $\ru(k)$ is independent of the initial measure considered \cite{gam3rapid,gam3}. 
Numerically, 
\begin{equation}
\ru(k)=1. - 0.35871134988 k^2 + 0.0535372882 k^4 -\dots \ .
\label{eq:ufp}
\end{equation} 

This function is related to the fixed point $f_{KW}(s^2)$ constructed 
in reference \cite{koch95} as follows
by the relation
\begin{equation}
\ru(k)\propto f_{KW}(({{c-4}\over{2c}})k^2)  \ .
\label{eq:trans}
\end{equation}
Extremely accurate values of the Taylor coefficients of $f_{KW}$ can be found in the file \verb+approx.t+
in \cite{koch95}. The constant of proportionality is fixed 
by the condition $\ru(0)=1$. The relation with the non-trivial fixed point $f^\star$ of equation (\ref{eq:kwrec}) is 
$f^\star (it)=f_{KW}(-t^2)$.

The first six eigenvalues of $M_{l,m}$ from reference \cite{gam3} are given in table \ref{table:eigen}.

\begin{table}
\caption{The fist six eigenvalues of the linearized RG transformation}
\vskip20pt
\label{table:eigen}
\centering
\begin{tabular}{||c|c||}
\hline
$n$&$\lambda_n$ \\ 
\hline
1&1.42717247817759\\  
2&0.859411649182006\\
3&0.479637305387532\\
4&0.255127961414034\\ 
5&0.131035246260843\\ 
6&0.0654884931298533\\
\hline
\end{tabular}
\end{table}
\noindent
Using $\gamma={\rm ln}(2/c)/{\rm ln} \lambda_1$ and $\Delta_s =-{\rm ln}\lambda_2/{\rm ln} \lambda_1$ from section \ref{sec:block}, we obtain
\begin{eqnarray}
\gamma &=& 1.299140730159\\
\Delta_s &=& 0.425946858988 \ .
\end{eqnarray}
These estimates of the exponents were in agreement \cite{gam3} with those obtained from fits of numerical data near criticality based on the parametrization 
of equation (\ref{eq:suspar}) with 12 significant digits for $\gamma$ and 6 significant digits for $\Delta_s$. 
These fits also provide the non-universal amplitudes. 
\subsection{The critical potential}
\label{subsec:critical}
\begin{figure}[t]
\begin{center}
\vskip90pt
\includegraphics[width=0.8\textwidth]{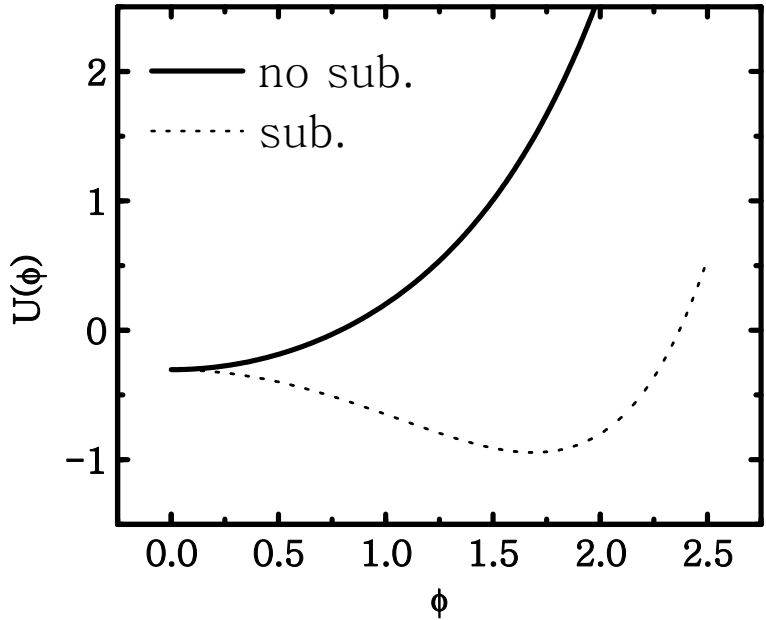}
\vskip-90pt
\caption{$U(\phi)$ defined in equation (\ref{eq:critpot}) (solid line); the subtracted potential $U(\phi)-(c/(2(2-c)))\phi^2$ (dotted line). }
\label{fig:critpot}
\end{center}
\end{figure}

It is possible to Fourier transform numerically $\ru(k)$ for a not too large value of the conjugate variable $\phi$. 
The result is that apparently, the Fourier transform is a positive bell-shaped function with only one maximum at $\phi=0$. We define
\begin{equation}
\label{eq:critpot}
	U(\phi)\equiv-{\rm ln}\widehat{\ru}(\phi)\ ,
\end{equation}
and we obtain an apparently convex function with only one minimum shown as the solid line in figure \ref{fig:critpot}.
In order to compare with calculations done with the $\kw _n$ formulation, we need to subtract $\ast \phi^2$ 
from this function. Since we have removed the $\beta$ dependence by reabsorbing it in the definition of $k$, 
we must calculate $\ast$ with $\beta =1$. Subtracting  
$(c/(2(2-c)))\phi^2\simeq 0.851 \phi^2$ from the rescaled potential then becomes a double-well 
potential with minima at $\phi\simeq \pm 1.688$. This is illustrated in figure \ref{fig:critpot}.
This explains that figures with a minimum away from the origin appear in references \cite{wilson72,baker77}.

It should be emphasized that $U(\phi)$ is a rescaled potential and not the effective potential. 
The relationship between the two can be worked out from equations (\ref{eq:twow}) and (\ref{eq:ft}).
Assuming we are exactly at the non trivial fixed point, $\myw$ remains $\widehat{\ru}(\phi)$.
One then obtains
\begin{equation}
\label{eq:zcrit}
	Z\propto R^{\star}(-iJ(4/c)^{n_{max}/2})\ ,
\end{equation}
and 
\begin{equation}
\label{eq:vcrit}
\nsites	V_{eff}(\magden)=U(\magden c^{n_{max}/2})\ .
\end{equation}
This result can be interpreted in the following way: at the fixed point, the only scale in the problem is the 
size of the system and all the quantities scale with it according to their dimension. If we perform Taylor expansions 
in equations (\ref{eq:zcrit}) and (\ref{eq:vcrit}), we see that when $n_{max}\rightarrow \infty$, all the coefficients 
of $Z$ blow up and the coefficients of $\magden ^{2r}$ blow up if $r>D/(D-2)$. In the same limit, the coefficients of $\magden ^{2r}$ tend to zero if $r<D/(D-2)$. The limiting case corresponds to $r_c=D/(D-2)$ 
where the coefficient is volume independent. For instance 
for $D=4$, $r_c=2$ and for $D=3$, $r_c=3$ which corresponds in both cases to the marginal direction in perturbation theory. It should however be noted that if we 
work at finite $\phi_c$, we are in fact probing $U$ at large value of its 
argument and the Taylor expansion may not be convergent (the radius of convergence 
should be determined by the complex zero of $\widehat{\ru}(\phi)$ closest to the origin). The above determination of $r_c$ implies 
that, 
\begin{equation}
\label{eq:uas}
U(\phi)\propto |\phi|^{2D/(D-2)}=|\phi|^{D/x_h}\ \ {\rm for} |\phi|\rightarrow\infty \ .
\end{equation}
For $D=3$, the $\phi^6$ behavior is compatible with the rigorous bound of equation 
(\ref{eq:rigbound}). 

Equation (\ref{eq:uas}) can be obtained from the scaling hypothesis discussed at the end of section 
\ref{subsec:basic}. First, we find from equation (\ref{eq:zcrit}) that 
\begin{equation}
\label{eq:gcrit}
	\scale^D G[J]=-\ln( R^{\star}(-iJ\scale^{(D+2)/2})) + {\rm constant}\ ,
\end{equation}
which implies that for large values of $|J|$ and $|\phi|$
\begin{equation}
	G[J]\propto |J|^{2D/(D+2)} =|J|^{D/\ye_h}\ ,
\end{equation}
and 
\begin{equation}
	|\phi| \propto |J|^{(D-2)/(D+2)}\ .
\end{equation}
Combining the two above equations and the Legendre transform, we recover 
equation (\ref{eq:uas}). 

\subsection{The low temperature phase}
\label{subsec:lt}
The calculation in the low temperature phase requires the introduction of 
a constant external magnetic field coupling linearly to $\phi_n$. 
Since 
 \begin{equation}
W_n(\phi_n,H)\propto W_n(\phi_n){\rm e}^{H\phi_n} \ ,
\end{equation}
after Fourier transforming and rescaling one obtains \cite{hyper}
\begin{equation}
{{R_n(k+iH(4/c)^{n/2})}\over{R_n(iH(4/c)^{n/2})}}= 
\sum_{q=0}^{\infty}{(-ik)^{q}\over q!}{\langle(\phi_n)^{q}
\rangle_{n,H}(c/4)^{qn/2}}\ .
\label{eq:genlt}
\end{equation}
The connected Green's functions can be obtained by taking the logarithm
of this generating function. The average are understood at non-zero $H$.
In order to observe the magnetization, it is essential 
to take the infinite volume limit {\it before} taking the limit $H \rightarrow 0$. 
For any non-zero $H$, no matter how small its
absolute value is, one can always find a $n$ large enough to have 
$|H(4/c)^{n/2}|\gg 1$. The non-linear effects are then important and 
linearization does not apply. It was checked \cite{hyper} that when such a $n$ is reached, the value
of the $\chi^{(q)}$ stabilizes at an exponential rate. One can then, {\it first}
extrapolate at infinite volume for a given magnetic field, and {\it then}
reduce the magnetic field in order to extrapolate a sequence of 
infinite volume
limits with decreasing magnetic field, 
toward zero magnetic field.

In the following, we use the notation 
\begin{equation}
\chi^{(q)}\propto(\beta-\beta_c)^{-\gamma_q} \ ,
\end{equation}
for the leading exponent. It is customary to use the notation $\gamma_1 = -\beta$, but 
we avoided it here because of a possible confusion with the inverse temperature. 
Obviously $\gamma_2 =\gamma$. 
In the symmetric phase and for $q$ even, we have the order of magnitude estimate
\begin{equation}
\chi^{(q)}\approx 2^{-n^\star}(4/c)^{qn^\star/2}\ .
\label{eq:bulk}
\end{equation}
with $n^\star$ defined as in section \ref{sec:block} by the relation $|\beta-\beta_c|\lambda^{n^\star}=1$.
Eliminating $n^\star$ and using the expression of $\gamma$ as in section \ref{sec:block}, we obtain
\begin{equation}
\gamma_q=\gamma \lbrack (q/2){\rm ln} (4/c)-{\rm ln}2\rbrack/{\rm ln}(2/c)\ .
\label{eq:gqhm}
\end{equation}
In the case $D=3$ $(c=2^{1/3})$, this becomes
\begin{equation}
\gamma_q=1.29914073\dots \times (5q-6)/4\ .
\label{eq:test}
\end{equation}
We will show below that this equation is a particular case of a more general relation 
that follows from the scaling hypothesis.
In reference \cite{hyper}, the following numerical results were obtained
\begin{eqnarray}
\gamma_1&=&-0.3247 \nonumber \\
\gamma_2&=&1.2997\\
\gamma_3&=&2.9237 \nonumber \ ,
\end{eqnarray}
which agrees with three significant digits with the prediction of equation (\ref{eq:test}).

Equation (\ref{eq:gqhm}) follows from a slightly stronger form of the scaling hypothesis. The basic scaling relation of equation (\ref{eq:scalinghyp}) is satisfied if we further assume 
that  
\begin{equation}
	G_s(t,J)=t^{D\nu}g(J/t^{\Delta_g})\ ,
\end{equation}
for a well-behaved function $g$ and 
with the gap exponent
\begin{equation}
\Delta_g\equiv \frac{\ye_h}{\ye_t}=\frac{(D+2-\eta)\nu}{2} \ .	
\end{equation}
By construction the argument of $g$ is invariant under the rescaling of equation (\ref{eq:scalinghyp}). Each derivative with respect to $J$ brings down a factor $t^{-\Delta_g}$. This implies that 
\begin{equation}
	\gamma_q=-D\nu +q\Delta_g\ .
\end{equation}
In the case of the HM, $\nu =\gamma/2$ and $\Delta_g=(D+2)\gamma/4$ and we recover 
equation (\ref{eq:test}) for $D=3$. For the HM in arbitrary dimension, we could also 
write 
\begin{equation}
\label{eq:yq}
	\gamma_q=-\frac{D-(q/2)(D+2)}{\ye_t} \ .
\end{equation}

\subsection{Practical aspects of the hierarchy problem}
\label{subsec:hier}
In absence of wave function renormalization, the square of the renormalized mass $m_R$ in units of the UV cutoff 
$\Lambda$
can be defined as the inverse susceptibility. 
Keeping the mass small when the cutoff increases requires a large susceptibility. 
In the calculations discussed above, a large susceptibility is obtained by fine-tuning 
$\beta$. However, we can also keep $\beta=1$ and fine tune another parameter 
such as the bare mass $m_B$ in a Landau-Ginzburg potential. In this case, we have 
\begin{equation}
m_R/\Lambda\sim(|m_{Bc}-m_B|/\Lambda)^{\gamma/2}\ ,
\label{eq:rat}
\end{equation}
In four dimensions, $\gamma=1$ and if we take 
$m=100$ GeV, a typical electroweak scale, 
and $\Lambda=10^{19}$ GeV of the order of the Planck mass, 
we need to fine-tune $m_B$
with 34 digits. This is often called the hierarchy problem and seen as an argument against fundamental 
scalars \cite{susskind78}. 
The main virtue of the RG approach is to 
separate the relevant and irrelevant part of the information contained in the partition
function. At each iteration, the information relevant to understand the 
large distance behavior is amplified, while the rest of the information
is discarded according to its degree of irrelevance. 
However if some ``noise'' is introduced in this process, for instance 
as round-off errors in the calculation, the error in the relevant 
direction will be amplified too. This may lead to situations 
where the amplified errors wipe out the final result.
In the case of the HM, the problem can be solved by increasing the 
arithmetic precision in the implementation of equation (\ref{eq:alg}). 
This is documented in reference \cite{numerr}.

\section{Perturbation theory with a  large field cutoff}
\label{sec:cutpt}

\subsection{Feynman rules and numerical perturbation theory}
An attractive feature of the HM is that it is possible to calculate perturbative 
series to large order by blockspinning numerically, order by order in an expansion parameter, for instance $\lambda$ for a $\lambda\phi^4$ perturbation. This method can be used analytically to reconstruct the Feynman rules \cite{ymunpublished,oktayphd}. 
In practice, the diagrammatic expansion is much more complicated than the numerical method. However, for comparison with calculations based on diagrams, it is useful to 
know the Feynman rules. 

For an initial measure of the form
\begin{equation}
		W_0(\phi)=\exp\Big(-(\ast +\frac{1}{2}m_B^2)\phi^2-\lambda \phi^4\Big)\ ,
\end{equation}
we obtain the usual Feynman rules for a $\lambda\phi^4$ theory with the following 
replacements:
\begin{eqnarray}
	\int \frac{d^D k}{(2\pi)^D} & \rightarrow & \sum _{n=0}^{\infty}2^{-n-1}\\ \nonumber
	\frac{1}{k^2+m_B^2} &\rightarrow & \frac{1}{k^2(n)+m_B^2}\ ,
\end{eqnarray}
with $k^2(n)=2\ast(c/2)^n$. The interpretation is quite simple, the integral over the 
momenta is replaced by a sum over momentum shells similar to those introduced by Wilson \cite{wilson71b,wilson72}. After one RG transformation the UV cutoff $\Lambda$ is lowered to $\Lambda/\scale=2^{-1/D}\Lambda$ and the volume of momentum space in $D$ dimension is reduced by a factor 2. 
The volume of the 0-th shell is 1/2, the volume of the 1st shell is 1/4 etc...
Similarly, $(c/2)^n=\scale^{-2n}$ represents the square of the momentum in the $n$-th shell.

\subsection{Perturbation theory with a large field cutoff}
It is well known \cite{leguillou90} that perturbative series are in general divergent.
Their zero radius of convergence is due to large field configurations \cite{pernice98,convpert}. 
However, the large field configurations have very little contributions to observables 
involving a few fields such as the magnetic susceptibility or the 4-point function.

This point was realized by the author in two different circumstances. The first is 
quantum mechanics, quantum field theory in 1+0 dimensions, where the field variable 
is usually denoted $x$. A large field cutoff can be implemented by imposing that 
the potential becomes $+\infty$ at $x=\pm x_{max}$. If the field cutoff 
$x_{max}$ is large enough, the effects on the low energy levels are exponentially 
small \cite{bacus,arbacc,tractable}. The second circumstance is the HM \cite{guide}. The numerical procedure described in section \ref{sec:num} is based on polynomial approximations and is purely algebraic, however, we need to input $R_0(k)$. So we need to 
do one integral numerically to start, namely the inverse Fourier transform of equation (\ref{eq:ft}). At the end, we need to Fourier transform if 
we want to extract the effective potential. In doing the initial integral numerically it is  
convenient to introduce a large field cutoff and then monitor the effect of this cutoff 
when it is increased. It is clear that for local measures that decay sufficiently fast, 
the effect is exponentially small for observables involving a few fields. 

\begin{figure}
\begin{center}
\includegraphics[width=0.8\textwidth]{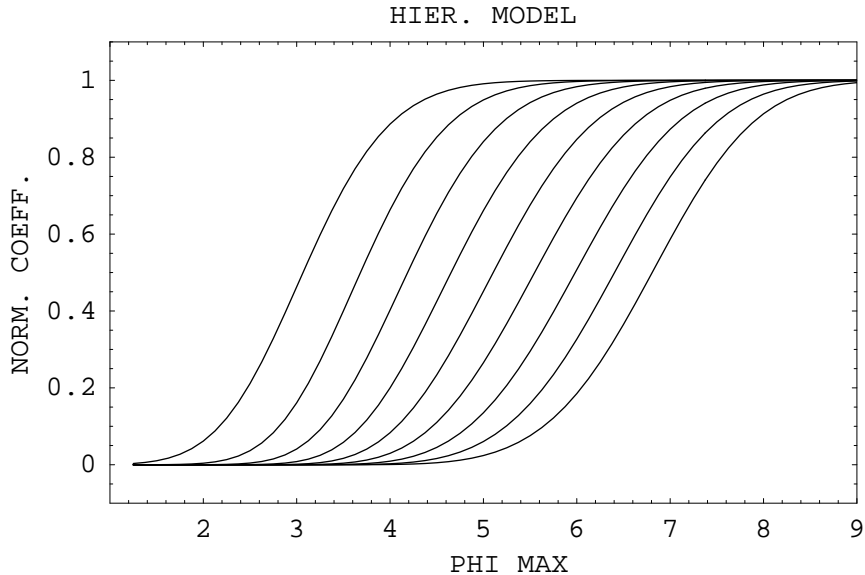}
\caption{First nine perturbative coefficients (left to right) 
for the two-point function in unit of their infinite field cutoff value, as a function of the field cutoff $\phi_{max}$.}
\label{fig:univcross}
\end{center}
\end{figure}

On the other hand, the large order of the perturbative series involve averages of large 
powers of the field and is sensitive to the field cutoff. This is illustrated in figure 
\ref{fig:univcross} where the perturbative coefficients of the zero-momentum two point function for $D=3$ are plotted in units of their value at infinite field cutoff as a function of the field cutoff.
One can see that for a fixed large field cutoff $\phi_{max}$, some low order coefficients may be close to their asymptotic values, a few coefficient may be in the 
crossover region and most coefficients are much smaller by several order of magnitude than their asymptotic value. 
These three regimes are reminiscent of the three regimes encountered when calculating renormalization group flows between two fixed points \cite{bagnuls01,pelissetto98,scalingjsp,tomboulis03}.
Note also that the shape of the transition seems 
universal as in the anharmonic oscillator case \cite{tractable,asymp06}.

This suggests a connection between the crossover observed in the 
behavior of the perturbative coefficients and the crossover behavior of the RG flows.
When we construct the RG flows starting 
near the Gaussian fixed point and let them evolve toward the high-temperature fixed point, it should be possible to describe the first iterations using the Gaussian scaling variables (see section \ref{sec:scaling}). On the other hand, after a large number of iterations, the scaling variables of the HT fixed point are 
the relevant ones. If we use regular perturbation theory, we expect that it will be impossible to find a region where the two expansions are valid due to the 
zero radius of convergence of the weak coupling expansions. On the other hand, if a field cutoff is introduced, the weak series have a nonzero radius of convergence and the direct calculations of critical amplitude as in reference \cite{scalingjsp} might be possible.
A generic feature that we then expect is that if 
we calculate the perturbative coefficients with a field cutoff, by blockspinning, 
the first coefficients should stabilize quickly, while the large order in perturbation should stabilize after more iterations. This property was verified in reference \cite{asymp06}. 

\subsection{Improved perturbative methods}
\begin{figure}
\begin{center}
\includegraphics[width=0.8\textwidth]{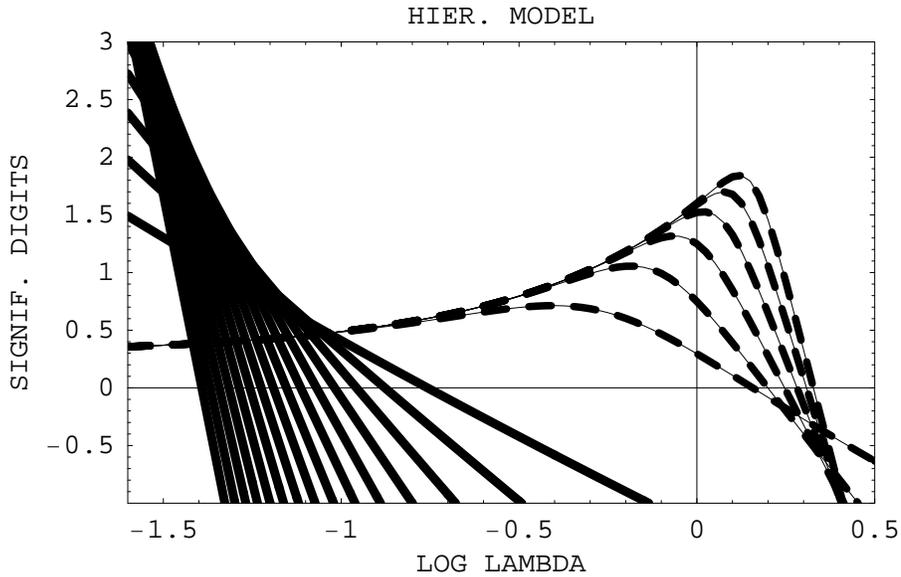}
\caption{Number of significant digits 
obtained with regular perturbation 
theory at order 1, 3, 5, ...., 15 (solid, turning clockwise with order) and
with $\phi_{max}$ = 2 (dash line),
at order 1, 3, ..., 11 (moving up with order) as a function of $\lambda$, 
for the two-point function of the HM.}
\label{fig:cutpt}
\end{center}
\end{figure}

The field cutoff significantly alter the accuracy of the perturbative series. 
This is illustrated in 
figure \ref{fig:cutpt} where the accuracy of perturbation theory for the 
two point function at various order is 
shown in regular perturbation theory and with a particular field cutoff.
The figure makes clear that at sufficiently large coupling, the modified series becomes 
more accurate than the regular series. It is also clear that for a given field cutoff, 
the accuracy peaks near a specific region of the coupling. 
It is likely that at a given coupling, it is possible to find an optimal field 
cutoff that can be determined approximately using a strong coupling expansion as 
in a simple integral discussed in reference \cite{optim}.

\subsection{Large field cutoff in ERGE}

Understanding and controlling the large field configurations is an issue that goes 
beyond the scope of perturbation theory. In particular, it appears in the context of 
the RG flows of the effective actions \cite{litim05}. It was noticed \cite{litim02b} that
the introduction of a background field suppresses large field contributions to the flows. 
We should also mention functional generalizations of the Callan-Symanzik equation 
\cite{alexandre01,polonyi04} where a running bare mass controls large fluctuations.

\section{Relation with the ERGE in the LPA approximation}
\label{sec:erge}
\subsection{Polchinski equation in the LPA}

As explained in the first sections, the RG transformation of the HM can be reduced to a 
a simple integral equation because of the very special form of the non-local interactions. In general, the 
real space RG seems difficult and one may prefer a formulation in terms of the 
Fourier transform of the fields. An UV cutoff can be introduced and lowered in variety of ways (sharp cutoff or smooth cutoff functions, ...). An effective action can be obtained by lowering the cutoff, or varying a parameter in 
the cutoff function in such a way that the large momentum components of the fields get more integrated. 
The derivative of the effective action with respect to the cutoff (or a related parameter) 
can then be expressed in terms of an integral over the momenta of a function of the action and its derivatives with 
respect to the Fourier transform of the fields. This equation  is exact and is often called an Exact Renormalization 
Group Equation (ERGE). This idea was introduced and developed in references \cite{wilson74,wegner72b,nicoll76,polchinski84} and has generated a large interest that is still ongoing. Progress have been reviewed for instance in references \cite{riedel86,morris93,bagnuls00,berges00,pawlowski05}. An ERGE can be rewritten as an infinite set of coupled partial differential equations. A simple starting point is to neglect 
the evolution of the terms in the effective action involving derivatives. This is called the Local Potential 
Approximation (LPA). 

A simple equation that can be written in this approximation for an ERGE with a smooth cutoff function is \cite{nicoll76}: 
\begin{equation}
\label{eq:nicoll}
	\frac{\partial V}{\partial t}=DV+(1-\frac{D}{2})\phi \frac{\partial V }{\partial \phi} 
	-(\frac{\partial V}{\partial \phi})^2 +\frac{\partial^2 V}{\partial \phi^2} \,
\end{equation}
where $V(t,\phi)$ is the effective potential and $t$ a parameter that increases when the cutoff decreases. 
\subsection{Infinitesimal form of Gallavotti's recursion formula}
We now consider the extension of Gallavotti's recursion formula equation (\ref{eq:galeq}).
We introduce the notations:
\begin{eqnarray}
	\gal _n (\phi)&=& {\rm e}^{-V(\Lambda,\phi)}\\
	\gal _{n+1}(\phi)&=& {\rm e}^{-V(\Lambda/\scale,\phi)}
\end{eqnarray}
In order to take the limit $\scale \rightarrow 1$, we need to fix the variance of the Gaussian weight and 
the overall normalization. We rewrite the extension of equation (\ref{eq:galeq}) for an arbitrary $\scale$ with an Gaussian weight parametrized in terms of a function $\rho(\scale)$ as
\begin{equation}
\label{eq:galinf}
\hskip-20pt
	\exp\Big(-V(\Lambda/\scale,\phi)\Big)=\int ^{+\infty}_{-\infty} \frac{d\xi}{\sqrt{\pi \rho(\scale)}}\  \exp\Big(-\frac{\xi^2}{\rho(\scale)}-\scale^D V(\Lambda,\scale^{1-D/2}\phi+\xi)\Big)\ .
\end{equation}
We require that $\rho(1)=0$ and $\rho'(1)\neq 0$, otherwise $\rho$ is arbitrary. In reference \cite{felder87}, we have 
$\rho(\scale)=2(\scale- 1)$.

The factor $(\pi \rho(\scale)) ^{-1/2}$ guarantees that 
when $\scale \rightarrow 1$, the two sides of equation (\ref{eq:galinf}) are equal. We now write $\scale = 1+\delta$ and expand in $\delta$. 
As mentioned, the terms of order 0 cancel. 
Terms of order $\delta^{1/2}$ appear but become zero after integration over 
$\xi$. Equating the terms of order $\delta$, we obtain 
\begin{equation}
\label{eq:felder}
	-\Lambda \frac{\partial V}{\partial \Lambda}=DV+(1-\frac{D}{2})\phi \frac{\partial V }{\partial \phi} 
	-\frac{\rho'(1)}{4} \Big[(\frac{\partial V}{\partial \phi})^2 -\frac{\partial^2 V}{\partial \phi^2}\Big]\ .
\end{equation}
If we write $\Lambda ={\rm e}^{-t}$ and change the $\phi$ scale in order to 
get rid of the factor $\frac{\rho'(1)}{4}$, we recover equation (\ref{eq:nicoll}). 

It should be pointed out that despite the fact that the differential equation comes as the coefficient of the 
order $\delta$ in the expansion of equation (\ref{eq:galinf}), equation (\ref{eq:felder}) is not a linearization of equation (\ref{eq:galinf}). Indeed, when we reabsorb $\rho$ in the integration variable $\xi$, we introduce a term of order $\sqrt{\rho}$ in the argument of the exponential. The terms that survive integration are quadratic in $\sqrt{\rho}$, 
namely the second derivative of $u$ and the obviously nonlinear square of the first derivative of $u$. This point was not fully understood in reference \cite{on06}. 

Note also that the Gaussian fixed point corresponds to $V=0$. The problem of finding the eigenvalues of the linearized RG transformation reduces to a time independent Schr\"{o}dinger problem. The $\phi\partial/\partial\phi$ term can be eliminated by minimal substitution, introducing a $\phi^2$ terms and the problem can be mapped into the problem of finding the eigenvalues of an harmonic oscillator. 

\subsection{The critical exponents of Polchinski's equation}
\label{subsec:polexp}
A related equation is the so-called Polchinski equation. It can be written for $N$ components as \cite{comellas97}. 
\begin{equation}
\label{eq:trav}
      \frac{\partial u}{\partial t}=\frac{2y}{N}u''+(1+\frac{2}{N}+(2-d)y-2yu)u'+(2-u)u \ ,
\end{equation}
with $y=\vec{\phi}\ . \ \vec{\phi}$, $u=2V'$ and the prime denotes derivatives with respect to $y$.
This equation can be derived \cite{ball94,comellas97} from an ERGE due to Polchinski \cite{polchinski84} 
using the LPA. For $N$=1, one can see that it follows from equation (\ref{eq:nicoll}), by reexpressing it in terms of 
$y$ and its derivatives and taking the derivative with respect to $y$ of the resulting equation. 
Equation (\ref{eq:nicoll}) is also obtained as the LPA of an ERGE due to Wilson \cite{bagnuls00}.
The exponents were calculated in reference \cite{comellas97}. In particular for $N=1$, they found $\gamma=1.2992$ which is close to the HM value. A more precise value 
$\gamma =1.29912$ was obtained in reference \cite{bervillier04}. 

In addition, Litim \cite{litim00,litim01b,litim02} proposed an optimized ERGE and 
suggested \cite{litim01,litim05} that it was equivalent to the Polchinski equation in the local potential approximation. The equivalence was subsequently proved by Morris \cite{morris05}. The value of $\gamma$ for the optimized ERGE \cite{litim02} in the case $N=1$ is 1.299124 and differ by 2 in the 5th decimal from the HM.  
More recently \cite{bervillier07}, the calculations using the optimized ERGE and Polchinski equation were 
both repeated with more accuracy and compared. The numerical difference between the 
exponents of the two (analytically equivalent) formulations was reduced to $10^{-14}$. 
Their final result is $\gamma =1.2991235477613$ which confirms 
the non-equivalence with the HM. This question is also discussed for $N>1$ in section \ref{sec:largen}.
\subsection{Infinitesimal form of Wilson approximate recursion formula}   
The infinitesimal form of Wilson approximate recursion formula
can be derived by following the same steps as for Gallavotti's recursion formula. 
First, we write $H_n[\phi]=\exp(-Q(\phi))$. We then use the arbitrariness of the scale of the fluctuations $\xi$ as previously. The only difference is that the term of order 
$\sqrt{\delta}$ disappear from 
\begin{equation}
Q((1+\delta)^{1-D/2}\phi+ \sqrt{\delta}\xi)+ Q((1+\delta)^{1-D/2}\phi- \sqrt{\delta}\xi)\ .
\end{equation}
Consequently, there seems to be no $(\partial Q/\partial \phi)^2$ term in the 
final equation. 
This point was also noticed in reference \cite{litim07}.
This suggests that the limit $\scale \rightarrow 1$ is Gaussian
($\gamma =1$). 
In reference \cite{fam}, a numerical calculation of $\gamma$ was done for 
values of $\scale = 2^{\zeta}$. $\zeta=1$ corresponds to Wilson's case while $\zeta=1/3$ 
corresponds to the HM. The 
limit we are interested in for the infinitesimal form is $\zeta \rightarrow 0$. 
Unfortunately, in this limit, the numerical procedure used in reference \cite{fam} 
becomes unstable because the errors in the integration routine become more important 
as we need to iterate more times the basic recursion formula. Figure 1 in reference \cite{fam}, indicates that $\gamma$ keeps increasing as $\zeta$ decreases. The 
last data point is for $\zeta=0.3$. For smaller values of $\zeta$ large errors bars 
develop as can be seen by repeating the calculation at closely chosen values of $\zeta$. 
For instance near $\zeta=0.15$, we found values of $\gamma$ as high as 1.34 and as low as 1.28. This calculation should be repeated with more accurate integration methods.

\subsection{Finite time singularities}
\label{subsec:finite}
It has been argued that some ERGE  in the LPA have finite time singularities \cite{hasenfratz88}. 
This is not surprising given that the solutions of the fixed point equation for equation (\ref{eq:trav}) are generically singular. More precisely, if we assume that $\partial u/\partial t=0$, we obtain a second-order differential 
equation for $u$. The solutions blow up at finite $y$ for generic ``initial'' conditions at $y=0$. 
This means that the derivative of the potential becomes singular at a finite value of the field. 
It has been shown numerically \cite{comellas97}, that regular solutions can be obtained for special initial values using the shooting method and that these solutions correspond to the non-trivial fixed point obtained with other 
formulations. Rigorous results concerning the existence of global stationary solutions of equation (\ref{eq:galinf}) can be found in reference \cite{felder87}. 
 
On the other hand, the possibility of having finite time singularities in 
Monte Carlo RG calculations for nearest neighbor models is controversial \cite{sokal94,bricmont01}. 
More generally, these reviews question the existence of renormalized or effective Gibbs measures defined by certain RG procedures. 

For the HM with the polynomial approximation discussed 
in section \ref{sec:num}, singularities after a finite number of iterations cannot appear (as they cannot appear 
for a finite dimensional quadratic map). More generally, it seems possible to prove the boundedness of $\kw_n$ or $\gal_n$ (defined in section \ref{sec:recursion}), for finite $n$, for a large class of initial functions. 

\subsection{Improvement of the LPA}
The improvement of the LPA for ERGE as a derivative expansion is a well-developed subject
\cite{morris93,bagnuls00,berges00,pawlowski05}. 
However progress is still needed in order to get estimates of the exponents which can compete in accuracy with 
the best methods available \cite{cberge}.
It is possible that the basic differential equations for the 
effective potential and the coefficients of terms involving derivatives of the fields could be 
worked backward, at finite $\scale$, in order to produce a set of manageable coupled integral equations.
The improvement of the hierarchical approximation is discussed with completely different methods in 
section \ref{sec:imp}. This could lead to $\scale\rightarrow 1$ equations that could be in turn 
compared with the existing ones. We hope that some communication between the two approaches will be 
developed in the future.

\section{The nonlinear scaling fields}
\label{sec:scaling}

\subsection{General ideas and definitions}
In the study of ordinary differential equations, a standard method \cite{arnold88} 
to go beyond the linearized approximation near a fixed point, 
consists in constructing new coordinates where the equations become linear. In the context of the RG method, these 
new coordinates are called the nonlinear scaling variables (or scaling fields) 
and were first introduced by Wegner \cite{wegner72,wegn}.

\subsection{The small denominator problem}
Rectification procedures are usually plagued with the ``small denominator problem" initially encountered 
by Poincar\'e in his study of perturbed integrable Hamiltonians \cite{poincare92}. 
In the RG case, this question needs to be discussed for each fixed point separately. 
To the best of our understanding, for the HM in $D=3$, the problem can be completely avoided (but in a non obvious way) for the HT 
fixed point, it is not present for the nontrivial fixed point and it is essential to generate logarithmic 
corrections to the scaling laws near the Gaussian fixed point \cite{wegner72,sonoda90}.

In the rest of this section, we will discuss in detail the case of the scaling variables associated with the 
HT fixed point and the nontrivial fixed point. Later we show that they can be combined in order to calculate 
non-universal critical amplitudes. The way the small denominator problem 
can be avoided for the scaling variables of the 
HT fixed point is interesting. At first sight, 
the construction seems impossible for $D=3$ and more generally for rational values of $D$, because some of the denominators are exactly zero. A numerical study in $D=3$ showed \cite{smalld} that for all zero denominators considered, a zero numerator miraculously appears. Explicit calculations in arbitrary dimensions and general 
arguments explaining why it should work to all orders were given in reference \cite{small03} and are summarized in subsection 
\ref{subsec:cancel}.

\subsection{The linear scaling variables of the HT fixed point}

As explained in section \ref{sec:num}, the RG transformation in the symmetric phase can be approximated very accurately in terms of a quadratic map in a $l_{max}$ dimensional 
space
\begin{equation}
a_{n+1, l} = \frac{u_{n,l}}{u_{n,0}} \ ,
\label{eq:aofu}
\end{equation}
with
\begin{equation}
u_{n,\sigma} = \Gamma_{\sigma}^{ \mu \nu} a_{n,\mu} a_{n,\nu} \ ,
\end{equation}
and
\begin{equation}
\Gamma_{\sigma}^{ \mu \nu}
= (c/4)^{\mu+\nu}\
\frac{(-1/2)^{\mu + \nu - \sigma}(2(\mu+\nu))!}{
(\mu+\nu-\sigma)!(2 \sigma)!}  \ ,
\label{eq:struct}
\end{equation}
for $\mu+\nu \geq\sigma$ and zero otherwise.
As in ``relativistic'' notations, 
the greek indices $\mu$ and $\nu$
go from $0$ to $l_{max}$, while latin indices $i$, $j$ go from 1 to
$l_{max}$. Repeated indices mean summation unless specified differently. 
With the normalization of equation (\ref{eq:aofu}), 
$a_n,0=1$ for any $n$ and is not a dynamical variable.
For small departure from the HT fixed point $\delta a_{n,i}$ the linear RG transformation reads 
\begin{equation}
\delta a_{n+1,i} \simeq {\mathcal M}_i^j\delta a_{n,j}\ ,
\end{equation}
with 
\begin{equation}
{\mathcal M}_i^j =2\Gamma^{j0}_i=2(\frac{c}{4})^j(-\frac{1}{2})^{j-i}\frac{(2j)!}{(2i)!(j-i)!}\ ,
\label{eq:lin}
\end{equation}
for $i\leq j$ and zero otherwise.

${\mathcal M}$ is of upper triangular form 
and the spectrum is given by the diagonal elements:
\begin{equation}
\tilde{\lambda}_{(r)}=2(c/4)^{r}=\scale^{D-r(D+2)} \ ,
\label{eq:hteigenv}
\end{equation}
in agreement with reference \cite{collet78}. Note that the quantity in the exponent of 
$\scale$ also appears in the exponent $\gamma_q$ when $q=2r$ in equation (\ref{eq:yq}).
The tilde refers to the HT variables, eigenvalues etc... in order to avoid confusion with the same quantities for the nontrivial fixed point. 
As we assume $c<2$ in order to have a well defined infinite volume limit 
(see section \ref{sec:motrig}),  
all the eigenvalues are less than 1 and the fixed point is completely attractive. As $r$ increases, the eigenvalues 
decrease and become more irrelevant. 
Equation (\ref{eq:hteigenv}) has a simple interpretation: the 2 stands for the volume increase and the $(c/4)^r$ for the rescaling of the $2r-$th power of the sum of the fields, which has a dimension $a^{D/2+1}$ in lattice spacing units, when the volume element $a^D$ is properly included, and of course assuming $\eta =0$.  The eigenvalues in equation (\ref{eq:hteigenv}) can also be seen in equation (\ref{eq:bulk})

We call ${\mathcal R}$ the matrix of right eigenvectors:
\begin{equation}
{\mathcal M}_{l}^{i} {\mathcal R}^r_i = \tilde{\lambda}_{(r)} {\mathcal R}^r_l \ ,
\label{eq:rdef}
\end{equation}
(with no summation over $r$). For convenience, the columns of ${\mathcal R}$ are ordered 
as the eigenvalues. 
We introduce the linear coordinates $\tilde{h}_{l}$ 
\begin{equation}
a_{n,l} = {\mathcal R}^r_l \tilde{h}_{n,r}\ ,
\label{eq:aofh}
\end{equation}
and which transform as 
\begin{equation}
{\tilde{h}}_{n+1,r}\simeq {\tilde{\lambda}}_{(r)}{\tilde{h}}_{n,r}
\label{eq:applin}
\end{equation}
in the linear approximation.
The matrix ${\mathcal R}^r_i$ and its inverse are also upper triangular and 
$\tilde{h}_{l}$ is of order $\beta ^l$, just as $a_{n,l}$ is.
We fix the normalization of the 
right eigenvectors in ${\mathcal R}$ in such way that all the diagonal elements are 1.
This guarantees that
$\tilde{h}_{l}=a_{n,l}+ {\mathcal O}(\beta ^{l+1})$. In reference \cite{small03} it was proved that 
for the upper diagonal elements ($j>i$),
\begin{equation}
{\mathcal R}_i^j=\Big(\frac{-c}{8-2c}\Big)^{j-i}\frac{(2j)!}{(2i)!(j-i)!}\ ,
\label{eq:pexp}
\end{equation}
and that
\begin{equation}
({\mathcal R}^{-1})_i^j=(-1)^{j-i}{\mathcal R}_i^j \ .
\end{equation}
Note that because of the HT selection rules (i. e., the upper triangular form of the matrices), 
the matrix elements do not depend on the choice of $l_{max}$. 

It is easy to rewrite the exact RG transformation in the $\tilde{h}_l$ coordinates. Starting with 
the basic equation (\ref{eq:aofu}), we replace $a_0$ by 1 and $a_l$ 
by ${\mathcal R}_l^p \tilde{h}_p$, we obtain a recursion formula of the form:

\begin{equation}
{\tilde{h}}_{n+1,l} = \frac{\tilde{\lambda}_{(l)} \tilde{h}_{l}
 + \Delta_{l}^{ p q} \tilde{h}_{p} \tilde{h}_{q} }
{1 + 2\Delta^{p0}_0 \tilde{h}_{p}
+ \Delta_{0}^{ p q} \tilde{h}_{p} \tilde{h}_{q}}\ ,
\label{eq:hrules}
\end{equation}
with coefficients calculable from equation (\ref{eq:struct}). For instance,
\[\Delta_{l}^{ p q} =({\mathcal R}^{-1})_l^{l'}\Gamma_{l'}^{p'q'}{\mathcal R}_{p'}^p
{\mathcal R}_{q'}^q \ .\]
Pursuing the relativistic analogy, upper roman indices transform with 
${\mathcal R}$ and the lower ones with $({\mathcal R})^{-1}$.
\subsection{The nonlinear scaling variables of the HT fixed point}
We now explain how to reexpress the linear variables $\tilde{h}_l$ 
in terms of the nonlinear scaling variables $\tilde{y}_l$ for which the 
approximate multiplicative transformation of equation (\ref{eq:applin}) is assumed to be exact: 
\begin{equation}
\label{eq:mult}
{\tilde{y}}_{n+1,r}=\tilde{\lambda} _{(r)}{\tilde{y}}_{n,r} \ .
\end{equation}
If we use ${\mathrm ln}(y_l)$ as our new coordinates, the RG flows become parallel
straight lines. All the dynamics is then contained in the mapping. The monotonicity of these functions 
suggests a possible connection with field theory entropy \cite{gaite96}, however the regularity near other 
fixed points may be an issue. 

As in reference \cite{wegner72,wegn}, we introduce the expansion
\begin{equation}
\tilde{h}_{l} = \tilde{y}_l+\sum_{i_{1},i_{2},\ldots} s_{l,i_{1} i_{2} \ldots} {\tilde{y}}_{1}^{i_{1}}
{\tilde{y}}_{2}^{i_{2}} \ldots \ ,
\label{eq:nlexph}
\end{equation}
where the sums over the $i$'s run from $0$ to infinity in each variable
with at least two non-zero indices.
In the following, we use the notation ${\mathbf{i}}$ for $(i_1,i_2,\dots)$. 
More generally, such vectors will be represented by boldface characters.
The unknown coefficients $s_{l,{\mathbf{i}}}$ in equation (\ref{eq:nlexph}) are obtained 
by matching two expressions of $\tilde{h}_{+1,l}$, one obtained from the RG transformation of the $h_l$ given in equation (\ref{eq:hrules}), the other obtained by evolving the scaling variables according to the exact multiplicative transformation equation (\ref{eq:mult}). 
The matching conditions can be written as: 
\begin{equation}
\tilde{h}_{n+1,l}(\tilde{{\mathbf{h}}}_n(\tilde{{\mathbf y}}))=\tilde{h}_{l}(\tilde{\lambda}_{1} {\tilde{y}}_{n,1},\tilde{\lambda}_{2} {\tilde{y}}_{n,2},
\dots ) \ .
\label{eq:match}
\end{equation}
and yield the conditions
\begin{equation}
s_{l, \mathbf{i}} = \frac{N_{l,\mathbf{i}}}{D_{l,{\mathbf i}}}
\ .
\label{eq:denom}
\end{equation}
with
\begin{eqnarray}
\label{eq:num}
N_{l, \mathbf{i}} =& \sum_{\mathbf{j}+\mathbf{k} = \mathbf{i}}
 ( -\Delta_{l}^{ p q} s_{p,\mathbf{j}}
 s_{q,\mathbf{k}} +s_{l,\mathbf{j}} \prod_{m} \tilde{\lambda}_{(m)}^{j_{m}}
2\Delta^{p0}_0 s_{p, \mathbf{k}} )\nonumber \\
&+\sum_{\mathbf{j}
+\mathbf{k}+\mathbf{r}=\mathbf{i}}
 s_{l, \mathbf{j}} \prod_{m} \tilde{\lambda}_{(m)}^{j_{m}} \Delta_{0}^{ p q}\ ,
 s_{p, \mathbf{k}} s_{q, \mathbf{r}} \ .
\end{eqnarray}
and
\begin{equation}
D_{l,{\bf i}}=\tilde{\lambda}_{(l)}-\prod_{m} \tilde{\lambda}_{(m)}^{i_{m}} \ .
\label{eq:deno}
\end{equation}
For a given set of indices $\mathbf{i}$, we
introduce the notation
\begin{equation}
{\mathcal{I}}_q ({\mathbf i})\equiv\sum_m i_m m^q \ .
\label{eq:indices}
\end{equation}
One sees that ${\mathcal{I}}_0$ is the degree of the associated 
product of scaling variables and ${\mathcal{I}}_1$
its order in the HT expansion (since $y_l$ is also of order $\beta ^l$).
Given that all the indices are positive and that at least one index
is not zero,
one can see that if ${\mathbf{j}}+{\mathbf{k}}={\mathbf{i}}$ then
${\mathcal I}_q({\mathbf{j}})<{\mathcal I}_q({\mathbf{i}})$ and
${\mathcal I}_q({\mathbf{k}})<{\mathcal I}_q({\mathbf{i}})$. Consequently,
equation (\ref{eq:num}) yields a solution order by order in ${\mathcal{I}}_0$ or
in ${\mathcal{I}}_1$ (since the r.h.s is always contains 
$s_{l,{\mathbf{i}}}$ of lower order in ${\mathcal{I}}_0$ and ${\mathcal{I}}_1$) 
provided that none of the denominators $D_{l,{\bf i}}$ are exactly zero.

We can now rewrite the denominators as 
\begin{equation}
D_{l,{\bf i}}=
2(\frac{c}{4})^l-2^{{\mathcal I}_0({\mathbf i})}
(\frac{c}{4})^{{\mathcal I}_1({\mathbf i})}\ .
\end{equation}
The parametrization 
of $c$ in equation (\ref{eq:cdim}), implies that a 
a zero denominator appears when
\begin{equation}
D-l(D+2)=D{\cal{I}}_0({\mathbf i}) -(D+2){\cal{I}}_1 ({\mathbf i})\ .
\end{equation}
Given that the ${\cal{I}}_q$ are integers, this can only occur at some rational 
values of $D$. Ignoring temporarily this set of values, we can say that for generic values 
of $c$, the denominators are not zero. 
In the spirit of dimensional regularization, 
we can perform, order by order in the HT order ${\mathcal I}_1$, the 
construction of the $s_{l,{\mathbf i}}$ for 
a generic value of $c$ and discuss the limit where $c$ takes some special value 
at the end of the calculation.
Since the linear problem is completely solved and we may 
assume ${\mathcal{I}}_0({\mathbf i}) >1$. In addition, since both $h_l$ and $y_l$ are of order $\beta^l$, 
we need ${\mathcal{I}}_1({\mathbf i}) \geq l$. At lowest non-trivial order in $\beta$, we 
have ${\mathcal{I}}_l({\mathbf i})= l$, and it has been shown \cite{small03} that, at that order, $s_{l,\bf i}$ has only an apparent pole of order $l$ at $c=0$ exactly canceled by a zero of the same order in the numerator.
If ${\mathcal I}_1({\mathbf i})>l$, we can write 
\begin{equation}
D_{l,{\mathbf i}}=2(\frac{c}{4})^l(c_{crit.})^{l-{\mathcal I}_1({\mathbf i})}\  T_{l,{\mathbf i}} ,
\end{equation}
with
\begin{equation}
T_{l,{\mathbf i}}=(c_{crit.}^{{\mathcal I}_1({\mathbf i})-l}-c^{{\mathcal I}_1({\mathbf i})-l})\ ,
\label{eq:tli}
\end{equation}
and
\begin{equation}
c_{crit.}=4\times 2^{(1-{\mathcal I}_0({\mathbf i}))/
({\mathcal I}_1({\mathbf i})-l)}\ .
\end{equation}
The only poles that we need to worry about are those where $0<c_{crit}<2$. 
An inspection \cite{small03} of the 175 terms up to order 
$\beta ^7$ shows that all the poles at $0<c<2$ were exactly canceled by zeros of the same order. 
Note that the maximally simplified rational expression for the $s_{l,{\mathbf i}}$ do have poles but at 
values of $c$ outside of the range $0<c<2$. A possible strategy for a proof would be to show inductively that for 
each term with $0<c_{crit}<2$, the cancellation occurs and so the undesired poles do not propagate to higher order.
Such an algebraic proof 
seems difficult because the $s_{l,{\mathbf i}}$ are rational expressions and the zeros 
at the numerator only appear after factorization of sums of such terms. 
It seems nevertheless reasonable to conjecture that $s_{l,{\mathbf i}}$ have no poles $0<c_{crit}<2$. 
If this conjecture is correct,
dimensional regularization provides a unique continuous expression for the coefficients for any $c$ with $0<c<2$ and 
the model is formally ``solvable" using the recursion for the coefficients given 
by equation (\ref{eq:num}).
The conjecture implies that for any value of $c$ in this interval, 
we can construct analytical expression of $a_{n,l}$ (which contains
all the thermodynamical quantities)
in terms
of $a_{0,l}$ (which depends on the initial energy density):
\begin{equation}
a_{n,l}=({\mathcal R}^{-1})_l^r\tilde{h}_r(\tilde{\lambda}_1^n\tilde{y}_1({\mathbf a}_0),\tilde{\lambda}_2^n\tilde{y}_2({\mathbf a}_0),
\dots)\ .
\label{eq:solve}
\end{equation}
We will see that the initial values of ${\mathbf y}({\mathbf a}_0)$ have a simple 
interpretation given in equation (\ref{eq:yint}). 

It is also possible to express the nonlinear scaling variables in terms of the linear variables. 
Writing, 
\begin{equation}
{\tilde{y}}_{l} = \tilde{h}_l+\sum_{{\mathbf i}} r_{l,{\mathbf i}} \prod _m \tilde{h}_m^{i_m} \ , 
\label{eq:nlexpy}
\end{equation}
we can determine order by order the unknown coefficients $r_{l,{\mathbf i}}$ of the expansion for generic 
values of $c$.
\subsection{Argument for the cancellation to all orders}
\label{subsec:cancel}
The 
generating function of the connected parts of the average values of the total field reads
\begin{equation}
{\rm ln}(R_n(k))=a^c_{n,1}k^2+a^c_{n,2}k^4+\dots \ , 
\label{eq:conngen}
\end{equation}
with 
\begin{equation}
a^c_{n,l}=\sum_{{\mathbf i}:{\mathcal I}_1({\bf i})=l}(-1)^{{\mathcal I}_0({\bf i})-1}({\mathcal I}_0({\bf i})-1)!\prod_m \frac{a_m^{i_m}}{i_m!}\ .
\label{eq:conna}
\end{equation}
We are working in the HT phase and that we do not need to subtract powers of 
the magnetization.
After a suitable rescaling of $k$ described in subsection \ref{subsec:pol}, we have 
\begin{equation}
a_{n,l}^c=(- \beta)^l \frac{1}{2l!}(\frac{c}{4})^{ln}\langle (\phin) ^{2l}\rangle ^c \ ,
\label{eq:acinter}
\end{equation}
We assume that the initial values $a_{0,l}$ are such that,
\begin{equation}
\lim_{n\rightarrow \infty }\chi^{(q)}_n=\chi^{(q)}\ ,
\label{eq:lim}
\end{equation}
is finite. 
In other words, we assume that $\beta<\beta_c$ and that all the $\chi^{(q)}$ are finite. 
As explained in section \ref{sec:motrig}, this statement can be proved rigorously for a Ising measure. 
From equation (\ref{eq:acinter}), it is then clear that 
for $n$ large enough, we have the leading scaling
\begin{equation}
a^c_{n,l}\propto\Big(2 (\frac{c}{4})^l\Big)^n=\tilde{\lambda}_{(l)}^n \ .
\label{eq:aclead}
\end{equation}
This suggests a simple relationship between $a^c_{n,l}$ and 
${\tilde{y}}_{n,l}$. Using M\"obius inversion formula \cite{polyzou80,kowalski81}, it has been shown that  
\begin{equation}
a_{l}^c=\tilde{y}_l+{\mathcal O}(\beta^{l+1})\ .
\label{eq:aclow}
\end{equation}
Equation (\ref{eq:aclow}) means that there are no nonlinear contributions of order $\beta ^l$ to $a_l^c$. For instance, there are no $y_1^3$ or $y_1y_2$ terms in $a_3^c$. 
This is expected because the nonlinear terms of order  $\beta ^l$ scale faster than 
$y_l$,  (assuming $0<c<2$). We say that a term ``scale faster", we mean that it 
goes to zero at a slower rate when $n$ becomes large.
In general, at each RG step, a term $\prod_m y_m^{i_m}$ of order  $\beta ^l$ is multiplied by 
\[2^{{\mathcal I}_0({\mathbf i})}(\frac{c}{4})^l > \tilde{\lambda}_{(l)}=2(\frac{c}{4})^l .\]
The strict inequality comes from the fact that for the nonlinear terms ${\mathcal I}_0({\mathbf i})>1$. 
It is thus clear that nonlinear terms of order $\beta^l$ would spoil the HT
scaling of equation (\ref{eq:aclead}) and contradict the existence of a infinite volume limit.

For higher order terms, the sign 
of the denominator $D_{l,{\mathbf i}}$ introduced in equation (\ref{eq:deno}) tells us 
whether or not the term scales faster or slower than the linear term. With our sign 
convention,
$c>c_{crit.}({l,{\mathbf i}})$, means $D_{l,{\mathbf i}}<0$ and the term spoils the 
HT scaling equation (\ref{eq:aclead}). Since the coefficients are rational functions 
of $c$, they cannot vanish suddenly when $c$ becomes larger than $c_{crit.}({l,{\mathbf i}})$. Consequently whenever $0<c_{crit.}({l,{\mathbf i}})<2$, the coefficient of the corresponding term 
is expected to vanish identically. 

We have checked that this argument is consistent with our previous explicit calculations. We have used equations (\ref{eq:conna}), (\ref{eq:aofh}) 
and the already calculated coefficients in equation (\ref{eq:nlexph}) to calculate
\begin{equation}
a_{l}^c = \tilde{y}_l+\sum_{{\mathbf i}: {\mathcal I}_1({\mathbf i})>l} t_{l,{\mathbf i}} {\tilde{y}}_{1}^{i_{1}}
{\tilde{y}}_{2}^{i_{2}} \ldots \ ,
\label{eq:nlexpac}
\end{equation}
up to order 7. For all the 50 terms 
with $0<c_{crit.}<2$, the corresponding $t_{l,{\mathbf i}}$ 
are identically zero.

The existence of an infinite volume limit implies that the small denominator problem can be evaded for 
any $c$ such that $0<c<2$. 
We have constructed the $a_l^c$ in terms of the 
$a_l$. However we could have proceeded directly, writing 
the $a^c_{n+1,l}$ in terms of the $a_{n,l}^c$:
\begin{equation}
a^c_{n+1,l}={\mathcal M}_l^k a^c_{n,k}+\sum_{k+q\geq l} v_l^{kq}a^c_{n,k}a^c_{n,q}+\dots	
\end{equation}
The coefficients $v_l^{kq}$ and the higher order ones can 
be obtained by using the expansion of equation (\ref{eq:conngen}) in the logarithm of equation (\ref{eq:rec})
and expanding order by order in ${\mathbf a}_n^c$.
The series does not terminate. The linear transformation is the same as before because
$a_l^c$ and $a_l$ only differ by nonlinear terms.
Using \begin{equation}
a_{n,l}^c = {\mathcal R}^r_l \tilde{h}_{r}^c\ ,
\label{eq:aofhc}
\end{equation}
we obtain
\begin{equation}
\tilde{h}^c_{n+1,l}=\tilde{\lambda}_{(l)} \tilde{h}^c_{n,l}+ \sum_{k+q\geq l}w_l^{kq}\tilde{h}^c_{n,k}\tilde{h}^c_{n,q}+\dots	
\end{equation}
We then introduce the expansion
\begin{equation}
\tilde{h}_{l}^c = \tilde{y}_l+\sum_{{\mathbf i}: {\mathcal I}_1({\mathbf i})>l}s^c_{l,{\mathbf i}} \prod _m \tilde{y}_m^{i_m} \ ,
\label{eq:hcnlexp}
\end{equation}
and obtain
\begin{equation}
s^c_{l, \mathbf{i}} = \frac{N^c_{l,{\mathbf i}}}{D_{l,{\mathbf i}}}
\label{eq:sc}
\ .
\end{equation}
with $N^c_{l,{\mathbf i}}$ given by a formula similar to equation (\ref{eq:num}), except
that it does not terminate. A detailed analysis shows that the 
two formulas have in common that the numerator depends 
only on coefficients of strictly lower orders in $\beta$, and equation (\ref{eq:sc}) can be used order by order 
in $\beta$ to construct the $s^c_{l, \mathbf{i}}$ for generic values of $c$.

Since ${\mathcal R}^{-1}$ is upper triangular, we see from equation (\ref{eq:aofhc}) that $\tilde{h}_l^c$ is equal to $a_l^c$ plus terms which go to zero faster. Consequently, 
for large $n$, the leading scaling is 
\begin{equation}
\tilde{h}^c_{n,l}\propto \tilde{\lambda}_{(l)}^n \ .
\label{eq:hclead}
\end{equation}
Following reasonings used before, this implies that terms 
in the expansion equation (\ref{eq:hcnlexp}) 
that scale faster than $y_l$ for any $0<c<2$ should have a vanishing coefficient.
In other words:
\[ 0<c_{crit.}(l,{\mathbf i})<2 \Rightarrow s^c_{l,{\mathbf i}}=0 \ .\]
Given the specific form of the $s^c_{l,{\mathbf i}}$ given in equation (\ref{eq:sc}), the $\tilde{h}_l^c$ have no poles for $0<c<2$. The $a_l^c$ being linear combinations of $\tilde{h}_l^c$ and the $a_l$ being linear combinations of products of 
$a_l^c$, we conclude that the expansion of the $a_l$ in terms of the 
scaling variables have also no poles for $0<c<2$, 

Again we see that there exists a unique continuous definition of the scaling variables that 
can be used at particular values of $c$ where the denominator is exactly zero.
From a practical point of view, the calculation at fixed $c$ of the $s^c_{l, \mathbf{i}}$ is easier
than the calculation of the $s_{l, \mathbf{i}}$, because 
no limit needs to be taken explicitly. The $s^c_{l, \mathbf{i}}$ being rational function of $c$ cannot be 
zero everywhere except at isolated values. Consequently, we can set to zero the $s^c_{l, \mathbf{i}}$ having 
$c_{crit.}(l,{\mathbf i})<2$ even at values of $c$ where $D_{l,{\mathbf i}}=0$.
 
The initial values $\tilde{{\mathbf y}}_0$ 
have a very simple interpretation. We know that 
${\tilde{y}}_{n,l}$ is the \textit{ only} leading term of $a_{n,l}^c$ when $n$ becomes large.
If at a given $0<c<2$, a nonlinear terms scales exactly like ${\tilde{y}}_{n,l}$, then by 
increasing $c$ slightly (but keeping $c<2$), we can make this term dominant 
in contradiction with the existence of the infinite volume limit.
Consequently,
\begin{equation}
\lim_{n\rightarrow\infty}\tilde{\lambda}_{l}^{-n}a^c_{n,l}=\lim_{n\rightarrow\infty}\tilde{\lambda}_l^{-n}
\tilde{y}_{n,l}=\tilde{y}_{0,l} \ .
\end{equation}
From equation (\ref{eq:acinter}), we see that
\begin{equation}
\tilde{y}_{0,l}=(- \beta)^l \frac{1}{2l!}\chi^{(2l)}\ .
\label{eq:yint}
\end{equation}
Furthermore \cite{hid}, we can consider the $\chi^{(2l)}$ as functions of the initial 
values $a_{0,l}$. If we now replace these initial values by the $m$-advanced values 
$a_{m,l}$, we find that 
$ \chi^{(2l)}(a_{m,l})=\tilde{\lambda}_{l}^{m}\chi^{(2l)}(a_{0,l})$. In this sense,  
the infinite volume limit quantities $\chi^{(2l)}$ can be seen as scaling variables. 
\subsection{The nonlinear scaling variables of the nontrivial fixed point}
The construction of the nonlinear scaling variables can be repeated 
verbatim \cite{jsp02} for the non trivial fixed point. 
However, unlike the HT case, all the calculations have to be performed numerically.
Starting from the basic quadratic map of equation (\ref{eq:aofu}), introducing new coordinates that are zero at the non-trivial fixed point, reexpressing these coordinates as linear 
combinations of the right eigenvectors and fixing the scale (for instance, by requiring that the 
HT fixed point is located at $(1,1,\dots)$), we obtain a RG transformation for the linear scaling variables 
(denoted $h$) that can be written in the form:
\begin{equation}
h_{n+1,r} = \frac{\lambda_{r} h_{n,r}
 + \Delta_{r}^{ p q} h_{n,p} h_{n,q} }
{1 + \Lambda^{p} h_{n,p} + \Delta_{0}^{ p q} h_{n,p} h_{n,q} } \ ,
\label{eq:hmdrules}
\end{equation}
As in the HT case, the first subscript refers to the number of iterations and the second is the 
index of the variable. 

As discussed in section \ref{sec:num}, for $D=3$ there is one and only one eigenvalue larger than 1. 
We can express 
the linear scaling variables in terms of 
the nonlinear scaling fields (denoted $y$):
\begin{equation}
h_{n,r} = \sum_{\mathbf{i}} t_{r,\mathbf{i}} \prod_{m} y_{n,m}^{i_{m}}\ ,
\end{equation}
and proceed as before. 
One can also find expansions of the scaling fields in terms of 
the linear variables, by setting 
\begin{equation}
y_{n,r} = \sum_{\mathbf{i}} u_{r,\mathbf{i}} \prod_{m} h_{n,m}^{i_{m}}\ ,
\end{equation}
The only potential problem comes from small denominators. 
In \cite{jsp02,smalld}, a partial numerical survey of possible small denominators was done. 
The worse case found was 
$\lambda_2^9\simeq\lambda_4$ with two parts in a thousand. 

A major difference with the HT case, is the existence of a relevant direction. Consequently, $y_{1}$ plays a very special role as the coordinate along the unstable direction. As $y_{_1}$ can be expressed in terms of the linear scaling variables which are themselves linear combinations of the original coordinates $\mathbf{a}$, we can have 
\begin{equation}
	y_{1}(\mathbf{a})=0
\end{equation}
as the equation defining the stable manifold. Similarly, the unstable manifold in the $\mathbf{a}$ coordinates corresponds to the 1-dimensional trajectory $\mathbf{a}(y_{1},0,0,\dots)$.
For practical purpose, one can expand the linear scaling variables to large order in $y_{1}$ and to low order in a few irrelevant variables.

\subsection{Convergence issues}

Up to now, it has been shown that it seems possible to construct formal expansion of the linear scaling variables 
in terms of the nonlinear scaling variables or vice-versa, for the HT fixed point or the nontrivial fixed point. 
This does not mean that these expansions define analytical functions. On the contrary, the existence of multiple 
fixed points, suggests that these expansions have at best a finite radius of convergence. 
Numerical experiments \cite{jsp02} testing the scaling of the nonlinear scaling variables suggest that the two expansions have overlapping region of convergence. 
This will be illustrated in subsection \ref{subsec:overlap}.

\subsection{The scaling variables of the Gaussian fixed point}

The eigenvalues of the parity preserving linearized RG transformation at the Gaussian fixed point \cite{collet78}
are
\begin{equation} 
\lambda_{Gj}=2c^{-j}=\scale^{D-j(D-2)}
\end{equation} 
for $j=1,\ 2,\dots$. The interpretation is simply the scaling of the parity invariant 
couplings in $g_{2j}\phi^{2j}$ interactions. For $D\geq 4$ there is only one relevant direction corresponding to the 
mass term. For $2<D<4$, there are at least two and at most a finite number of relevant directions. 

This problem can be reformulated in term of the evolution operator of the harmonic oscillator 
\begin{equation}
	H=\frac{p^2}{2m}+\frac{m\omega^2x^2}{2} \ ,
\end{equation}
during a finite euclidean time $t=-i\tau$. The correspondence is $c=\exp(2\omega\tau)$ and $m=\beta (c-1)/(2-c)$ in $\hbar=1$ units. The connection with the $\scale \simeq 1+\delta $ limit discussed in section \ref{sec:erge} is 
$\omega \tau =\delta(D-2)/2$.

There are many zero denominators in integer
dimensions,
 e.g., $\lambda_1=\lambda_2^2$ for $D=3$. 
If the numerators are not zero, one
can modify \cite{wegner72} the situation by considering $n$-dependent
coefficients. This 
generates logarithmic corrections which are 
necessary. These can be observed in the large order
of the HT temperature expansion in reference \cite{ht4}.
Another possibility is to use the idea 
of dimensional regularization \cite{thooft73} as already explained in the HT case. 
This might help reinterpreting the connection between the $1/\epsilon$ poles in the $D=4-\epsilon$ regularization and the logarithmic divergences in a cutoff regularization \cite{thooft76}. 

Practical constructions of the nonlinear scaling variables remain to be developed. As the radius of convergence of perturbative series is zero it is not clear that the procedure 
would work as in the  two other cases. Modified perturbative methods where a cutoff in field space is introduced 
\cite{convpert,tractable,asymp06} might work better.

We are not aware of any explicit construction of the scaling variables for the low-temperature fixed point.

\section{Interpolation between fixed points and critical amplitudes}
\label{sec:global}
\subsection{Global RG flows}

The critical amplitudes are in general non-universal and their calculation requires that we go beyond the linear approximation and interpolate among fixed points. 
In the rest of this section, we consider mostly the flows from the non-trivial fixed point to the HT fixed point. 

\begin{figure}[t]
\begin{center}
\vskip90pt
\includegraphics[width=0.8\textwidth]{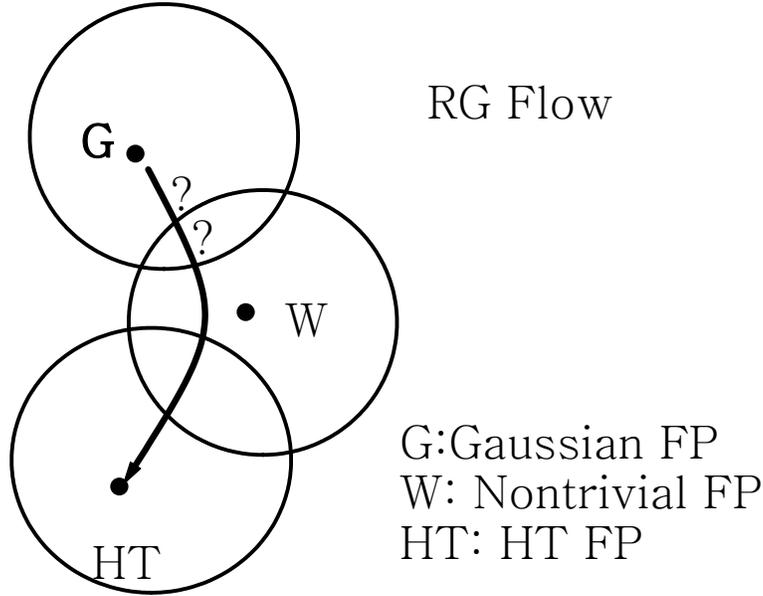}
\vskip-90pt
\caption{A qualitative description of a RG flow starting near the Gaussian fixed point, passing by the nontrivial fixed point and ending on the HT fixed point.  }
\label{fig:global}
\end{center}
\end{figure}

Given the success of field theoretical methods based on perturbation theory \cite{brez,zinnjustinbook,pelissetto00b}, 
it would certainly be desirable to extend the construction of nonlinear scaling variables with initial condition near the Gaussian 
fixed point. As explained in the previous section, this construction of the nonlinear scaling variables 
remains to be done. 
The basic picture is that the Gaussian fixed point is located on the stable manifold of the nontrivial fixed point and that it is possible, for $D<4$, to use the unstable directions of the Gaussian fixed point to reach the nontrivial fixed point. For this reason, the Gaussian fixed point is often called the UV fixed point and 
the nontrivial fixed point the IR fixed point. 
The situation is depicted in figure \ref{fig:global}. 
\subsection{Critical amplitudes and RG invariants}
In section \ref{sec:scaling}, we have seen that the construction of the coordinates $\mathbf{a}$ in terms of the 
nonlinear HT scaling variables provides a formal solution to the problem of the RG flows. However, this is not the end of the story since it is not clear how accurate finite order expansions can be. Also the initial values were identified up to a factor $(- \beta)^r /2r!$ with the infinite volume susceptibilities $\chi ^{(2r)}$, the very quantities that we would like to compute! 

Indeed, our goal is to compute $\chi ^{(2r)}$ or equivalently $\tilde{y}_{0,r}$ for initial conditions near the 
nontrivial fixed point where it is easy to use the other scaling variables $\mathbf{y}$. For simplicity, we will assume that the initial conditions are exactly on the unstable direction and close to the nontrivial fixed point: 
\begin{eqnarray}
	y_{0,1}&=&u \ ,\nonumber \\
		y_{0,2}&=&0 \ ,\nonumber \\
			y_{0,3}&=&0 \ \dots\nonumber
\end{eqnarray}
More complicated cases are discussed in reference \cite{jsp02}. Under a RG transformation, $u\rightarrow\lambda_1 u$
while the HT nonlinear scaling variables transform multiplicatively according to equation (\ref{eq:mult}).  
From equations (\ref{eq:bulk}) and (\ref{eq:hteigenv}), we can rewrite 
\begin{equation}
	\gamma_{2r}=-\frac{{\rm ln}\tilde{\lambda}_{(r)}}{{\rm ln}\lambda_1}
\end{equation}
 This implies that 
\begin{equation}
	 C_r\equiv {\tilde{y}}_{n,r}(y_{n,1})^{\gamma_{2r}}\ ,
\end{equation}
 are $n$-independent or in other words, RG-invariant. These relations suggest that 
the two fixed points are in some approximate sense dual 
\cite{dual} to each others. 
We can now rewrite the basic quantities that we want to calculate as 
\begin{equation}
	{\tilde{y}}_{0,r}={\tilde{y}}_{0,r}(y_{0,1})^{\gamma_{2r}}(y_{0,1})^{-\gamma_{2r}}=C_r(u)^{-\gamma_{2r}}
	\label{eq:trade}
\end{equation}
The constant $C_r$ are not universal. They depend on the normalization choice for $y$, however equation (\ref{eq:trade}) is independent of this choice. It is possible \cite{osc2} to relate $u$ to $\beta-\beta_c$ for a particular model 
and consequently, calculating $C_r$ provides an estimate of the critical amplitudes. The remaining question 
is: can we calculate $C_r$?

\subsection{Overlapping regions of convergence}
\label{subsec:overlap}

\begin{figure}[t]
\vskip100pt
\begin{center}
\includegraphics[width=0.9\textwidth]{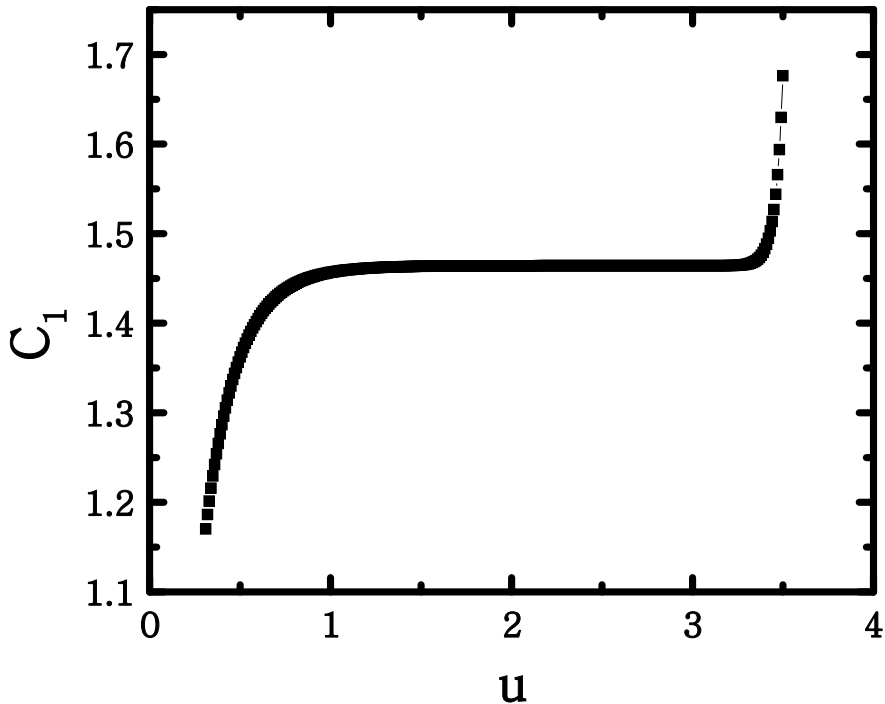}
\vskip-100pt
\caption{$C_1(u)$ defined in equation (\ref{eq:cone}) versus $u$. }
\label{fig:cone}
\end{center}
\end{figure}

In principle, the RG invariants $C_r$ can be calculated for any $n$. In practice, low $n$ calculation fail because of the low accuracy of the HT expansion and large $n$ calculations fail because of the low accuracy of the 
expansion in the nontrivial scaling variables. The accuracy of the two expansions in intermediate regions can be tested empirically by monitoring the stability of the estimates of $C_r$. In figure \ref{fig:cone}, we have displayed  
\begin{equation}
\label{eq:cone}
C_1(u)=y_1(\mathbf{h}(\mathbf{h}_W(u))u^{\gamma}\ ,
\end{equation}
with the HT scaling variable $y_1(\mathbf{h})$ calculated up to order $\beta^{11}$ and $\mathbf{h}_W(u)$ calculated up to order 80 in $u$.
The RG invariance only implies that $C_1(\lambda u)=C_1(u)$ (we remind that $\lambda\simeq 1.427$). 
A wide plateau is observed in figure \ref{fig:cone} for $1<u<3$, indicating good convergence properties 
in overlapping regions. RG invariance only forces periodicity on a log scale, but apparently the log-periodic 
oscillations are very small. 
They can be seen better in simplified models \cite{osc2}. Log-periodic oscillation are discussed into more detail below in subsection \ref{subsec:lp}. 
\subsection{Approximately universal ratios of amplitudes}
\label{subsec:approx}
Using the first $l_{max}$ HT nonlinear scaling variables, it is 
possible to construct $l_{max}-1$ constants of motion:
\begin{equation}
G_r\equiv- (2r)!\frac{{\tilde{y}}_{n,r}}{(-2{\tilde{y}}_{n,1})^{(r-1)(D/2)+r}}
\end{equation}
These quantities are RG invariants.
We can evaluate them at $n=0$. Using equation (\ref{eq:yint}), we obtain
\begin{equation}
\label{eq:gr}
G_r=(-1)^{r+1}\frac{\beta ^{\frac{D}{2} (1-r)}\chi^{(2r)}}{(\chi^{(2)})^{(r-1)(D/2)+r}}	
\end{equation}
If the conjecture \cite{glimm87} that $ (-1)^{r+1}\chi^{(2r)}>0 $ is 
correct, then $G_r>0$.

We now evaluate the universal ratio on the unstable direction of the unstable fixed point. 
We call $G_r(u)$ the corresponding value, we have 
\begin{equation}
G_r(\lambda u)=G_r(u)\ ,
\end{equation}
Consequently, we have the Fourier expansion:
\begin{equation}
	G_r(u)=\sum_qA_{r,q}u^{i q \omega} =\sum_qA_{r,q}{\rm e}^{i q \omega \ln u}\ ,
\end{equation}
with
\begin{equation}
\omega =\frac{2\pi}{\ln \lambda_1}\ .
\end{equation}
The function is clearly periodic in ln($u$) and we call the oscillations due to the non-zero Fourier modes ``log-periodic''. The coefficients can be calculated as 
\begin{equation}
	A_{r,q}=\frac{1}{\lambda_1}\int_1^{\lambda_1}\frac{du}{u}u^{-i q \omega}G_r(u)\ .
\end{equation}

The oscillatory terms are  
very small, as noticed in references \cite{osc1,osc2,dual} and we have the approximate
universal ratios 
\begin{equation}
G_r(u)\simeq A_{r,0}\ .
\end{equation}
These constants can be estimated using the methods discussed in subsection \ref{subsec:overlap}.
The smallness of the nonzero Fourier modes also applies the non-universal function $C_l(u)$ discussed in subsection \ref{subsec:overlap}.

\subsection{More about log-periodic corrections}
\label{subsec:lp}
The possibility of log-periodic terms 
were first discussed in references \cite{wilson72,
niemeijer76}. They were 
identified in the high-temperature expansion \cite{osc1,osc2} of the HM. 
The amplitudes $A^{(2l)}_{per.}$ 
are however quite small, typically, they affect the 16-th significant 
digit of the susceptibility and it takes a special effort to resolve them 
numerically. They are amplified \cite{osc1,osc2} by estimators of critical exponents 
such as the extrapolated slope $\hat{S}_m$ designed \cite{nickel80} to remove subleading corrections in estimation of the critical exponents at successive order in the HT expansion. This prevents an accurate determination 
of $\gamma$ from HT expansion \cite{osc1,osc2}.

Limit cycles often appear in two dimensional ordinary differential equations. 
However, their stability in higher dimensions is an issue debated in 
the dynamical system community. For instance, the Landau scenario for the onset of turbulence, 
based on the appearance of limit cycles, is not considered viable. For the case of interest 
here, the log-periodic oscillations reflect the discrete invariance of the original hamiltonian 
of equation (\ref{eq:ham}). This symmetry protects the periodicity even in the continuum limit. 
However, this remnant of the discrete structure is very small numerically. 

\begin{figure}[t]
\begin{center}
\vskip100pt
\includegraphics[width=0.8\textwidth]{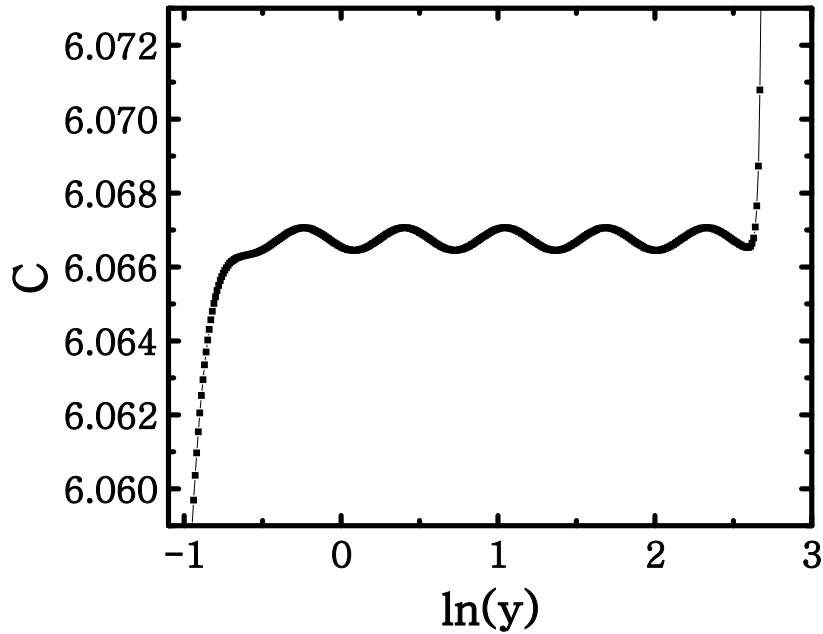}
\vskip-80pt
\caption{The analog of $C_1$ in figure \ref{fig:cone}, for the quadratic map of equation (\ref{eq:hmap}) with $\xi=0.1$, versus the natural logarithm of the scaling variable $y$.   }
\label{fig:osc}
\end{center}
\end{figure}

It is possible to design toy models \cite{osc2} where the effect is larger for instance, the quadratic map 
\begin{equation}
\tilde{h}_{n+1} = \xi \tilde{h}_{n} + (1-\xi) \tilde{h}_{n}^{2} \ ,
\label{eq:hmap}\end{equation}
for small values of $\xi$.
The first nonzero Fourier mode is quite visible in the analog of $C_1$ defined in equation (\ref{eq:cone}) as shown in figure \ref{fig:osc} for $\xi=0.1$.

\section{Nontrivial Continuum limits}
\label{sec:nontrivial}
\subsection{The infinite cutoff limit}
\label{subsec:infinite}
In this section, we apply 
the general procedure outlined by Wilson in reference \cite{wilson72} for the approximate recursion formula of equation 
(\ref{eq:warf}) to the HM .
We consider a sequence $K=1,2\dots$ of models with 
$\beta=\beta _c  -\lambda_1 ^{-K} u$
where 
$u$ is positive but not too large and $\lambda_1$ the only relevant eigenvalue as in section \ref{sec:global}.  
$\beta_c$ depends on the particular choice of initial local measure $W_0$.  
We introduce the increasing sequence of UV cutoffs 
\begin{equation}
\Lambda_K=2^{\frac{K}{D}}\Lambda_R \ , 
\end{equation}
with $\Lambda_R$ a scale of reference. 
We define the renormalized mass
\begin{equation}
\label{eq:mr}
m_R^2 =\frac{\Lambda_K^2}
{\chi^{(2)}(\beta _c  -\lambda ^{-K} u)}\ ,
\end{equation}
where ${\chi^{(2)}(\beta _c  -\lambda_1 ^{-K} u)}$ means the susceptibility at $\beta=\beta _c  -\lambda_1 ^{-K} u$.
Given that
\begin{equation}
\lambda_1^\gamma ={2^{\frac{2}{D}}},
\end{equation}
the dependence on the UV cutoff disappears at leading order and one obtains
\begin{equation}
m_R^2=\frac{\Lambda_R^2 u^\gamma}{A^{(2)}_0+A^{(2)}_1u^{\Delta}(\frac{\Lambda_R}
{\Lambda_K})^{\frac{2\Delta}{\gamma}}+LPC+\dots}
\end{equation}
with the log-periodic corrections 
\begin{equation}
LPC=A^{(2)}_{per.}\cos \left(\omega \left(\ln u+\frac{2}{\gamma}\ln \left(\frac{\Lambda_R}{\Lambda_K}\right)\right)+\phi^{(2)}\right) \ .
\end{equation}
In the infinite cut-off limit $(K\rightarrow\infty)$, the subleading corrections 
disappear. On the other hand, the LPC do not, and we are in presence of a limit 
cycle with a cutoff dependence quite similar to references \cite{wilson03,braaten03}.
Strictly speaking the infinite cutoff limit does not exist, however, for practical 
purpose, the effects of the oscillations are so small that it introduces uncertainties
that are smaller than the accuracy with which we establish the universality.

We could now define renormalized coupling constants in cutoff units using the higher order $\chi^{(q)}$ 
just as we have done in equation (\ref{eq:mr}) for the renormalized mass. However, it is usually more convenient 
\cite{parisi88} to use the value of these couplings in units of the renormalized mass. We thus consider dimensionless couplings of the form
\begin{equation}
U^{(q)}\propto \chi^{(q)}(\beta _c  -\lambda_1 ^{-L} u)m_R^{q(1+D/2)-D}\ .
\end{equation}
The fact that the quantity is dimensionless implies that the UV cutoff dependence disappears and we 
are left with a quantity proportional to the ratios of susceptibilities defined in equation (\ref{eq:gr}). 
For $D=3$, we define
\begin{equation}
U^{(2l)}\equiv {\rm lim}_{\beta\rightarrow\beta_c}(-1)^{l+1}\chi^{(2l)}\;(\chi^{(2)})^{(3-5l)/2}\;\beta^{3(1-l)/2} \ .
\label{eq:lamdastar}
\end{equation}
In subsection \ref{subsec:approx}, it has been argued that along the unstable manifold, this quantity is approximately  
universal. Indeed by taking the limit $\beta\rightarrow\beta_c$, the flow starts on the 
stable manifold but then 
ends up on the unstable manifold and the $U^{(2l)}$ should be universal in the approximation where the very small log-periodic 
oscillations are neglected. This approximate universality means that once we have picked the renormalized mass, 
all the other renormalized couplings are completely fixed.  

A counterpart of this discussion for the LPA of ERGE can be found in sections 2.10
and 3.4 of reference  \cite{bagnuls00} where a discussion of the various continuum limits that can be constructed near the Gaussian fixed point can also be found. 
Field theoretical approaches of IR stable 
trajectories are also discussed in references \cite{bagnuls90,bagnuls97,bagnuls00b}. 
It would be interesting to see how the notions developed in these articles (for 
instance the ``large river effect'') can be used for the HM. 

\subsection{Numerical estimates of the universal ratios}
\begin{figure}[t]
\begin{center}
\vskip100pt
\includegraphics[width=0.8\textwidth]{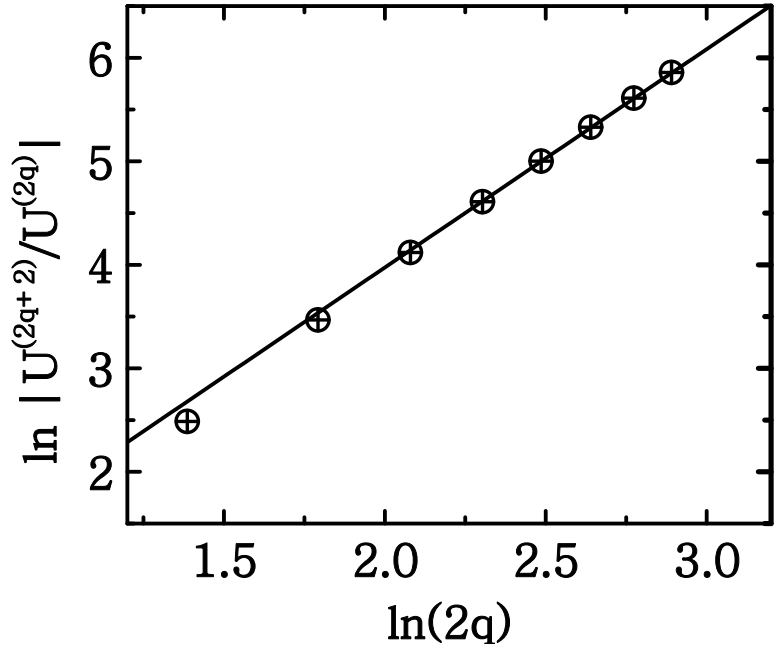}
\vskip-80pt
\caption{ ln($U^{(2q+2)\star}/U^{(2q)\star}$) versus ln($2q$). The straight line has a slope 2.11.}
\label{fig:lnrat}
\end{center}
\end{figure}

These expectations have been checked numerically in reference \cite{universality03}, by calculating 
the $U^{(2l)\star}$ for four different measures. The results were consistent with universality with six or seven 
significant digits. The results are given in table \ref{table:uq} with uncertainties of order one in the last printed digit. 
\begin{table}
\centering
\caption{\label{table:uq}Universal values of $U^{(2l)\star}$ .}
\vskip20pt
\begin{tabular}{||c|c||}
\hline
$2l$ & $U^{(2l)\star}$ \\
\hline
4 & 1.505871 \\ 6 & 18.10722 \\ 8 & 579.970 \\ 10 & 35653.8 \\
 12 & ${\displaystyle 3.57769\, {10}^6 }$ \\ 14 & ${\displaystyle 5.31763\,{10}^8}$ \\ 
 16 & ${\displaystyle1.09720\, {10}^{11}}$ \\ 
 18 & ${\displaystyle 3.00025\,{10}^{13} }$\\ 20 & ${\displaystyle 1.04998\,{10}^{16}}$ \\  
 \hline
\end{tabular}
\end{table}  
Various fits of the asymptotic behavior were performed in reference \cite{universality03} and it was concluded 
that the leading 
growth is consistent with 
\begin{equation}
\label{eq:ugrowth}
U^{(q)\star}\approx q !   \ .
\end{equation}
This is illustrated in figure \ref{fig:lnrat} where ln($U^{(2q+2)\star}/U^{(2q)\star}$) is plotted versus ln($2q$). A factorial growth as in equation (\ref{eq:ugrowth}) would imply a straight line with slope 2, which is very close 
to the slope 2.1 of the linear fit in figure \ref{fig:lnrat}.

This factorial growth is similar 
to what is found in reference  \cite{goldberg90,cornwall90,zakharov91,voloshin92}
for other models studied in the context of 
multiparticle production. Note that the generating function of the connected 
$2l$-points function has a $1/(2l)!$ factor at order $2l$ (see equation (\ref{eq:conn})) 
which means that the expansion of the generating function of the connected functions in powers of an 
external field has a finite radius of convergence.  

\subsection{Other universal ratios}
\label{subsec:other}
\begin{figure}
\begin{center}
\vskip100pt
\includegraphics[width=0.8\textwidth]{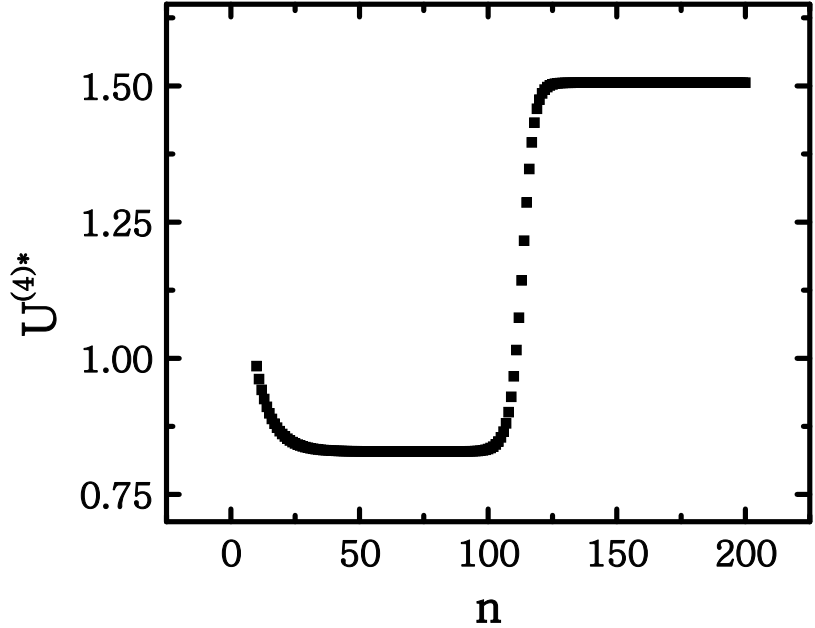}
\vskip-70pt
\caption{$U^{(4)\star}$ estimates as the number of iterations $n$ increases. }
\label{fig:u4}
\end{center}
\end{figure}

The $U^{2r\star}$ were calculated by picking an arbitrary measure, finding $\beta_c$ corresponding to that 
measure and then calculating $R_n$ until the ratios stabilize with sufficient accuracy. 
In this process, it can be noticed that when $R_n\simeq R^{\star}$, typically after 20 or 30 iterations, the 
$U^{2r}$ temporarily stabilize at different values than the final ones. 
We call these temporary values $\bar{U}^{2r\star}$. 
This is illustrated in figure \ref{fig:u4} for $r=2$. This figure shows the importance 
of going sufficiently far from the nontrivial fixed point to get the correct answer. 
Similar behavior with temporary plateaus can found for the LPA of ERGE in reference \cite{bagnuls00b}.
$\bar{U}^{2r\star}$ can be calculated from the the numerical coefficients of $R^{\star}$. 
For instance, 
\begin{equation}
	\bar{U}^{4\star}\simeq 24\times(0.053537 - 0.35871^2/2)/(2\times 0.35871)^{7/2}\simeq 0.8287
\end{equation}
This number was calculated by expanding the logarithm of equation (\ref{eq:ufp}) and then fixing the normalizations as in equation (\ref{eq:lamdastar}).
A few other values are given in table \ref{table:uqbar}. They are clearly different from 
the $U^{2r\star}$. From values for $r$ up to 10, their growth seems consistent with a simple factorial. 
\begin{table}
\centering
\caption{\label{table:uqbar}Universal values $\bar{U}^{(2l)\star}$ .}
\vskip20pt
\begin{tabular}{||c|c||}
\hline
$2l$ & $\bar{U}^{(2l)\star}$ \\
\hline
4 &  0.828719\\ 6 & 4.17757\\ 8 &49.3335 \\ 10 & 1033.20 \\
 \hline
\end{tabular}
\end{table}

\subsection{The critical potential of the symmetric phase}

Following section \ref{sec:block}, the effective potential can now be expressed as a function of $\chi^{(2)}$, 
$\magden$ and the $U^{2r\star}$. 
To simplify the notations, we define $m\equiv (\chi^{(2)})^{-1/2}$, the renormalized mass in cutoff units.
For $D=3$, the effective potential reads 
\begin{equation}
\label{eq:veffm}
	V_{eff}(\magden)=m^3 F(\magden m^{-1/2})\ .
\end{equation}
with 
\begin{equation}
	F(x)=\sum_{r=1}^{\infty} \frac{f_{2r}}{2r!}x^{2r}\ ,
\end{equation}
and  $f_2=1$, $f_4=-U^{4\star}$, $f_6=10(U^{4\star})^2-U^{6\star}$ etc...
Equations (\ref{eq:vcrit}) and (\ref{eq:veffm}) have the same form. The correspondence 
is $m\leftrightarrow \scale^{-n_{max}}$. If we calculate the universal ratios, the 
parameter $m$ or $\scale^{-n_{max}}$ disappear and we obtain numerical values 
$\bar{U}^{(2l)\star}$ for equation (\ref{eq:vcrit}) and the distinct 
values 
$U^{(2l)\star}$ for equation (\ref{eq:veffm}). 
Consequently, the function $F$ is distinct from the function $U$ in equation (\ref{eq:vcrit}). 

In reference \cite{universality03}, rescaling in both coordinates were applied in order to 
compare with a parametrization introduced by Campostrini, Pelissetto, Rossi and Vicari
 \cite{Campostrini99} where the effective potential is expressed in terms of a universal function 
\begin{equation}
A(z)=z^2/2+z^4/24+\sum_{l\geq 3}\frac{r_{2l}}{(2l)!}z^{2l}\ ,
\label{eq:aofz}	
\end{equation}
We have 
\begin{equation}
r_{2l}=f_{2r}/f_4^{r-1}\ ,
\end{equation}
and these quantities 
can be trivially reexpressed in terms of the $U^{2r\star}$, for instance, 
\begin{equation}
r_6=10-\frac{U^{(6)\star}}{(U^{(4)\star})^2}\ .
\end{equation}
Numerical values in table \ref{table:ru} are not very far from those 
calculated for nearest neighbor models. Approximate relations among the $r_{2l}$ \cite{guida96,Campostrini99} were also checked in reference \cite{universality03}. 
The first 4 coefficients of the field expansion are positive and there is no convexity issue near the origin. The discussion of the convexity at arbitrary field strength, 
requires some understanding of the radius of convergence of the expansion. 
The convexity of the effective potential has been demonstrated for ERGE, using spectral representations for the RG flows, in reference \cite{litim06} where a connection between this issue and the finite time singularity 
discussed above in subsection \ref{subsec:finite} is 
also drawn.  
\begin{table}
\centering
\caption{Universal values of $r_{2l}$ calculated numerically and 
compared to values obtained for the nearest neighbor Ising universality class. \label{table:ru}}
\vskip20pt 
\begin{tabular}{||c|c|c||}
\hline
2l & $r_{2l}$ & $r_{2l}$ for other models\\
\hline
 6 & 2.0149752 & 2.048(5)\cite{Campostrini99} \\ 
 8 & 2.679529 & 2.28(8)\cite{Campostrini99}\\ 
 10 & -9.60118 & -13(4)\cite{Campostrini99}\\ 
 12 & 10.7681 & 20(12) \cite{Morris96}\\
 14 & 763.062& 560(370)  \cite{Morris96}\\
\hline 
\end{tabular}
\end{table}  

\section{The large-N limit}
\label{sec:largen}

\subsection{Calculations at finite $N$}
If we keep the $O(N)$ symmetry unbroken, the Fourier transform of the local measure depends only on $\vec{k} .\vec{k} \equiv u$. Here $\vec{k}$ is a source conjugated to the local field variable $\vec{\phi}$. Replacing $k$ by $u$ and the second derivative by the $N$-dimensional Laplacian in equation (\ref{eq:rec}), we obtain the RG transformation for the Fourier transform of the local measure:
\begin{equation}
R_{n+1,N}(u)\propto 
\exp\Big( -\frac {1}{2} \beta ( 4u
\frac{\partial^2 }{\partial u^2}+
2N \frac{\partial }{\partial u} )
\Big)\Big( R_{n,N}( c u/4 ) \Big)
^2 \ , \label{eq:recursionN}
\end{equation}

The values of the exponents $\gamma$, $\Delta$ and the inverse critical temperature for a measure generalized 
Ising measure (or nonlinear sigma model measure) $\delta(\vec{\phi}.\vec{\phi} -1)$ calculated in reference 
\cite{on06} are given 
in table \ref{table:two}. To facilitate the comparison, we also display $\nu =\gamma /2$ (since $\eta =0$ here) and  $\omega = \Delta/\nu $ in table \ref{table:four}. The HM results coincide with the 4 digits given in 
column (2) of table 3 (for $\nu$) and 4 (for $\omega$) in \cite{comellas97} for Polchinski equation. They coincide with the six digits for $\nu$ given in the line $d=3$ of table 8 of \cite{gottker99} for 
the HM with  $N$= 1, 2, 3, 5 and 10. 
As in the case $N=1$ discussed before, we found discrepancies of order 1 in the fifth digit of $\nu$ and slightly 
larger for $\omega$ with the values found in table 1  of  \cite{litim02}. 
For $N=1$, the same discrepancy can be found in references \cite{bervillier04,bervillier07}.  
Our estimated errors are of order 1 in the 9-th digit. For $N=1$, this is confirmed by an independent method \cite{gam3}. For $N=$ 2, 3, 5, and 10, this is confirmed up to the sixth digit \cite{gottker99}. Consequently, 
a discrepancy in the 5-th digit cannot be explained by numerical errors. 
It seems clear that the two models are inequivalent. 
Note also that for  $N\geq 2$, $\alpha$, the specific heat exponent shown in table 
\ref{table:four}, is more negative than for nearest neighbor models \cite{pelissetto00b,zinnjustinbook}.

\begin{table}
\centering
\caption{$\gamma$, $\Delta$ and $\beta_c/N$ for $N=1\dots 20$. \label{table:two}}
\vskip20pt

\begin{tabular}{||c|c|c|c||} 
\hline
$N$& $\gamma$ & $\Delta$ & $\beta_c/N$ \cr
\hline
1 & 1.29914073 & 0.425946859 & 1.179030170 \cr 2 & 1.41644996 & 0.475380831 & 1.236763288 \cr 3 & 
   1.52227970 & 0.532691965 & 1.275794011 \cr 4 & 1.60872817 & 0.590232008 & 1.302790391 \cr 5 & 1.67551051 & 
   0.642369187 & 1.322083069 \cr 6 & 1.72617703 & 0.686892637 & 1.336351901 \cr 7 & 1.76479863 & 0.723880426 & 
   1.347244235 \cr 8 & 1.79469274 & 0.754352622 & 1.355791342 \cr 9 & 1.81827105 & 0.779508505 & 1.362657559 \cr 10 & 
   1.83722291 & 0.800424484 & 1.368284407 \cr 11 & 1.85272636 & 0.817977695 & 1.372974325 \cr 12 & 1.86561092 & 
   0.832855522 & 1.376940318 \cr 13 & 1.87646998 & 0.845589221 & 1.380336209 \cr 14 & 1.88573562 & 0.856588705 & 
   1.383275590 \cr 15 & 1.89372812 & 0.866171682 & 1.385844022 \cr 16 & 1.90068903 & 0.874586271 & 
   1.388107107 \cr 17 & 1.90680338 & 0.882027998 & 1.390115936 \cr 18 & 1.91221507 & 0.888652409 & 
   1.391910870 \cr 19 & 1.91703752 & 0.894584429& 1.393524199 \cr 20 & 1.92136121 & 0.899925325 & 1.394982051 \cr 
   $\infty$ & 2& 1& $\frac{2-c}{2(c-1)}=1.42366..$ \cr
\hline
\end{tabular}
\end{table}

\begin{table}
\centering
\caption{$\nu$, $\omega$ and $\alpha$ for $N=1\dots 20$. \label{table:four}}
\vskip15pt
\begin{tabular}{||c|c|c|c||} 
\hline
$N$ & $\nu =\gamma /2$ & $\omega = \Delta/\nu $& $\alpha=2-3\nu$\cr
\hline
 1 & 0.649570 & 0.655736  & 0.051289  \cr 
 2 & 0.708225 & 0.671229  & -0.124675   \cr 
 3 & 0.761140& 0.699861   &  -0.283420   \cr 
 4 & 0.804364 & 0.733787  &   -0.413092 \cr 
 5 & 0.837755 & 0.766774  &   -0.513266\cr 
 6 & 0.863089 & 0.795854  &  -0.589266\cr 
 7 & 0.882399 & 0.820355  &  -0.647198\cr 
 8 & 0.897346 & 0.840648  &  -0.692039\cr 
 9 & 0.909136 & 0.857417  &  -0.727407\cr 
 10 & 0.918611 & 0.871342 &    -0.755834\cr 
\hline
\end{tabular}
\end{table}

\subsection{Ma's equation}
The basic RG equation in the large $N$ limit was first derived by Ma \cite{ma73}. 
It can be used for conventional $O(N)$ sigma models or for the $O(N)$ version of the HM.
We consider the partition function
\begin{equation}
Z(\vec{J})=\prod _x\int_{-\infty}^{+\infty} d^N\phi_x {\rm e}^{-S
+\sum_x\vec{J}_x\vec{\phi}_x}\ ,
\end{equation}
with
\begin{equation}
\label{eq:naction}
S=-{1\over 2}\sum_{xy}
\vec{\phi}_x\Delta_{xy}\vec{\phi}_y+\sum_xV_o(\phi^2_x)\ .
\end{equation}
We use the notation $\phi^2_x \equiv \vec{\phi}_x.\vec{\phi}_x$ and 
$\Delta_{xy}$ is a symmetric
matrix with negative eigenvalues. We assume that $\sum_x\Delta_{xy}=0$. 
This condition is not satisfied by the quadratic form appearing in equation (\ref{eq:ham}).
Using results of subsection \ref{subsec:ultram}, it is possible to show that a term $\ast\sum_x\phi_x^2$ must 
added to $\beta H$, the non-local part of the action,  in order to satisfy this condition. In order to keep 
the original partition function invariant, this term must be subtracted from the original potential. 
Consequently, the potential in equation (\ref{eq:naction}) is related to the local measure introduced in section 
\ref{sec:model} by the relation 
\begin{equation}
\label{eq:subv}
	V_0(\phi^2)=-{\rm ln}W_0(\phi)-\ast \phi^2\ .
\end{equation}
One sees that $V_0=0$ for the Gaussian fixed point. 
Defining the rescaled potential
\begin{equation}
V_0(X)=NU_0({X\over N})\ ,
\label{eq:u0}
\end{equation}
and using a saddle point calculation of the partition function in the large $N$ limit, 
it is possible to show  \cite{ma73,david84} that
$M^2\equiv 2\partial V_{eff}/\partial \phi^2_c$ obeys the self-consistent
equation 
\begin{equation}
2U_0'(\phi_c^2+f_{\Delta}(M^2))=M^2\ ,
\end{equation}
where $f_{\Delta}(M^2)$ is the one-loop integral corresponding to the
quadratic form $\Delta$ and a mass term $M^2$. 
The prime denotes the derivative with respect to $\phi^2$. $M^2$ was denoted $u$ in subsection \ref{subsec:polexp}.
For the HM, it is shown in section \ref{sec:cutpt} that the function $f$ 
\begin{equation}
f_{HM}(z)=\sum_{n=0}^{\infty}{{2^{-n-1}}\over{2\ast(c/2)^n+z}}\ ,
\label{eq:fhm}
\end{equation}
For comparison, for a sharp cutoff model (SCM) in 3 dimensions, we have
\begin{equation}
f_{SCM}(z)=\int _{|k|\leq 1}{{d^3 k}\over {(2\pi)^3}}{1\over {k^2+z}}\ .
\label{eq:fscm}
\end{equation}

For a generalization of the Ising model (nonlinear sigma models) with a local measure $\delta(\phi^2-1)$ 
as in reference \cite{on06}, 
the saddle point condition simplifies to $f(M^2)=1/N$. For the HM, $\beta_c$ can then be determined from the 
condition $f_{HM}(0)=1/N$. This implies $\beta_c=N(2-c)/(2(c-1))$. This result was verified in reference \cite{on06}
and is consistent with table \ref{table:two}.

Let us consider two models, the first one with a rescaled potential
$U_0$, a UV cutoff $\Lambda$  and a quadratic form $\Delta$ 
and a second  model with a rescaled potential
$U_{0,S}$, a UV cutoff $\Lambda/S$ and a quadratic form $\Delta_S$. 
For $D=3$ and in the large-$N$ limit, the two models have the same 
dimensionful zero-momentum Green's functions provided that:
\begin{eqnarray}
\label{eq:rg}
&U&'_{0,S}(\phi^2)= \\ 
&S&^2U_0' \biggl(\bigl(\phi^2-f_{\Delta_S}(2U'_{0,S}(\phi^2)\bigr)/S+
f_{\Delta}\bigl((2/S^2) U'_{0,S}(\phi^2)\bigr)\biggr)\nonumber
\end{eqnarray}
In the two cases considered above
$f_{\Delta}=f_{\Delta_\scale}\equiv f$ and the fixed point equation becomes
very simple.
In addition, $\beta$ will be set to 1 in the rest of this section.

Following references \cite{ma73,david84}, we introduce the inverse function: 
\begin{equation}
F(2U'_0(\phi^2))=\phi^2\ ,
\end{equation}
and the function $H(z)\equiv F(z)-f(z)$.
With these notations, 
the fixed point equation corresponding to equation (\ref{eq:rg}) is simply 
\begin{equation}
H(z)=SH(z/S^2)\ .
\label{eq:fpeqh}
\end{equation}

For the SCM, $S$ is allowed to vary continuously in equation (\ref{eq:fpeqh})
and the general solution is 
\begin{equation}
F(z)=f_{SCM}(z)+Kz^{1/2}\ .
\end{equation}
For the HM, with $D=3$, $S$ can only be an integer power of $\scale=2^{1/3}$ and the
general solution has an infinite number of free parameters:
\begin{equation}
F(z)=f_{HM}(z)+\sum_qK_q z^{1/2+iq\omega}\ ,
\label{eq:hmfp}
\end{equation}
with
\begin{equation}
\omega\equiv {3\pi\over{{\rm ln}2}}\simeq 13.6\ ,
\label{eq:omega}
\end{equation}
and $q$ runs over positive and negative integers.
 There exists a unique choice of the $K_q$ in equation
(\ref{eq:hmfp}) which cancels exactly the singular part of $f_{HM}$.
It was shown in reference \cite{complexs} that the fixed point corresponds to 
\begin{equation}
F^{\star}(z)=f_{HM,reg.}={1\over{4\ast}}\sum_{l=0}^{\infty}\biggl({-z\over
  2\ast}\biggr)^l{1\over{1-c^{2l-1}}}\ .
\label{eq:fexp}
\end{equation}
This expansion has a radius of convergence $2\ast c^2\simeq 2.7024$ for $\beta=1$. 

\subsection{Singularities of the critical potential $U_0^{\star}$}
\begin{figure}[t]
\begin{center}
\vskip100pt
\includegraphics[width=0.8\textwidth]{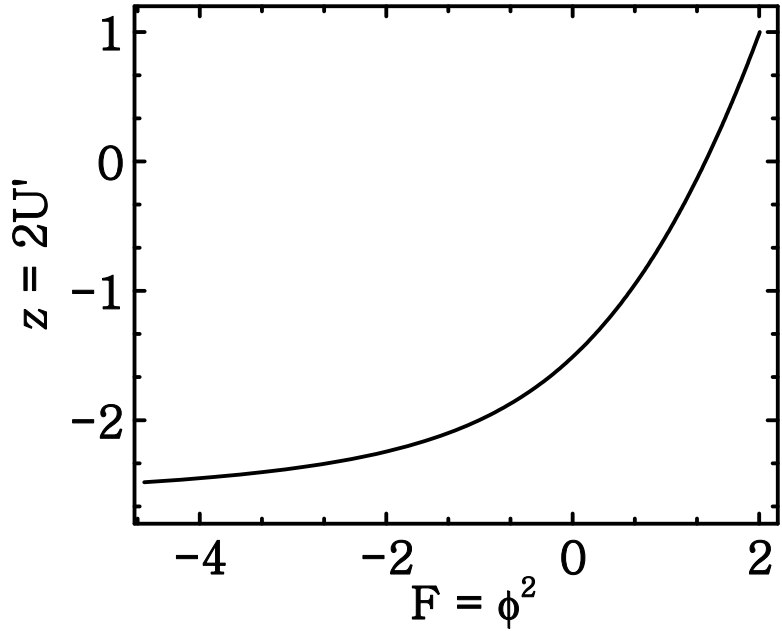}
\vskip-70pt
\caption{$z$ versus $F_{HM}^{\star}(z)$. }
\label{fig:foz}
\end{center}
\end{figure}

\begin{figure}[t]
\begin{center}
\vskip100pt
\includegraphics[width=0.6\textwidth]{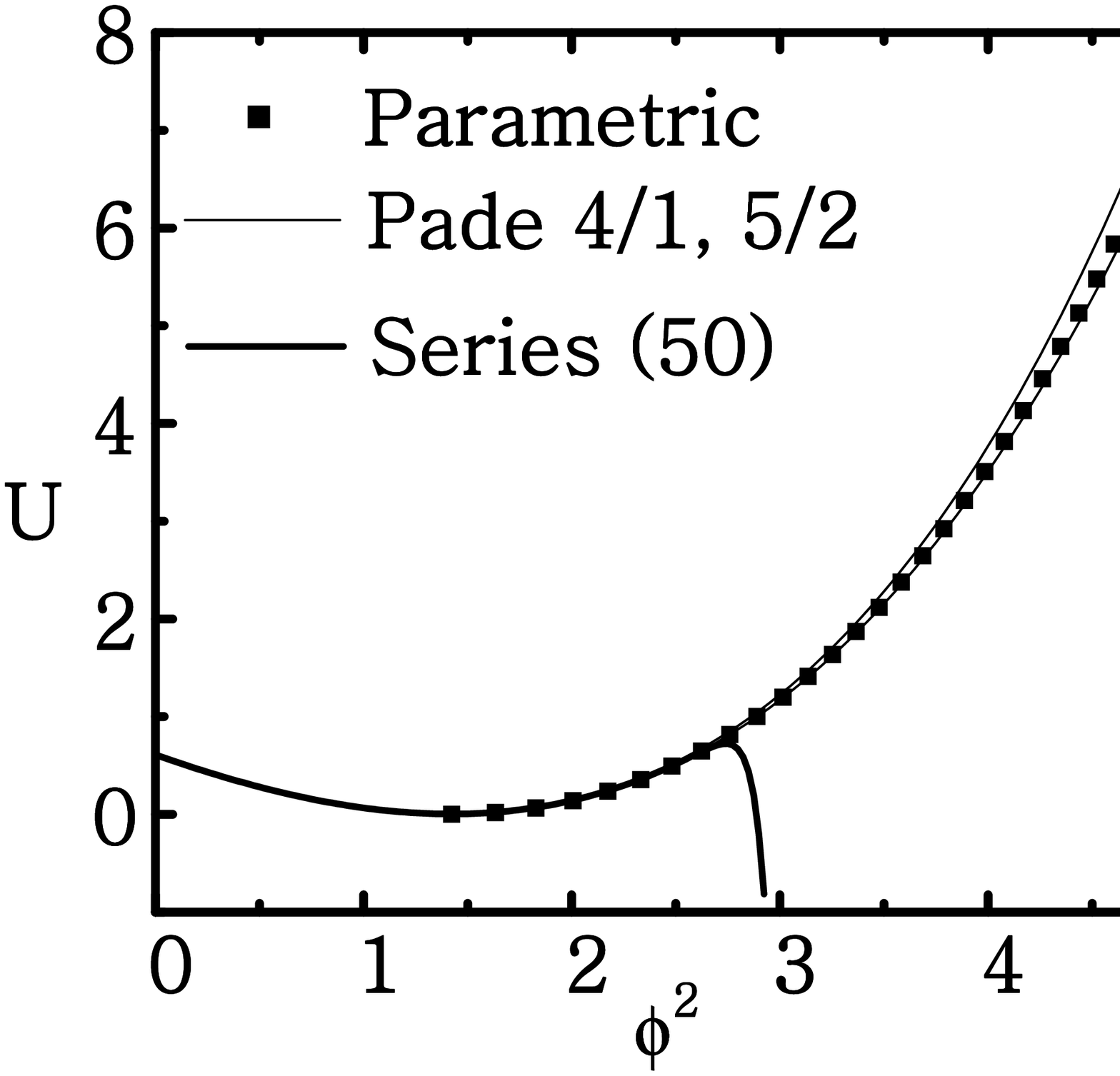}
\includegraphics[width=0.6\textwidth]{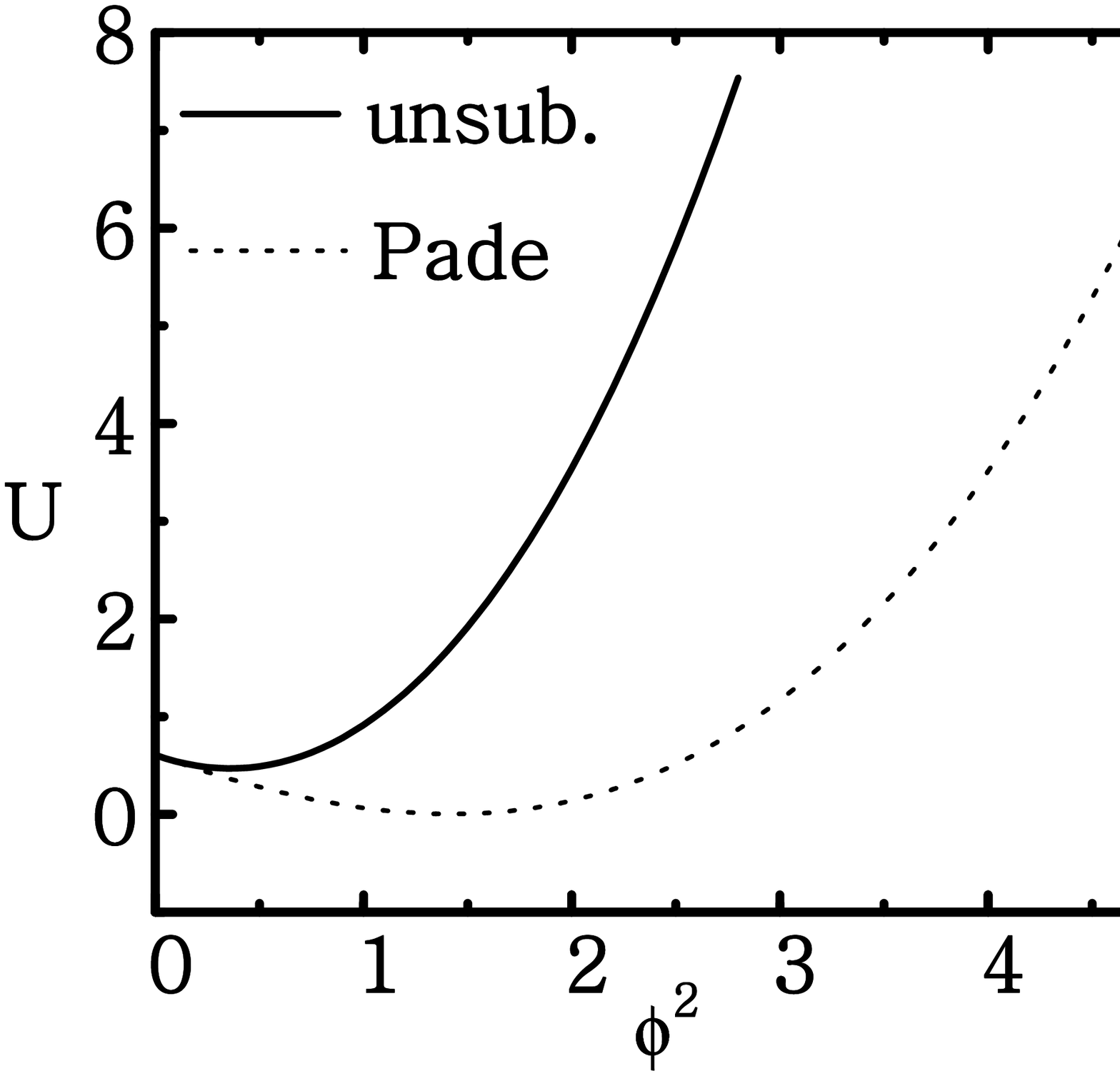}
\vskip-70pt
\caption{$U_0^{\star}(\phi ^2)$ for the HM 
with a parametric plot (filled squares), the
series truncated at order 50 (thick solid line) and Pad\'e
approximants [4/1] (thin line slightly above the squares) and 
 [5/2] (thin line closer to the squares). The constant has been fixed
 in such a way that the value at the minimum is zero (upper graph). 
 In the lower graph, we have added $\ast \phi^2$ to the [5/2] Pad\'e.}
\label{fig:uuu}
\end{center}
\end{figure}

Following reference \cite{complexs}, we can use equation (\ref{eq:fexp}) to define $F(z)$ on the negative real axis. 
We remind that $F$ plays the role of $\phi^2$. 
As we move toward more negative values of $z$, $F$ becomes 
zero within the radius of convergence of the expansion. The situation
is illustrated in figure \ref{fig:foz}. 
Numerically, $F^{\star}(-1.5107\dots)=0$. We then reexpand the series
about that value of $z$ (which corresponds to $F=\phi^2=0$) and invert
it. The resulting series is an expansion of $2U_0^{\star}$' in $\phi^2$.
After integration, and up to an arbitrary constant $u_0$, we obtain a 
Taylor series for the critical potential $U_0^{\star}$. We denote the expansion as 
\begin{equation}
U_0^{\star}(\phi^2)=\sum_{n=0}^{\infty} u_n (\phi^2)^n\ .
\label{eq:uudef}
\end{equation}
The absolute value of the coefficients appears to grow at an
exponential rate. Linear fits 
suggest a radius of convergence of order  2.5. The signs follow the periodic pattern + + - -. This suggests singularities  along the imaginary axis. This analysis is confirmed by an analysis \cite{complexs} of the poles 
of Pad\'e approximants.  
As $\phi^2$ exceeds the critical values estimated in the previous
section, the power series is unable to reproduce the expected function
$U_0^{\star}$. The situation is illustrated in figure \ref{fig:uuu}. For comparison, we have also added $\ast\phi^2$
to $U_0^{\star}$
in order to undo the subtraction of equation (\ref{eq:subv}) and obtain a function that has a simple 
relation with the effective potential as in equation (\ref{eq:vcrit}). Again, the resulting function 
appear to be convex. The numerical values of $U_0^{\star}$ in figure \ref{fig:uuu} 
have been calculated
using a parametric representation discussed in reference \cite{complexs}. Finite radius of convergence have been observed in expansions in the power of the fields based on ERGE in the LPA \cite{morris94,bervillier07}. More generally, it would be interesting to 
compare the large-$N$ limit for the HM and in the ERGE approach as discussed, for instance, in references \cite{tetradis93,litim95,tetradis95,comellas97,dattanasio97,kubyshin02}.

\subsection{Open problems}
The phase diagram of $3D$ models with a conventional kinetic term shows interesting features such as a line 
of tricritical points ending at the so-called BMB point
\cite{bmb,david84,david85}. A similar study should be done for the HM. 
A general interpretation of the complex singularities of the critical potential 
would be interesting.
From the numerical values of the exponents, it is possible to estimate the low order coefficients of the $1/N$ expansion \cite{cookpro}.  
The method of calculation of the coefficients using numerical values, was developed with 
the Sterling series for which we were able the first 7 coefficients  accurately.
These results are consistent with the hypothesis that the $1/N$ expansions 
considered are asymptotic but Borel summable (no indication for poles on the positive real axis).

\section{The improvement of the hierarchical approximation}
\label{sec:imp}
In this section, we describe the HM as a spin model on the 2-adic line. The main 
motivation for this new way to look at the model is that it suggests a way to modify 
the model in order to approximate nearest neighbor models in $D$-dimensions.
\subsection{Scalar models on ultrametric spaces}
\label{subsec:ultram}

It is possible to reformulate the HM as a scalar model on the 2-adic line \cite{lerner89}. 
We give here a presentation \cite{meuricejmp95} that does not require a detailed knowledge of the $p$-adic 
numbers. We will rewrite the hamiltonian of the HM given in equation (\ref{eq:ham}) using a function 
$v(x,y)$ which specifies 
the level $l$ at which $x$ and $y$ start to differ.
More precisely, if $x$ and $y$ are distinct,
$v(x,y)=l$ when $x_m=y_m$ for all $m$ such that
$n\geq m>l>0$
and $x_l\neq y_l$. At coinciding arguments, we define $v(x,x)=0$. 
Referring to figure \ref{fig:boxes}, we can see that $2^{v(x,y)}$ is the size of the smallest 
block containing both $x$ and $y$. 
$H$ can then be rewritten as
\def\bk{{\bf K}}
\def\bkx{{\bf K}_{xy}}
\def\cov{{c \over 4}}
\begin{equation}
\label{eq:hquad}
H=-{1\over 2}\Big(\sum _{x,y} \bkx \phi_x \phi_y + {\bf L}\sum _x
\phi_x ^2 \Big)\ ,
\end{equation}
where
\begin{equation}
\bkx = \cases{ ((\cov )^{v(x,y)}-(\cov )^{n+1})(1-\cov )^{-1}
\ {\rm if}\ x \neq y \cr
0 \ \ \ \ \ \ \ \ \ \ {\rm if }\ x=y \cr }
\end{equation}
and
\begin{equation}
{\bf L}=((\cov )-(\cov )^{n+1})(1-\cov )^{-1}\ .
\end{equation}

As made clear by the above equations, the strength
of the interaction
between two fields $\phi _x$ and $\phi _y$ depends only on the
value of $v(x,y)$. Consequently, the invariance of $v(x,y)$ under a group
of transformation 
implies the invariance of $H$ under corresponding
transformations.

For the reader familiar with the $p$-adic numbers, 
The function $2^{v(x,y)}$ is a regularized
version of the 2-adic distance. Namely,
\begin{equation}
2^{v(x,y)} = \cases{  |x-y|_2 \ {\rm if}\ |x-y|_2 > 1  \cr
                     1\ {\rm if} \ |x-y|_2 \leq 1 } 
                     \end{equation}
The 2-adic norm satisfies a relation stronger than the triangle inequality, namely 
\begin{equation}
	|x+y|_2\leq {\rm Max}(|x|_2,|y|_2)\ .
\end{equation}
Normed spaces for which this stronger inequality holds are called ultrametric spaces.
These concepts are explained in more detail for instance in references \cite{cuerna,freund}.                    

In order to describe the invariance of $v(x,y)$, we 
associate with the sequence of 0's  and 1's $x_n ..... x_1$, introduced in section \ref{sec:model} for 
$2^n$ sites,
a rational number of the form
\begin{equation}
\label{eq:frac}
x=\sum_{m=1}^n x_m 2^{-m}\ .
\end{equation}
The reason for this ``inversion'' is that $|2|_2=2^{-1}$ and similarly in momentum space, the largest shell will 
correspond to odd integers, the next shell by the multiple of 2 of these odd integers and so on.
If two numbers $x$ and $y$ have this form 
then $x+y$ can also be written in this form provide that we drop the integer part of the sum.
Equivalently, we can write $x=q/2^n$ and $y=r/2^n$ with $q$ and $r$ integers
between 0 and $2^n -1$ and add $q$ and $r$ $modulo$ $2^n$.
Since the integers $modulo\ 2^n$ form an additive group,
the set of fractions associated
with the sites form a group for the
addition modulo 1. 
The odd integers $modulo$ $2^n$ form a multiplicative
group. In the limit $n\rightarrow\infty$ the group is called the 2-adic units. We can pick a canonical form for the representatives
of such integers as
\begin{equation}
\label{eq:units}
u=1+2z
\end{equation}
where $z$ is a positive integer between 0 and $2^{n-1} -1$.
Obviously, if $x$ has the form of equation (\ref{eq:frac}) then $ux$ has
also this form after discarding its integer part.

We are now in position to define a group of transformation
acting on the fractions associated with the sites. If $x$ and $a$
have the form of equation (\ref{eq:frac}) and $u$ has the form of equation (\ref{eq:units}),
we define a transformation of $x$ depending on $a$ and $u$ and denoted
$x[u,a]$ which reads
\begin{equation}
x[u,a]=ux+a \ ,
\end{equation}
where the r.h.s is understood $modulo \ 1$.
It is clear that these transformations form a (non-abelian) group that we could 
call the Poincar\'e group of the HM.
We can
interpret $x[0,a]$ as a
translation and $x[u,0]$ as a rotation like in
the usual Poincar\'e groups. In that sense, this is a ``global''
group of transformation. This is in contrast with the symmetries
noted by Dyson \cite{dyson69} which consists in interchanging
$x_m.....x_{l+1} 1 x_{l-1}......x_1$ and $x_m.....x_{l+1} 0 x_{l-1}......x_1$
``locally'' which are discussed in section \ref{sec:model}.
It is possible to prove that 
\begin{equation}
v(x[u,a],y[u,a])=v(x,y)\ .
\end{equation}
Since $\bkx $ depends only on $v(x,y)$, this implies that
\begin{equation}
\bkx = {\bf K}_{x[u,a] y[u,a]} \ .
\end{equation}
that $H$ is  invariant under  the  transformation
\begin{equation}
\label{eq:delphi}
\phi_x \longrightarrow \phi_{x[u,a]}
\end{equation}
\noindent
We can then prove that
\begin{equation}
<\phi_{x[u,a]}\phi_{y[u,a]}>\ = \ <\phi_x \phi_y\ >\ .
\end{equation}
\subsection{Improvement of the hierarchical approximation}
\label{subsec:imp}
In this subsection, we follow references \cite{meurice93,marseille93}.
We consider a Gaussian model on a finite one-dimensional
lattice with $2^n$ sites.
The Fourier modes
of the scalar field are denoted
$\Phi _k $ where $\Phi _k^{\star} =\Phi _{-k}$ and $k$ are integers
used to express the momenta
in ${2\pi \over {2^n}}$ units.
These integers are understood modulo $2^n$ in the following (periodicity
in momentum space).
The action reads
\begin{equation}
S={1\over 2^{n+1}} \sum_{k=1}^{2^n} g(k) \Phi_k \Phi_{-k} 
\end{equation}
where $g(k)$ is even, real and positive .

We proceed in three steps. First, we relabel the momenta in a way which is
convenient to read their
``shell'' assignment as in Wilson \cite{wilson71b}. This
relabeling allows us to introduce
a group of transformation whose orbits are precisely these shells. 
The group is the multiplicative group of the 2-adic units and its 
representations are known \cite{taibleson}.
Completeness can be used to expand the kinetic term , i.e.,
the function $g(k)$, for
each of the shells.
This solves the bookkeeping problem. Nicely enough, the
classification of the representations
of the group mentioned above comes with an index indicating its resolution
power (called the degree
of ramification). This naturally provides the successive orders of our
perturbative expansion.
We then show that if we only retain the trivial representation in the
expansion, we obtain
the hierarchical approximation. In this limit, the group
of transformation is a
symmetry of the action which can be identified with the symmetry group of the
HM mentioned above.

We can relabel the momenta $k$. For this purpose, we use a set orthonormal
functions which is a discrete version of the Walsh system \cite{fine}. 
We first define
\begin{equation}
\Psi _0 (k) \equiv \cases{  1\ {\rm{if}} \ k\ =\ -2^{n-2} +1,....,2^{n-2} -1
\cr
                               \omega \ {\rm{if}}\ k\ =\ 2^{n-2} \cr
                              \omega ^{\ast} \ {\rm if} \ k \ = \ -2^{n-2} \cr
                               0 \ {\rm otherwise} } 
                               \end{equation}
and

\begin{equation}
\Psi _1 (k)\equiv 1- \Psi _0 (k)\  
\end{equation}
with the notation $\omega \equiv {{1+i}\over 2}$.

\noindent
For a given integer
$a=a_0 + a_1 2^1 + a_2 2^2+........+a_{n-1} 2^{n-1}$ with $a_l =\ 0\ {\rm or }\
 1$
we define
\begin{equation}f_a(k)\equiv \prod_{l=0}^{n-1} \Psi _{a_l} (2^l k)\ . \end{equation}
It is clear that $f_a^{\star}(k) =f_a(-k)$ and we can check that
\begin{equation}\sum_k f_a(k) f_b ^{\star}(k) = \delta _{a,b}\ .\end{equation}
A more detailed analysis shows that $f_a (k)$ is non-zero only when
$k=\pm k[a]$ for a function $k[a]$ which will
be specified. More precisely, it is possible to write
\begin{equation}f_a(k)= \omega \delta _{k,k[a]} + \omega ^{ \star} \delta_{k,-k[a]}\
.\end{equation}
This relation {\it defines} a one-to-one map $k[a]$.
In the following, $a$ will also be treated as an integer modulo $2^n$.
We can now expand
\begin{equation}\Phi _k = \sum_{a=1}^{2^n} c_a f_a (k)\ .\end{equation}
We can then rewrite
\begin{equation}S={1\over 2^{n+1}} \sum_{a=1}^{2^n} \tilde g(a) c_a ^2 \end{equation}
where $\tilde g (a) \equiv  g(k[a])$.
By construction, the $c_a$ are real field variables. For convenience
we shall also use their complex form
\begin{equation}
\label{eq:defsig}
\sigma_a \equiv {1\over 2}(c_a +c_{-a}) + {i \over 2} (c_a - c_{-a})
\ .\end{equation}

We can now explain the correspondence between this relabeling and Wilson's cell
decomposition. Clearly, if $a_0=1$, $f_a (k)$ is supported
in the high momentum
region. More precisely, the 0-th shell, i.e, the one integrated first
in the RG procedure, consists in configurations which can be expanded
in terms of the $f_{1+a_1 2+...} (k)$.
Similarly, the modes corresponding tho the $l$-th shell are made out of the
the $f_{ 2^l + a_{l+1} 2^{l+1} +....} (k)$.

\subsection{The Hierarchical Approximation and its Systematic Improvement}

In the previous section, we have introduced new field variables
$c_a$ corresponding to the $l$-th shell
when $a$ can be divided by $2^l$ but not by $2^{l+1}$.
This property is not affected if $a$ is multiplied (modulo $2^n$)
by any odd number. As explained in subsection \ref{subsec:ultram}, odd numbers form an abelian group
with respect to the multiplication modulo $2^n$.
The orbit of this group within the integers modulo $2^n$ are precisely
the sets of numbers that we have put in correspondence with the shells.
The representations of this group have been studied and classified \cite{taibleson}.
In order to use this classification, we have
to embed the labels introduced above and denoted $a$, in the
$2$-adic integers. 
When $a$ can be divided by $2^l$ but not by $2^{l+1}$, we say that the
2-adic norm, noted $|a|_2$, is $2^{-l}$. In the infinite volume limit,
or in other words when $n$ tends to infinity, the multiplicative group
of the odd numbers is the 2-adic units. The representations of
this group will be denoted $\Pi _s$. This means that
if $u_1$ and $u_2$ are 2-adic units,
then $\Pi _s (u_1 u_2) = \Pi _s (u_1) \Pi _s (u_2)$.
The label $s$
specifies the representation in a way which will be explained below.

It is easy to construct explicitly the representations $\Pi _s$.
A 2-adic
unit can be written \cite{serre}
as $u=\pm Exp(4z)$ where $z$ is a (2-adic) integer and $Exp$ the 2-adic
exponential.
$\Pi _s (u)$ is even or odd under multiplication by $- 1$.
On the other hand, $z$ is an additive parametrization and
the $z$ dependence of $\Pi _s$ will be of the
form $e^{i{2\pi z  q\over {2^r}}}$ where $q$ is an odd
integer and $r$ a positive integer. Taibleson \cite{taibleson} calls $r+2$
the degree of ramification.
In summary, the label $s$ is a short notation for the parity,
$r$ and $q$ an odd integer modulo $2^r$.

We can now use these representations to expand the the kinetic
term function $\tilde g (a)$ in each shell. For a given shell $l$,
the $a$ have the form $2^l u$ (so $|a|_2=2^{-l}$) and we can write,

\begin{equation} \label{eq:sumrep}\tilde g (2^l u) = \sum _s g{\ \atop {l,s}}  \Pi _s (u)  \end{equation}
At finite volume, i.e., at finite $n$, the units are understood
modulo $2^{n-l}$ and consequently
the sum over the representations
$s$ is restricted to $r \leq
n-l-2$. The numerical coefficients $g {\ \atop {l,s}}$ are easily calculable
using the orthogonality relations among the representation.

The hierarchical approximation is obtained by retaining only the trivial
representation in the expansion equation (\ref{eq:sumrep}). In this approximation,
and using the definition introduced in equation (\ref{eq:defsig}), the action reads
\begin{equation}S={1\over 2^{n+1}} \sum_{l=0}^{n-1} g{\ \atop {l,+,0}} \sum_{a:|a|_2 =
2^{-l}}
\sigma _a \sigma _{-a} \end{equation}
After a Fourier transform, we obtain a hierarchical model having the
general form (with arbitrary $b_l$ couplings at level $l$) (see section \ref{sec:motrig}).

The classification of the representations of the 2-adic units suggest
that we improve the hierarchical approximation by taking into
account the additional terms in the equation (\ref{eq:sumrep}) ${\it
 order\ by\ order\ in\ the\ degree\ of\ ramification}$. Intuitively, this
corresponds to the fact that the degree of ramification measures
the ``power of resolution'' of the representation. Numerically, this
works reasonably well: in a simple example, the coefficients become smaller as the degree increases \cite{meurice93}. 

The actions written above take a more familiar form after Fourier transformation
in $a$.
In general, the $l$-th momentum shell is responsible for
interactions among the averages of the Fourier transformed fields
inside boxes of size $2^l$. In the hierarchical approximation, the interactions
depend
only on the position of these boxes inside boxes of size $2^{l+1}$. When the
corrections are
introduced up to $r=r_{max}$, the interactions depend on the position inside
boxes of
size $2^{l+2+r_{max}}$. 

\subsection{The Improvement of the Hierarchical Approximation As a Symmetry
Breaking Problem}

It is important to realize that in the hierarchical approximation, $S$ 
is invariant under the transformation
\begin{equation}
\sigma_a \longrightarrow \sigma_{ua} 
\end{equation}
for any odd number $u$.
When the other terms of the expansion are incorporated, this symmetry
is broken by each term in a definite way.
This allows us to use Ward identities techniques.
We discuss the simplest case below.

Suppose we want to calculate the 2-point function, using the perturbative
expansion described
in the previous section.
First we use the new variables $c_a$ and the inverse of the map $k[a]$
defined in section 2
to write
\begin{equation}
<{\Phi _k \Phi_{-k} }> = <c^2 _{a[k]} > \ . \end{equation}

In the hierarchical approximation, the value of this expression depends
only on the momentum shell specified by $|a|_2$. In other words,
\begin{equation}
 <c^2 _{ua}>  _0 = <c^2 _{a}>_0  
\end{equation} 
where $<...> _0$ means that the quantity is evaluated, at order zero,
or in other words, in the
hierarchical approximation.

Suppose that we now include a correction $\delta \tilde g(a) $ to
the approximation of $\tilde g(a)$.
Then in first order in this perturbation, we
recover the a momentum dependence within the shells given by
\begin{equation}
< c^2 _{ua} > _1 = <c^2 _{a}> _1-{1 \over 2^{n+1}}
\sum _b (\delta \tilde
g(ub)- \delta \tilde g(b) ) <c_b^2 c_a^2> _0 
\end{equation}
In the model considered here, the  $<c_b^2 c_a^2> _0$ contribution
can be evaluated straightforwardly and we can check that we recover
the first term in the expansion of $(1/\tilde g(ua))-(1/\tilde g(a)))$.
The important point is that the corrections are evaluated using the
unperturbed action.
It is clear that similar methods can be used for higher point functions
and in the interacting case.

An extension to $D$-dimension can be constructed easily by noticing
the approximate correspondence between the integration over successive shells
and the block-spin method. On a $D$-dimensional cubic lattice,
we can decompose the block-spin procedure into $D$ steps (one in each
directions). 

\subsection{Other applications}
The 2-adic formulation of the HM can be used for other purposes. For instance, it is possible to understand 
the absence of certain diagrams in the approximate recursion formula \cite{wilson74}. The lack of diagrams with 
three lines having large momenta coming out of a vertex can be understood from the fact that the sum of two 
2-adic integers with 2-adic norm 1 is a 2-adic integer with a strictly lower norm. In simpler words, the sum of 
two odd number is even. 2-adic analysis was also used to study the Symanzik representation of the HM \cite{sym91}. 
The basic ingredient \cite{lucio90} being that the quadratic form in $H$ defines a random walk with a Hausdorff 
dimension $2/D$. 

\section{Models with approximate supersymmetry}
\label{sec:susy}

In subsections \ref{subsec:hier} and \ref{subsec:infinite}, we made clear that the continuum limit requires a 
fine-tuning and that this procedure maybe be seen as unnatural. 
This feature is clearly related to the existence of an unstable direction. In four dimensions, there are 
several ways to get rid of the unstable directions and limit the flow to marginal directions. 
One possibility is to impose a gauge symmetry that forbids a mass term for the gauge fields. 
Another possibility is to introduce new degrees of freedom with opposite statistics that 
partially cancel the quantum fluctuations. This possibility has been exploited in the perturbative 
treatment of supersymmetric models. In the following, we follow this second idea and construct models 
with approximate supersymmetry. Other hierarchical models involving fermions can be found in the literature 
\cite{dorlas90,lerner94}. 
  
We consider a free action for $N$ massless scalar fields 
$\phi_x^{(i)}$ and  fermion fields $\psi_x^{(i)}$ and $\bar{\psi}_x^{(i)}$: 
\begin{equation}
S_{free}={1\over 2}\sum_{x,y,i}\phi_x^{(i)}D^2_{xy}\phi_y^{(i)}\ +\sum_{x,y,i}\bar{\psi}_x^{(i)}D_{xy}\psi_y^{(i)}\ ,
\end{equation}
where $x$ and $y$ run over the sites and $i$ from 1 to $N$. 
The $\psi_x^{(i)}$ and  $\bar{\psi}_x^{(i)}$ are Grassmann numbers 
integrated with a measure 
\begin{equation}
\int\prod_{x,i}d\psi_x^{(i)}d\bar{\psi}_x^{(i)} \ .
\end{equation}
We require that $D_{xy}^2$ has positive eigenvalues and that we can write
\begin{equation}
D_{xy}^2=\sum_zD_{xz}D_{zy} \ .
\label{eq:sq}
\end{equation}

The free action $S_B^{free}+S_F^{free}$ is invariant at first order 
under the transformation
\begin{eqnarray}
\nonumber
\delta\phi^{(i)}_x=\epsilon\bar{\psi}_x^{(i)}+{\psi_x^{(i)}}\bar{\epsilon}\\ 
\delta\psi_x^{(i)}=\epsilon\sum_xD_{xy}\phi_y^{(i)}\\
\nonumber
\delta\bar{\psi}_x^{(i)}=\bar{\epsilon}\sum_xD_{xy}\phi_y^{(i)} \ . \\ 
\nonumber
\end{eqnarray}
The $\epsilon$ and $\bar{\epsilon}$ are Grassmann numbers.
Integration by part or Leibnitz's rule cannot be used for $D_{xy}$ and the 
order $\epsilon\bar{\epsilon}$ variations do not cancel.

We now give the explicit form for the bosonic part at finite volume.
\begin{eqnarray}
S_B^{free} &=&
-\frac{\beta_B}{2}\sum_{n=1}^{n_{max}}({c_B\over4})^n\sum_{x_{n_{max}},...,x_{n+1},i} 
(\sum_{x_n,...,x_1}\phi^{(i)}_{(x_{n_{max}},...x_1)})^2\\ \nonumber
&\ &+{\beta_B c_B\over{2(2-c_B)}}\sum_{x_{n_{max}},...,x_{n+1},i}( \phi^{(i)}_{(x_{n_{max}},...x_1)})^2\ ,
\label{eq:hambose}
\end{eqnarray}
with $c_B=c=2^{1-2/D}$. The fermionic part reads:
\begin{eqnarray}
\hskip-80pt
S_F^{free} =
-{\beta_F}\sum_{n=1}^{n_{max}}({c_F\over4})^n\sum_{x_{n_{max}},...,x_{n+1},i} 
(\sum_{x_n,...,x_1}\bar{\psi}^{(i)}_{(x_{n_{max}},...x_1)})
(\sum_{x_n,...,x_1}\psi^{(i)}_{(x_{n_{max}},...x_1)})
\\
+{\beta_Fc_F\over{2-c_F}}\sum_{x_{n_{max}},...,x_{n+1},i}
\bar{\psi}^{(i)}_{(x_{n_{max}},...x_1)}\psi^{(i)}_{(x_{n_{max}},...x_1)}\ ,
\label{eq:hamf}
\end{eqnarray}
with $c_F=2^{1-1/D}$. We have the simple relation $c_B/2=(c_F/2)^2$.
Using the techniques explained in \cite{meuricejmp95}, one can show that the fermionic
operator is the square root of the bosonic operator (see equation (\ref{eq:sq}))
provided that 
\begin{equation}
{\beta_Fc_F\over{2-c_F}}=( {\beta_B c_B\over{2-c_B}})^{1\over 2}
\end{equation}

We can introduce local interactions.
The Grassmann nature of the fermionic fields restricts severely the type
of interactions allowed. For instance, for one flavor ($N=1$), the most
general bosonic local measure is 
\begin{equation}
{\cal{W}}(\phi,\psi,\bar{\psi})=W(\phi)+\psi\bar{\psi}A(\phi)
\end{equation}
For convenience, we can reabsorb the local quadratic terms,in the local measure.
In the following, 
$W(\phi)$ will take 
the Landau-Ginzburg (LG) form:
\begin{equation}
W(\phi)\propto \exp\Big(-( {\beta_B c_B\over{2(2-c_B)}}+{1\over 2}m_B^2 )\phi^2-
\lambda_B\phi^{4}\Big) \ .
\label{eq:lg}
\end{equation}
If the two functions $W$ and $A$ are proportional, the fermionic degrees
of freedom decouple.
The renormalization group transformation takes the form
\begin{eqnarray}
W&\rightarrow & 2 A\star W \\
A&\rightarrow &2\beta_F A\star W +({4\over{c_F}})W\star W
\end{eqnarray}
where the $\star$ operation is defined as
\begin{equation}
\Big(A\star B\Big)(\phi)\equiv  e^{{\beta_B \over 2} ( \phi ^2)}
\int d\phi ' A({{(\phi2c_B^{-{1\over 2}} -\phi ')}\over 2})
B({{(\phi2c_B^{-{1\over 2}} +\phi ')}\over 2}) \ ,
\label{eq:bsp}
\end{equation}
The introduction of a Yukawa coupling can be achieved 
by having $A(\phi)$ linear. Such a term breaks explicitly the $Z_2$ symmetry of the 
LG measure. 
Models with two flavors ($i=1,2$) with the 
type of bilinear coupling appearing in the Wess-Zumino\cite{wess74} model can be written as 
\begin{eqnarray}
\nonumber
{\cal{W}}(\phi^{(i)},\psi^{(i)},\bar{\psi}^{(i)})=
W(\phi^{(i)})
+A(\phi^{(i)})(\bar{\psi}^{(1)}\psi^{(1)}+\bar{\psi}^{(2)}\psi^{(2)})+\\
B(\phi^{(i)})\psi^{(1)}\psi^{(2)}-B^{\star}(\phi^{(i)})
\bar{\psi}^{(1)}\bar{\psi}^{(2)}+
T(\phi^{(i)})\bar{\psi}^{(1)}\psi^{(1)}\bar{\psi}^{(2)}\psi^{(2)} \ .
\end{eqnarray}
This is not the most general measure, however it closes under 
the renormalization group transformation which takes the form
\begin{eqnarray}\nonumber
W&\rightarrow & (W\star T +A\star A +B\star B^{\star} )\equiv W'\\ 
\nonumber
A&\rightarrow &\beta_F W' +{4\over{ c _F}}A\star T\\ 
B&\rightarrow &{4\over{ c _F}}B\star T\\ 
\nonumber
T&\rightarrow & {8\over{ c _F^2}}T\star T+ \beta_F {8\over{ c _F}}A\star T+
(\beta_F)^2 W' \ .\\
\nonumber
\end{eqnarray}
In addition, if we impose that the function $B$ has the following form:
\begin{equation}
B(\phi^{(i)})=(\phi^{(1)}+i\phi^{(2)})P((\phi^{(1)})^2+(\phi^{(2)})^2) \ ,
\label{eq:bdef}
\end{equation}
while $W$, $A$ and $T$ are $O(2)$-invariant,
the model is then invariant under the R-symmetry
\begin{eqnarray}
\nonumber
(\phi^{(1)}+i\phi^{(2)})&\rightarrow & e^{i\theta}(\phi^{(1)}+i\phi^{(2)})\\
\psi^{(j)}&\rightarrow &e^{-i{\theta\over 2}}\psi^{(j)}\\ 
\nonumber
\bar{\psi}^{(j)}&\rightarrow &e^{i{\theta\over 2}}\bar{\psi}^{(j)} \ .
\end{eqnarray}

\begin{figure}
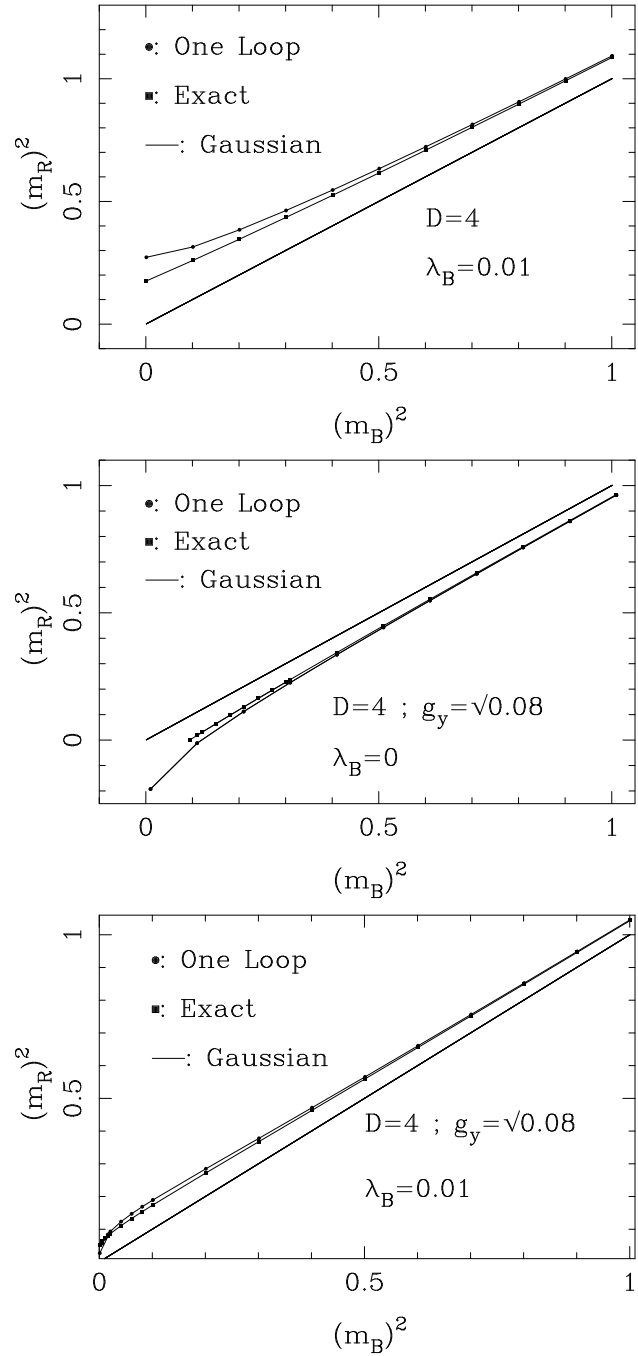

\vskip50pt
\centering
\includegraphics[width=2.3in,angle=270]{one.ps}
\vskip5pt
\includegraphics[width=2.3in,angle=270]{two.ps}
\vskip5pt
\includegraphics[width=2.3in,angle=270]{three.ps}
\vskip10pt
\caption{
The renormalized mass as a function of the bare mass in a bosonic $O(2)$
model with bare quartic coupling fixed to 0.01 (top), with bare quartic coupling fixed to 0 and a Yukawa coupling equal to
$\sqrt{0.08}$ (middle) and with bare quartic coupling fixed to 0.01 and a Yukawa coupling equal to
$\sqrt{0.08}$ (bottom).}
\label{fig:susy}
\end{figure}

We summarize three numerical calculations \cite{susy99} performed for the second model
with $D=4$.
First, the case where the fermions decouple from the bosons was considered.
$W$ takes the form
\begin{equation}
W(\phi)\propto \exp\Big(-((( {\beta_B c_B\over{2(2-c_B)}})+{1\over 2}m_B^2 )
\sum_i(\phi^{(i)})^2
+ 
\lambda_B(\sum_i(\phi^{(i)})^2)^2) \Big)\ .
\end{equation}
The value of $m_R^2$, defined as the 
inverse of the zero-momentum two-point function, is shown in the top part of figure \ref{fig:susy} as 
a function of $m_B^2$. These quantities are expressed in 
cut-off units. For reference we have also displayed the one-loop
perturbative result and the trivial Gaussian result. One sees that the 
scalar self-interaction moves $m_R^2$ up and $m_R^2\simeq 0.2$ when 
$m_B^2$ goes to zero. The one-loop result is quite good when
$m_R^2$ is large enough but deteriorates when this quantity becomes smaller.

The second calculation was done in a bosonic model with 
a bare mass $m_B$ and $\lambda_B=0$ 
coupled to a fermion with the following couplings:
\begin{eqnarray}
\nonumber
A&=&(-1-m_B)W\\
P&=&g_yW\\
\nonumber
T&=&((-1-m_B)^2 + g_y^2((\phi^{(1)})^2+(\phi^{(2)})^2))W \ .
\nonumber
\end{eqnarray}
The results are shown in the middle part of figure \ref{fig:susy} for $g_y=\sqrt{0.08}\simeq0.28$. 
One sees that the Yukawa coupling moves $m_R^2$ down.
For $m_B^2\simeq 0.094$, $m_R$ becomes 0 and for smaller of $m_B^2$, we enter
the broken symmetry phase.

Finally, the previous calculation was repeated
with $\lambda_B=0.01$ 
instead of 
0. 
In perturbation theory, the one-loop quadratic divergence cancel when
$m_B=0$ and 
\begin{equation}
8\lambda_B=g_y^2 \ ,
\end{equation}
which justifies the choice of coupling constant.
The results are shown in figure \ref{fig:susy}.
One sees that the Yukawa coupling in part cancels the effects of the scalar 
self-interaction, however, the cancellation is not as good as in the one-loop
formula where $m_R$ goes to zero when $m_B^2$ goes to zero. Instead, we 
found numerically that $m_R^2\simeq 0.044$ when $m_B^2$ goes to zero.
It is of course possible to fine-tune $g_y$ in order to get $m_R=0$.
\section{Conclusions}

In conclusion, we have shown that the calculation of the effective potential 
at the nontrivial fixed point (equation (\ref{eq:vcrit})) and for a massive theory in the symmetric phase (equation (\ref{eq:veffm}) can be performed 
numerically with great accuracy. 
The second calculation 
requires the ability to follow the RG flows all the way from the nontrivial point to the HT attractive 
fixed point. For the HM, this can be accomplished numerically or by constructing the nonlinear scaling variables. 
The situation could be compared to a quantum mechanical problem which can be solved accurately and consistently 
by different methods but for which there is no closed form solution. 
There remain many open problems: the construction of the scaling variables near the Gaussian fixed point and in the 
broken symmetry phase, the exploration of the phase diagram for models with $N$ components, the explicit calculation 
of the $1/N$ expansion of the critical exponents and most importantly, the improvement of the 
hierarchical approximation. 

The HM is a good laboratory to explore new ideas. In particular the idea of introducing a large field cutoff 
\cite{convpert} in order to produce a perturbative series with a better large order behavior or to generate  
a new type of RG flows when the field cutoff is lowered \cite{ymerge} according to the simple scheme
\begin{equation}
\int_{||\phi||<\phi_{max}} D\phi\  {\rm e}^{-S}=\int_{||\phi||<\phi_{max}/\xi} D\phi\ {\rm e}^{-S'(\xi)}\ .
\end{equation}

Our review emphasizes the connection with the ERGE and the possibility to move among improvements 
of the LPA by taking the limit $\scale \rightarrow 1$ or discretizing continuous RG flow equations. 
It should be reminded that before approximations are made, ERGE can be formulated several equivalent ways. For instance the exact Polchinski equation is related to 
flow equations for the effective action \cite{wetterich92,ellwanger93,morris93} by a 
Legendre transform. However, their derivative expansions are inequivalent and there is 
room for optimization \cite{litim05}. In contrast, for the HM, despite a program of systematic improvement, exact equations for lattice models remain to be found. 
We hope that some communication can be established between these approaches in the future. 

Except for section \ref{sec:susy}, most of numerical work was done for $D=3$. Considering models with 
bosons and fermions producing effects of opposite signs on the effective potential may help resolve controversial 
issues \cite{holland04}  regarding the triviality and the stability of the effective potential in $D=4$.

Finally, it would be desirable to apply similar methods for gauge theories. 
In  quantum chromodynamics, understanding how asymptotic freedom and confinement can 
be smoothly connected in a single theory amounts to constructing the renormalization group (RG) flows of the theory far away from the two fixed points of interest. In this example it is expected that nothing dramatic takes 
place as we interpolate between the two regimes, however, understanding confinement in terms of 
the weak coupling variables remains a challenge. 
Proving the existence of a mass gap in Yang-Mills theories remains one of the 
Clay Millennium Prize Problems. Recent efforts have made to understand the basic problems by using decimation procedures on 
the lattice \cite{tomboulis03,tomboulis05,tomboulis06} by using ERGE functional methods 
\cite{litim98,pawlowski03,pawlowski05,tomboulis06,dlerge}. As the Large Hadron Collider 
is almost in operation, a more solid understanding of gauge theories should 
be a priority in the theory community.

\ack
This research was supported in part by the Department of Energy
under Contract No. FG02-91ER40664 and also by a 
Faculty Scholar Award at The University of Iowa and a residential
appointment at the Obermann Center for Advanced Studies at the
University of Iowa. 
We thank the participants of the joint Math-Physics seminar at the 
University of Iowa for interesting suggestions. We thank the organizers and the participants of 
the third ERG conference in Lefkada for providing many motivations and ideas. We thank P. Wittwer and G. Gallavotti 
for explaining their work to us. We thank C. Bervillier and D. Litim for sharing results prior to publication. 
We thank J. Gaite for stimulating questions. We thank H. Sonoda for many suggestions 
to improve the manuscript in particular regarding discussions of the scaling hypothesis. 
We thank B. Oktay, M. H. Reno, A. Abdesselam, C. Bervillier, D. Litim and J. Pawlowski for comments on earlier versions of this manuscript.


\begin{thebibliography}{100}
\expandafter\ifx\csname url\endcsname\relax
  \def\url#1{{\tt #1}}\fi
\expandafter\ifx\csname urlprefix\endcsname\relax\def\urlprefix{URL }\fi
\providecommand{\eprint}[2][]{\url{#2}}

\bibitem{wilson71a}
Wilson K 1971 {\em Phys.\ Rev.\ D\/} {\bf 3} 1818

\bibitem{wilson71bone}
Wilson K~G 1971 {\em Phys. Rev. B\/} {\bf 4} 3174--3183

\bibitem{wilson71b}
Wilson K 1971 {\em Phys. Rev. B\/} {\bf 4} 3184--3205

\bibitem{wilson74}
Wilson K and Kogut J 1974 {\em Phys.\ Rep.\/} {\bf 12} 75

\bibitem{dyson69}
Dyson F 1969 {\em Comm.\ Math.\ Phys.\/} {\bf 12} 91

\bibitem{baker72}
Baker G 1972 {\em Phys.\ Rev.\ B\/} {\bf 5} 2622

\bibitem{kaufman81}
Kaufman M and Griffiths R~B 1981 {\em Phys. Rev. B\/} {\bf 24} 496--498

\bibitem{bagnuls00}
Bagnuls C and Bervillier C 2001 {\em Phys. Rept.\/} {\bf 348} 91
  (\textit{Preprint} \eprint{hep-th/0002034})

\bibitem{berges00}
Berges J, Tetradis N and Wetterich C 2002 {\em Phys. Rept.\/} {\bf 363}
  223--386 (\textit{Preprint} \eprint{hep-ph/0005122})

\bibitem{pawlowski05}
Pawlowski J~M 2005  (\textit{Preprint} \eprint{hep-th/0512261})

\bibitem{polchinski84}
Polchinski J 1984 {\em Nucl. Phys.\/} {\bf B231} 269--295

\bibitem{hasenfratz86}
Hasenfratz A and Hasenfratz P 1986 {\em Nucl. Phys.\/} {\bf B270} 687--701

\bibitem{felder87}
Felder G 1987 {\em Commun. Math. Phys.\/} {\bf 111} 101--121

\bibitem{gottker99}
Gottker-Schnetmann J 1999  (\textit{Preprint} \eprint{cond-mat/9909418})

\bibitem{litim00}
Litim D~F 2000 {\em Phys. Lett.\/} {\bf B486} 92--99 (\textit{Preprint}
  \eprint{hep-th/0005245})

\bibitem{oktayphd}
Oktay M~B 2001 Nonperturbative methods for hierarchical models (Ph. D. thesis),
  UMI-30-18602

\bibitem{bervillier04}
Bervillier C 2004 {\em Phys. Lett.\/} {\bf A332} 93--100 (\textit{Preprint}
  \eprint{hep-th/0405025})

\bibitem{on06}
Godina J~J, Li L, Meurice Y and Oktay M~B 2006 {\em Phys. Rev.\/} {\bf D73}
  047701 (\textit{Preprint} \eprint{hep-th/0511194})

\bibitem{bervillier07}
Bervillier C, Juttner A and Litim D~F 2007  (\textit{Preprint}
  \eprint{hep-th/0701172})




\bibitem{meurice93}
Meurice Y 1993  (\textit{Preprint} \eprint{hep-th/9307128})

\bibitem{marseille93}
Meurice Y 1994 {\em Proceedings of the International Europhysics Conference on
  High-Energy Physics, Marseille, France, Jul 22-28, 1993\/} ed Carr J and
  Perrottet M (Editions Frontieres) p~89 presented at International Europhysics
  Conference on High Energy Physics, Marseille, France, 22-28 Jul 1993

\bibitem{kbook}
Kleinert H and Schulte-Frohlinde V 2001 {\em Critical properties of
  phi**4-theories\/} (River Edge, USA)

\bibitem{pelissetto00b}
Pelissetto A and Vicari E 2002 {\em Phys. Rept.\/} {\bf 368} 549--727
  (\textit{Preprint} \eprint{cond-mat/0012164})

\bibitem{bagnuls90}
Bagnuls C and Bervillier C 1990 {\em Phys. Rev. B\/} {\bf 41} 402--406

\bibitem{bagnuls97}
Bagnuls C and Bervillier C 1997 {\em J. Phys. Stud.\/} {\bf 1} 366

\bibitem{bagnuls00b}
Bagnuls C and Bervillier C 2000 {\em Condensed Matter Phys.\/} {\bf 3} 559
  (\textit{Preprint} \eprint{hep-th/0002254})

\bibitem{osc1}
Meurice Y, Ordaz G and Rodgers V~G~J 1995 {\em Phys.\ Rev.\ Lett.\/} {\bf 75}
  4555

\bibitem{osc2}
Meurice Y, Niermann S and Ordaz G 1997 {\em J.\ Stat.\ Phys.\/} {\bf 87} 363

\bibitem{jogi98}
Jogi P, Sornette D and Blank M 1998 {\em Phys. Rev. E\/} {\bf 57} 120--134

\bibitem{collet78}
Collet P and Eckmann J 1978 {\em A Renormalization Group Analysis of the
  Hierarchical Model in Statistical Mechancs\/} (Berlin: Springer-Verlag)

\bibitem{wegner72}
Wegner F 1972 {\em Phys. Rev. B\/} {\bf 3} 4529

\bibitem{wilson72}
Wilson K 1972 {\em Phys.\ Rev.\ D\/} {\bf 6} 419

\bibitem{brez}
Brezin E, Guillou J~L and Zinn-Justin J 1976 {\em Phase Transition and Critical
  Phenomena\/} vol~6 (London: Academic Press)

\bibitem{zinnjustinbook}
Zinn-Justin J 2002 {\em Int. Ser. Monogr. Phys.\/} {\bf 113} 1--1054

\bibitem{cberge}
Bervillier C 2006 Status of the derivative expansion in the exact
  renormalization group equation talk given at the 3rd International Conference
  on ERG, http://www.cc.uoa.gr/~papost/Bervillier.pdf
  
\bibitem{litim07}
Litim D~F 2007  (\textit{Preprint}
  \eprint{arxiv:0704.1514})

\bibitem{litim98}
Litim D~F and Pawlowski J~M 1998  (\textit{Preprint} \eprint{hep-th/9901063})

\bibitem{tomboulis03}
Tomboulis E~T 2004 {\em Nucl. Phys. Proc. Suppl.\/} {\bf 129} 724--726
  (\textit{Preprint} \eprint{hep-lat/0309006})

\bibitem{pawlowski03}
Pawlowski J~M, Litim D~F, Nedelko S and von Smekal L 2004 {\em Phys. Rev.
  Lett.\/} {\bf 93} 152002 (\textit{Preprint} \eprint{hep-th/0312324})

\bibitem{tomboulis05}
Tomboulis E~T 2006 {\em PoS\/} {\bf LAT2005} 311 (\textit{Preprint}
  \eprint{hep-lat/0509116})

\bibitem{tomboulis06}
Tomboulis E~T and Velytsky A 2006  (\textit{Preprint} \eprint{hep-lat/0609047})

\bibitem{dlerge}
Litim D 2006 Functional flows for qcd talk given at the 3rd International
  Conference on ERG, http://www.cc.uoa.gr/~papost/LITIM.pdf

\bibitem{fukuda74}
Fukuda R and Kyriakopoulos E 1975 {\em Nucl. Phys.\/} {\bf B85} 354

\bibitem{wegner76}
Wegner F~J  In *Phase Transitions and Critical Phenomena, Vol.6*, London 1976,
  7-124

\bibitem{niemeijer76}
Niemeijer T and van Leeuwen J 1976 {\em Phase Transitions and Critical
  Phenomena, vol. 6\/} ed Domb C and Green M (New York: Academic Press)

\bibitem{cardy96}
Cardy J~L  Cambridge, UK: Univ. Pr. (1996) 238 p. (Cambridge lecture notes in
  physics: 3)

\bibitem{kadanoff66}
Kadanoff L 1966 {\em Physics\/} {\bf 2} 263

\bibitem{bell75}
Bell T~L and Wilson K~G 1975 {\em Phys. Rev. B\/} {\bf 11} 3431--3444

\bibitem{kw86}
Koch H and Wittwer P 1986 {\em Commun. Math. Phys.\/} {\bf 106} 495--532

\bibitem{koch91}
Koch H and Wittwer P 1991 {\em Commun. Math. Phys.\/} {\bf 138} 537--568

\bibitem{gallavotti78}
Gallavotti G 1978 {\em Memorie dell' Accademia dei Lincei\/} {\bf 15} 23

\bibitem{wegner74}
Wegner F 1974 {\em J. Phys. A\/} {\bf 7} 2098

\bibitem{fam}
Meurice Y and Ordaz G 1996 {\em J. Phys. A (Letter to the Editor)\/} {\bf 29}
  L635

\bibitem{kw88}
Koch H and Wittwer P 1988 {\em Nonlinear Evolution and Chaotic Phenomena\/} vol
  176 ed Gallavotti G and Zweifel P (Plenum)

\bibitem{gam3rapid}
Godina J, Meurice Y and Oktay M 1998 {\em Phys. Rev. D\/} {\bf 57} R6581

\bibitem{gam3}
Godina J, Meurice Y and Oktay M 1999 {\em Phys. Rev. D\/} {\bf 59} 096002

\bibitem{litim02}
Litim D~F 2002 {\em Nucl. Phys.\/} {\bf B631} 128--158 (\textit{Preprint}
  \eprint{hep-th/0203006})

\bibitem{thouless69}
Thouless D 1969 {\em Phys. Rev.\/} {\bf 187} 732--733

\bibitem{spencer82}
Frohlich J and Spencer T 1981 {\em Commun. Math. Phys\/} {\bf 84} 87--101

\bibitem{simon81}
Simon B and Sokal A 1981 {\em J. Stat. Phys.\/} {\bf 25} 679

\bibitem{freund}
Brekke L and Freund P~G~O 1993 {\em Phys. Rept.\/} {\bf 233} 1--66

\bibitem{missarov88}
Missarov M~D 1988  CPT-88/P-2151

\bibitem{cuerna}
Meurice Y 1989  Proc. of 3rd Mexican School of Particles and Fields, Oaxtepec,
  Morales, Mexico, Dec 5-16, 1988

\bibitem{lucio90}
Lucio J~L and Meurice Y 1991 {\em Mod. Phys. Lett.\/} {\bf A6} 1199--1202

\bibitem{lerner89}
Lerner E~Y and Missarov M~D 1989 {\em Theor. Math. Phys.\/} {\bf 78} 177--184

\bibitem{meuricejmp95}
Meurice Y 1995 {\em Jour. Math. Phys.\/} {\bf 36} 1812

\bibitem{bleher75}
Bleher P and Sinai Y 1975 {\em Comm. Math. Phys.\/} {\bf 45} 247

\bibitem{collet77}
Collet P and Eckmann J~P 1977 {\em Commun. Math. Phys.\/} {\bf 55} 67--96

\bibitem{abd06}
Abdesselam A 2006 A complete renormalization group trajectory between two fixed
  points
  \urlprefix\url{http://www.citebase.org/abstract?id=oai:arXiv.org:math-ph/061%
0018}

\bibitem{kupiainen83}
Gawedzki K and Kupiainen A 1983 {\em Commun. Math. Phys.\/} {\bf 89} 191--220

\bibitem{koch94}
Koch H and Wittwer P 1994 {\em Commun. Math. Phys.\/} {\bf 164} 627--647

\bibitem{koch95}
Koch H and Wittwer P 1995 {\em Math. Phys. Electr. Jour.\/} {\bf 1} Paper 6

\bibitem{baker77}
Baker G~A and Golner G~R 1977 {\em Phys. Rev. B\/} {\bf 16} 2081--2094

\bibitem{kim77}
Kim D and Thompson C~J 1977 {\em J. Phys.\/} {\bf A10} 1579--1598

\bibitem{collet77b}
Collet P, Eckmann J~P and Hirsbrunner B 1977 {\em Phys. Lett.\/} {\bf B71}
  385--386

\bibitem{pinn94}
Pinn K, Pordt A and Wieczerkowski C 1994 {\em J. Statist. Phys.\/} {\bf 77} 977
  (\textit{Preprint} \eprint{hep-lat/9402020})

\bibitem{hyper}
Godina J~J, Meurice Y and Oktay M 2000 {\em Phys. Rev. D\/} {\bf 61} 114509

\bibitem{guide}
Godina J, Meurice Y, Oktay M and Niermann S 1998 {\em Phys. Rev. D\/} {\bf 57}
  6326

\bibitem{scalingjsp}
Meurice Y and Niermann S 2002 {\em J. Statist. Phys.\/} {\bf 108} 213--246
  (\textit{Preprint} \eprint{cond-mat/0105380})

\bibitem{susskind78}
Susskind L 1979 {\em Phys. Rev.\/} {\bf D20} 2619--2625

\bibitem{numerr}
Meurice Y and Oktay B 2001 {\em Phys. Rev. D\/} {\bf 63} 016005

\bibitem{ymunpublished}
Meurice Y unpublished

\bibitem{leguillou90}
LeGuillou J~C and Zinn-Justin J 1990 {\em Large-Order Behavior of Perturbation
  Theory\/} (Amsterdam: North Holland)

\bibitem{pernice98}
Pernice S and Oleaga G 1998 {\em Phys. Rev. D\/} {\bf 57} 1144

\bibitem{convpert}
Meurice Y 2002 {\em Phys. Rev. Lett.\/} {\bf 88} 141601 (\textit{Preprint}
  \eprint{hep-th/0103134})

\bibitem{bacus}
Bacus B, Meurice Y and Soemadi A 1995 {\em J. Phys. A\/} {\bf 28} L381

\bibitem{arbacc}
Meurice Y 2002 {\em J. Phys.\/} {\bf A35} 8831--8846 (\textit{Preprint}
  \eprint{quant-ph/0202047})

\bibitem{tractable}
Li L and Meurice Y 2005 {\em J. Phys. A\/} {\bf 38} 8139--8153
  (\textit{Preprint} \eprint{hep-th/0506038})

\bibitem{bagnuls01}
Bagnuls C and Bervillier C 2002 {\em Phys. Rev.\/} {\bf E65} 066132
  (\textit{Preprint} \eprint{hep-th/0112209})

\bibitem{pelissetto98}
Pelissetto A, Rossi P and Vicari E 1998 {\em Phys. Rev.\/} {\bf E58} 7146--7150

\bibitem{asymp06}
Li L and Meurice Y 2006 {\em J. Phys.\/} {\bf A39} 8681--8698
  (\textit{Preprint} \eprint{hep-th/0507196})

\bibitem{optim}
Kessler B, Li L and Meurice Y 2004 {\em Phys. Rev.\/} {\bf D69} 045014
  (\textit{Preprint} \eprint{hep-th/0309022})

\bibitem{litim05}
Litim D~F 2005 {\em JHEP\/} {\bf 07} 005 (\textit{Preprint}
  \eprint{hep-th/0503096})

\bibitem{litim02b}
Litim D~F and Pawlowski J~M 2002 {\em Phys. Lett.\/} {\bf B546} 279--286
  (\textit{Preprint} \eprint{hep-th/0208216})

\bibitem{alexandre01}
Alexandre J, Polonyi J and Sailer K 2002 {\em Phys. Lett.\/} {\bf B531}
  316--320 (\textit{Preprint} \eprint{hep-th/0111152})

\bibitem{polonyi04}
Polonyi J and Sailer K 2005 {\em Phys. Rev.\/} {\bf D71} 025010
  (\textit{Preprint} \eprint{hep-th/0410271})

\bibitem{wegner72b}
Wegner F~J and Houghton A 1973 {\em Phys. Rev.\/} {\bf A8} 401--412

\bibitem{nicoll76}
Nicoll J~F, Chang T~S and Stanley H~E 1976 {\em Phys. Rev. A\/} {\bf 13}
  1251--1264

\bibitem{riedel86}
Riedel E, Golner G~R and Newman K~E 1985 {\em Annals Phys.\/} {\bf 161}
  178--238

\bibitem{morris93}
Morris T~R 1994 {\em Int. J. Mod. Phys.\/} {\bf A9} 2411--2450
  (\textit{Preprint} \eprint{hep-ph/9308265})

\bibitem{comellas97}
Comellas J and Travesset A 1997 {\em Nucl. Phys.\/} {\bf B498} 539--564
  (\textit{Preprint} \eprint{hep-th/9701028})

\bibitem{ball94}
Ball R~D, Haagensen P~E, Latorre Jose I and Moreno E 1995 {\em Phys. Lett.\/}
  {\bf B347} 80--88 (\textit{Preprint} \eprint{hep-th/9411122})

\bibitem{litim01b}
Litim D~F 2001 {\em Phys. Rev.\/} {\bf D64} 105007 (\textit{Preprint}
  \eprint{hep-th/0103195})

\bibitem{litim01}
Litim D~F 2001 {\em Int. J. Mod. Phys.\/} {\bf A16} 2081--2088
  (\textit{Preprint} \eprint{hep-th/0104221})

\bibitem{morris05}
Morris T~R 2005 {\em JHEP\/} {\bf 07} 027 (\textit{Preprint}
  \eprint{hep-th/0503161})

\bibitem{hasenfratz88}
Hasenfratz A and Hasenfratz P 1988 {\em Nucl. Phys.\/} {\bf B295} 1

\bibitem{sokal94}
Sokal A~D, van Enter A~C~D and Fernandez R 1994 {\em J. Statist. Phys.\/} {\bf
  72} 879--1167 (\textit{Preprint} \eprint{hep-lat/9210032})

\bibitem{bricmont01}
Bricmont J, Kupiainen A and Lefevere R 2001 {\em Phys. Rept.\/} {\bf 348} 5--31

\bibitem{arnold88}
Arnold V 1988 {\em Geometrical Methods in the Theory of Ordinary Differential
  Equations\/} (New York: Springer-Verlag)

\bibitem{wegn}
Wegner F 1976 {\em Phase Transition and Critical Phenomena\/} vol~6 ed Domb C
  and Green M (London: Academic Press)

\bibitem{poincare92}
Poincar\'e H 1892 {\em Les Methodes Nouvelles de la Mecanique Celeste\/}
  (Paris: Gauthier-Villars)

\bibitem{sonoda90}
Sonoda H 1991 {\em Nucl. Phys.\/} {\bf B352} 585--600

\bibitem{smalld}
Meurice Y 2001 {\em Phys. Rev. E\/} {\bf 63} 055101(R)

\bibitem{small03}
Meurice Y 2004 {\em Phys. Rev.\/} {\bf E69} 056108 (\textit{Preprint}
  \eprint{cond-mat/0312188})

\bibitem{gaite96}
Gaite J and O'Connor D 1996 {\em Phys. Rev. D\/} {\bf 54} 5163

\bibitem{polyzou80}
Polyzou W 1980 {\em Jour. Math. Phys.\/} {\bf 21} 506

\bibitem{kowalski81}
Kowalski K, Polyzou W and Redish E 1981 {\em Jour. Math. Phys.\/} {\bf 22} 1965

\bibitem{hid}
Sonoda H private communication

\bibitem{jsp02}
Meurice Y and Niermann S 2002 {\em Jour. Stat. Phys.\/} {\bf 108} 213

\bibitem{ht4}
Godina J~J, Meurice Y and Niermann S 1998 {\em Nucl. Phys. B\/} {\bf 519} 737

\bibitem{thooft73}
't~Hooft G 1973 {\em Nucl. Phys.\/} {\bf B61} 455--468

\bibitem{thooft76}
't~Hooft G 1976 {\em Phys. Rev.\/} {\bf D14} 3432--3450

\bibitem{dual}
Meurice Y and Niermann S 1999 {\em Phys. Rev. E\/} {\bf 60} 2612

\bibitem{glimm87}
Glimm J and Jaffe A 1987 {\em Quantum Physics\/} (New York: Springer-Verlag)

\bibitem{nickel80}
Nickel B 1982 {\em Phase Transitions, Cargese 1980\/} (New York: Plenum Press)

\bibitem{wilson03}
Glazek S~D and Wilson K~G 2002 {\em Phys. Rev. Lett.\/} {\bf 89} 230401
  (\textit{Preprint} \eprint{hep-th/0203088})

\bibitem{braaten03}
Braaten E and Hammer H~W 2003 {\em Phys. Rev. Lett.\/} {\bf 91} 102002
  (\textit{Preprint} \eprint{nucl-th/0303038})

\bibitem{parisi88}
Parisi G 1988 {\em Statistical Field Theory\/} (New York: Addison Wesley)

\bibitem{universality03}
Meurice Y and Oktay M~B 2004 {\em Phys. Rev.\/} {\bf D69} 125016
  (\textit{Preprint} \eprint{hep-th/0401144})

\bibitem{goldberg90}
Goldberg H 1990 {\em Phys. Lett.\/} {\bf B246} 445--450

\bibitem{cornwall90}
Cornwall J~M 1990 {\em Phys. Lett.\/} {\bf B243} 271--278

\bibitem{zakharov91}
Zakharov V~I 1991 {\em Phys. Rev. Lett.\/} {\bf 67} 3650--3653

\bibitem{voloshin92}
Voloshin M~B 1992 {\em Nucl. Phys.\/} {\bf B383} 233--248

\bibitem{Campostrini99}
Campostrini M, Pelissetto A, Rossi P and Vicari E 1999 {\em Phys. Rev.\/} {\bf
  E60} 3526--3563 (\textit{Preprint} \eprint{cond-mat/9905078})

\bibitem{guida96}
Guida R and Zinn-Justin J 1997 {\em Nucl. Phys.\/} {\bf B489} 626--652
  (\textit{Preprint} \eprint{hep-th/9610223})

\bibitem{litim06}
Litim D~F, Pawlowski J~M and Vergara L 2006  (\textit{Preprint}
  \eprint{hep-th/0602140})

\bibitem{Morris96}
Morris T~R 1997 {\em Nucl. Phys.\/} {\bf B495} 477--504 (\textit{Preprint}
  \eprint{hep-th/9612117})

\bibitem{ma73}
Ma S~K 1973 {\em Phys. Lett.\/} {\bf A43} 475--476

\bibitem{david84}
David F, Kessler D~A and Neuberger H 1984 {\em Phys. Rev. Lett.\/} {\bf 53}
  2071

\bibitem{complexs}
Meurice Y 2003 {\em Phys. Rev.\/} {\bf D67} 025006 (\textit{Preprint}
  \eprint{hep-th/0208181})

\bibitem{morris94}
Morris T~R 1994 {\em Phys. Lett.\/} {\bf B334} 355--362 (\textit{Preprint}
  \eprint{hep-th/9405190})

\bibitem{tetradis93}
Tetradis N and Wetterich C 1994 {\em Nucl. Phys.\/} {\bf B422} 541--592
  (\textit{Preprint} \eprint{hep-ph/9308214})

\bibitem{litim95}
Litim D and Tetradis N 1995  (\textit{Preprint} \eprint{hep-th/9501042})

\bibitem{tetradis95}
Tetradis N and Litim D~F 1996 {\em Nucl. Phys.\/} {\bf B464} 492--511
  (\textit{Preprint} \eprint{hep-th/9512073})

\bibitem{dattanasio97}
D'Attanasio M and Morris T~R 1997 {\em Phys. Lett.\/} {\bf B409} 363--370
  (\textit{Preprint} \eprint{hep-th/9704094})

\bibitem{kubyshin02}
Kubyshin Y~A, Neves R and Potting R 2002 {\em Int. J. Mod. Phys.\/} {\bf A17}
  4871--4902 (\textit{Preprint} \eprint{hep-th/0202199})

\bibitem{bmb}
Bardeen W~A, Moshe M and Bander M 1984 {\em Phys. Rev. Lett.\/} {\bf 52} 1188

\bibitem{david85}
David F, Kessler D~A and Neuberger H 1985 {\em Nucl. Phys.\/} {\bf B257}
  695--728

\bibitem{cookpro}
Cook J, Li L and Meurice Y work in progress

\bibitem{taibleson}
Taibleson M 1975 {\em Fourier Analysis on Local Fields\/} (Princeton University
  Press)

\bibitem{fine}
Fine N 1949 {\em Trans. Amer . Math.\/} {\bf 65} 372

\bibitem{serre}
Serre J 1973 {\em A Course in Arithmetic\/} (New York: Springer)

\bibitem{sym91}
Meurice Y 1991 {\em Phys. Lett.\/} {\bf B265} 377--381

\bibitem{dorlas90}
Dorlas T~C 1991 {\em Commun. Math. Phys.\/} {\bf 136} 169--194

\bibitem{lerner94}
Lerner E~Y and Missarov M~D 1994 {\em Theor. Math. Phys.\/} {\bf 101}
  1353--1360

\bibitem{wess74}
Wess J and Zumino B 1974 {\em Nucl. Phys.\/} {\bf B70} 39--50

\bibitem{susy99}
Meurice Y 1999 Non-perturbative fine-tuning in approximately supersymmetric
  models (\textit{Preprint} \eprint{hep-lat/9906033})

\bibitem{ymerge}
Meurice Y 2006 Rg analysis of the interpolation between the weak and strong
  coupling talk given at the 3rd International Conference on ERG,
  http://www.cc.uoa.gr/~papost/Meurice.pdf

\bibitem{wetterich92}
Wetterich C 1993 {\em Phys. Lett.\/} {\bf B301} 90--94

\bibitem{ellwanger93}
Ellwanger U 1994 {\em Z. Phys.\/} {\bf C62} 503--510 (\textit{Preprint}
  \eprint{hep-ph/9308260})

\bibitem{holland04}
Holland K 2005 {\em Nucl. Phys. Proc. Suppl.\/} {\bf 140} 155--161
  (\textit{Preprint} \eprint{hep-lat/0409112})

\end{thebibliography}
\section*{References}
\providecommand{\newblock}{}

\end{document}